\documentclass[11pt,a4paper,x11names]{article}
\usepackage{jheppub}

\usepackage{amsfonts}
\usepackage{tikz-cd,tikz}
\usepackage{physics}
\usepackage{float}
\usepackage{listings}
\usepackage{calc}
\usepackage{multicol}

\usepackage{subcaption}

\DeclareFixedFont{\ttb}{T1}{txtt}{bx}{n}{12}
\DeclareFixedFont{\ttm}{T1}{txtt}{m}{n}{12}
\DeclareFixedFont{\ttms}{T1}{txtt}{m}{n}{9}
\DeclareFixedFont{\ttmss}{T1}{txtt}{m}{n}{7}

\definecolor{deepblue}{rgb}{0.0, 0.0, 0.55} 
\definecolor{shadecolor}{rgb}{0.96,0.96,0.91}
\definecolor{shadecolor2}{rgb}{1,1,0.99}

\newcommand{\mcomment}[1]{}
\newcommand\mathstyle{\lstset{
language=Mathematica,
basicstyle={\scriptsize\def\fvm@Scale{.5}\fontfamily{fvm}\selectfont},
otherkeywords={self},
keywordstyle=\ttb\scriptsize\color{deepblue},
emph={MyClass,__init__},
emphstyle=\ttb\color{deepred},
backgroundcolor=\color{pink!20!white},
stringstyle=\color{deepgreen},
commentstyle=\color{SkyBlue3!70!PaleGreen4},
frame=tb,
showstringspaces=false
}}

\newwrite\todofile
\immediate\openout\todofile=\jobname.tdo

\newcounter{todocounter}

\newcommand{\printtodos}{
        \section*{To-Do List}
        \immediate\closeout\todofile
        \input{\jobname.tdo}
}
\lstnewenvironment{mathematica}[1][]
{
\mathstyle
\lstset{#1}
}
{}

\newcommand{\la}[1]{\label{#1}}
\newcommand{\eq}[1]{(\ref{#1})}
\newcommand{\nn}{\nonumber}

    \newcommand{\beq}{\begin{equation}}
    \newcommand{\eeq}{\end{equation}}
    \newcommand\beqa{\begin{eqnarray}}
    \newcommand\eeqa{\end{eqnarray}}
\newcommand\bea{\begin{array}}
\newcommand\eea{\end{array}}

\newcommand{\ii}{i}
\newcommand{\bfB}{\mathbf{B}}

\newcommand{\algsu}{\mathfrak{su}}

\newlength{\widthLOne}
\newlength{\widthLTwo}
\newcommand{\bP}{\mathbf{P}}

\newcommand{\bQ}{\mathbf{Q}}

\newcommand{\bB}{\textbf{B}}

\newcommand{\betheQ}{\mathbb{Q}}
\renewcommand{\bfB}{\mathbf{B}}
\newcommand{\bfR}{\mathbf{R}}

\newcommand{\AdS}{\text{AdS}}
\newcommand{\CFT}{\text{CFT}}
\newcommand{\Sphere}{\mathrm{S}}
\newcommand{\Torus}{\mathrm{T}}

\newcommand{\sL}{\mbox{\tiny L}}
\newcommand{\sR}{\mbox{\tiny R}}
\newcommand{\sI}{\mbox{\tiny I}}

\newcommand{\sRL}{\sR\sL}
\newcommand{\sLL}{\sL\sL}

\newcommand{\fQ}{{Q}}

\newcommand{\es}{\emptyset}
\newcommand{\sigmaBES}{\sigma_{\text{BES}}}

\newcommand{\Li}{\text{Li}}
\newcommand{\FS}{\text{FS}}

\newcommand{\algpsu}{\mathfrak{psu}}

\newcommand{\mcS}{\mathcal{S}}
\newcommand{\mcSd}{\dot{\mathcal{S}}}

\newcommand\IntUpperCC[1][1]{
    \begin{tikzpicture}[scale=0.2]
        \coordinate (center) at (1.1,0.15);
        \draw[black,-{>[scale=0.6]}] (0, 0.0) + (center) arc (0:180:0.5);
       $\int$
    \end{tikzpicture}
    }
\newcommand\IntUpperC[1][1]{
    \begin{tikzpicture}[scale=0.2]
        \coordinate (center) at (0.1,0.15);
        \draw[black,-{>[scale=0.6]}] (0, 0.0) + (center) arc (180:0:0.5);
       $\int$
    \end{tikzpicture}
    }
    \newcommand\IntLowerC[1][1]{
    \begin{tikzpicture}[scale=0.2]
        \coordinate (center) at (1.1,0.55);
        \draw[black,-{>[scale=0.6]}] (0, 0.0) + (center) arc (0:-180:0.5);
       $\int$
    \end{tikzpicture}
    }
    \newcommand\IntLowerCC[1][1]{
    \begin{tikzpicture}[scale=0.2]
        \coordinate (center) at (0.1,0.55);
        \draw[black,-{>[scale=0.6]}] (0, 0.0) + (center) arc (180:360:0.5);
      $\int$
    \end{tikzpicture}}

\def\[{\left[}
\def\]{\right]}
\def\({\left(}
\def\){\right)}
\def\<{\left<}
\def\>{\right>}

\title{Demystifying the Massless Sector in AdS${}_3$ Quantum Spectral Curve}
\author{Simon Ekhammar${}^{\bullet \gamma}$, Nikolay Gromov${}^\bullet$ and Bogdan Stefa\'nski, jr.${}^\circ$}
\affiliation{
    $\bullet$ Mathematics Department, King's College London,
    The Strand, London WC2R 2LS, UK\\
    $\circ$ Centre for Mathematical Science, City St. Geroge's, University of London, Northampton Square, EC1V 0HB London, UK\\
    $\gamma$ Department of Physics and Astronomy, Uppsala University,
Box 516, SE-751 20 Uppsala, Sweden 
}\emailAdd{simon.ekhammar@kcl.ac.uk}
 \emailAdd{nikolay.gromov@kcl.ac.uk}
 \emailAdd{bogdan.stefanski.1@city.ac.uk}

\abstract{
We show that
in the asymptotic large-volume limit, the original proposal for Quantum Spectral Curve for $\AdS_3\times \Sphere^3\times \Torus^4$ with R-R flux
has a wider class of solutions, than studied previously. We argue that in this limit the QSC reduces to a finite set of Bethe equations for both massive and massless  particle types. We also find that the QSC imposes more constraining 
conditions on the dressing phases than previously known and we present solutions of those equations.
}

\begin{document}

\maketitle
\newpage
\section{Introduction}

The $\AdS_3/\CFT_2$ correspondence with small $(4,4)$ superconformal symmetry~\cite{Maldacena:1997re} remains a
significant challenge in the study of exact holography. While much is known about BPS quantities, nonprotected sectors remain poorly understood at generic points in the moduli space. As in certain higher-dimensional dual pairs~\cite{Beisert:2010jr}, there is strong evidence that integrable holographic methods will also apply to this setting. Light-cone quantisation of the Green-Schwarz superstring action on $\AdS_3\times \Sphere^3\times \Torus^4$ with R-R flux
near the BMN vacuum~\cite{Berenstein:2002jq} has been used to find the exact 2-body worldsheet S matrix together with crossing equations for the so-called dressing factors~\cite{Borsato:2013qpa,Borsato:2014exa,Borsato:2014hja}. Since this 2-body S-matrix satisfies the Yang-Baxter equation, the large-volume spectrum of worldsheet theory is then encoded into Asymptotic Bethe Ansatz (ABA) equations~\cite{Borsato:2016kbm,Borsato:2016xns}. However, unlike higher-dimensional integrable holographic models, $\AdS_3/\CFT_2$ duals have massless excitations in their spectrum, which makes it harder to apply large-volume methods, such as the ABA, to explicitly determine the spectrum because massless and mixed-mass wrapping effects are no longer exponentially suppressed~\cite{Abbott:2015pps}.

To overcome these challenges, one needs to develop exact finite-volume methods, such as the Quantum Spectral Curve (QSC) or the Thermodynamic Bethe Ansatz (TBA). The QSC was first introduced in~\cite{Gromov:2014caa} in the context of maximally supersymmetric $\AdS_5/\CFT_4$ and has since been used to precisely determine many non-protected observables in theories such as planar ${\cal N}=4$ super Yang-Mills (SYM) theory or ABJM theory~\cite{Aharony:2008ug}.
The QSC is a set of functional equations for so-called Q-functions, which are
additionally constrained by analyticity properties.
Recently, a QSC was conjectured for string theory  on $\AdS_3\times \Sphere^3\times \Torus^4$ with R-R flux~\cite{Cavaglia:2021eqr,Ekhammar:2021pys}. An important difference of this QSC, unlike its higher-dimensional cousins, is the presence of non-square-root branch points~\cite{Borsato:2013hoa}.\footnote{This  generalisation of the allowed analytic properties of Q-functions significantly expands the space of possible QSCs and so deserves further detailed study as part of a programme for a classification of QSCs.} As a concrete application of this formalism, the first numerical predictions for energies of non-protected Konishi-like states and their higher-spin generalisations were obtained using this QSC in~\cite{Cavaglia:2022xld} to high order in a weak-coupling analytic expansion.\footnote{
More recently some weak coupling predictions, for different states, became available via an alternative TBA formalism~\cite{Brollo:2023pkl,Brollo:2023rgp}.}

It is believed that solutions of the QSC should be in one-to-one correspondence with super-conformal primaries.
In~\cite{Cavaglia:2021eqr,Ekhammar:2021pys}, a large-volume limit was found, in which a certain class of solutions of the QSC reduced to the \textit{massive} sector Bethe equations~\cite{Borsato:2013qpa}. However, the massless sector remained largely unexplored even though the first steps to identify the corresponding QSC solutions
were proposed in \cite{Ekhammar:2021pys}.
In this paper we will study a new class of solutions of the $\AdS_3$ QSC,
to a certain extent inspired by the recent discovery of massless modes in the AdS$_5\times $S$^5$ QSC in the Regge regime~\cite{Ekhammar:2024neh},
which in the large volume limit gives the full set of Bethe equations, including the massless excitations~\cite{Borsato:2016xns}. In this limit, the QSC analyticity properties reduce to simple discontinuity constraints on certain scalar factors that appear in the Q-functions. As in higher-dimensional models, the more familiar S-matrix dressing phases can be constructed from these QSC building blocks. The QSC discontinuity constraints impose stricter restrictions on the corresponding dressing factors leading to a unique solution, ruling out any CDD-like factors~\cite{Castillejo:1955ed}.

These results are important for several reasons. Firstly, they demonstrate that the initial $\AdS_3$ QSC proposals are complete, in the sense that they include all types of particles in the large volume limit. Secondly, our findings provide asymptotic expressions for the constituents of the QSC, enabling both perturbative and numerical studies of these states using precise QSC tools. Finally, our results allow for the determination of all the CDD factors considered in previous studies and help resolve some discrepancies found in the existing literature.

This paper is organised as follows. In section~\ref{sec:qsc-gen} we review the R-R $\AdS_3$ QSC proposal~\cite{Cavaglia:2021eqr,Ekhammar:2021pys}, while in section~\ref{sec:aba-gen} we summarise the ABA equations for this theory~\cite{Borsato:2016xns}.
In Section~\ref{sec:MasslessABALimit}, we examine the large-volume limit of QSC for the case involving both massive and massless modes, discussing the form of the P and Q functions in this limit. Section~\ref{sec:DerivingBAE} derives all massive and massless Bethe equations from QSC. Section~\ref{sec:dressingpases} explores the crossing relations and presents explicit expressions for the dressing phases. Finally, Section~\ref{sec:conclusions} concludes the paper with a summary of our findings and their implications for further analysis.

\section{QSC generalities}
\label{sec:qsc-gen}
The structure of the QSC is governed to a large extent by the symmetries of the system. 
The global part of the small $(4,4)$ superconformal symmetry algebra  is $\algpsu(1,1|2)_{\sL}\times \algpsu(1,1|2)_{\sR}$, with L and R denoting the left and right sectors of the dual $\CFT_2$. The four Cartan generators consist of two $\algsu(1,1)_{\sI}$ charges $\Delta_{\sI}$ and two $\algsu(2)_{\sI}$ charges $J_{\sI}$, with I $=$ L, R. These can be conveniently combined into
\begin{equation}
    \Delta=\Delta_{\sL}+\Delta_{\sR}\,,\quad S=\Delta_{\sL}-\Delta_{\sR}\,,\quad J=J_{\sL}+J_{\sR}\,,\quad K=J_{\sL}-J_{\sR}\,,
\end{equation}
where $\Delta$ is the dimension, $S$ the spin and, by convention, $J$ the angular momentum along $\Sphere^3$ in which the Berenstein-Maldacena-Nastase (BMN)~\cite{Berenstein:2002jq} geodesic extends.

Correspondingly, the $\AdS_3$ QSC is built from two $\algpsu(1,1|2)$ Q-systems, each consisting of $2^4$ functions of the spectral parameter $u$. We will denote the left Q-system by
\beqa
\la{Qfull}&&\fQ_{\es|\es}= Q_{12|12} = 1,\\
&&\fQ_{a|\es}\equiv\bP_{a},\quad\quad\;\,\fQ_{\es|k}\equiv\bQ_{k},\\
\la{Qupdef}&&\fQ_{a|k},\quad\quad\quad\quad\quad Q^{a|k}\equiv\epsilon^{ab}\epsilon^{kl}Q_{b|l}\\
&&\fQ_{12|k}\equiv \bQ^l\epsilon_{lk},\quad\fQ_{a|12}\equiv \bP^b\epsilon_{ba}\;.
\eeqa
Above, $a,b,k,l=1,2$ and $\epsilon_{12}=\epsilon^{12}\equiv 1$ is the anti-symmetric Levi-Civita tensor so that $\epsilon_{ab}\epsilon^{bc} = -\delta^{c}_a$. The Q-functions satisfy conventional QQ-relations
\beqa \label{eq:QQrelations}
\nn \fQ_{A a \mid I} \fQ_{A \mid I i} & =&\fQ_{A a \mid I i}^{+} Q_{A \mid I}^{-}-Q_{A a \mid I i}^{-} Q_{A \mid I}^{+}\;, \\
\label{QQsys} Q_{12 \mid I} Q_{\emptyset \mid I} & =&Q_{1 \mid I}^{+} Q_{2 \mid I}^{-}-Q_{1 \mid I}^{-} Q_{2 \mid I}^{+}\;, \\
\nn Q_{A \mid 12} Q_{A \mid \emptyset} & =&Q_{A \mid 1}^{+} Q_{A \mid 2}^{-}-Q_{A \mid 1}^{-} Q_{A \mid 2}^{+}\;,
\eeqa
where $A$ and $I$ are multi-indices, for example:
\beq\label{eq:QQtwoeq}
Q_{a|i}^+-Q_{a|i}^- = \bQ_a \bP_i\,,\qquad\qquad
-\bQ^2 \bQ_1 = Q_{1|1}^+ Q_{2|1}^- - Q_{1|1}^- Q_{2|1}^+
\;.
\eeq
The identity $Q_{12|12}=\det Q_{a|b}$ is a consequence of the QQ-relations. Together with \eq{Qfull} and \eq{Qupdef} it leads to
\beq\label{Qupdown}
Q^{a|k}Q_{b|k}=\delta_b^a\;.
\eeq 
The right copy of the Q-system will be distinguished by using dotted indices, for example as $\fQ_{\dot a|\dot k}$ or $\bP_{\dot a}$. All Q-functions are assumed to be analytic in the upper-half $u$-plane on the defining sheet.

\paragraph{Asymptotics.}
We require all Q-functions to have power-like large-$u$ asymptotics  on their  defining sheet dictated by the quantum numbers of the state under consideration:
\begin{equation}\label{Ppow}
    \bP_{a} \simeq \mathbb{A}_a \,u^{M_{a}}\,,
    \quad
    \bP^{a} \simeq \mathbb{A}^{\dot{a}} \, u^{-M_a-1}
    \quad
    \bQ_{k} \simeq \mathbb{B}_k\,u^{\hat{M}_{k}}\,,
    \quad
    \bQ^{k} \simeq \mathbb{B}^k\,u^{-\hat{M}_{k}-1}\,,
\end{equation}
with  
\begin{align}\label{eq:QuantumNumbersAsymptotics}
    M_{a} &= \left\{-\frac{J}{2}-\frac{K}{2}-1,\frac{J}{2}+\frac{K}{2}\right\}-\frac{\hat{B}}{2}\,,
    \quad
    \hat{M}_{k} = \left\{\frac{\Delta}{2}+\frac{S}{2},-\frac{\Delta}{2}-\frac{S}{2}-1\right\}+\frac{\hat{B}}{2}\,,
    \\
    \label{eq:QuantumNumbersAsymptoticsDot}
    M_{\dot{a}} &= \left\{-\frac{J}{2}+\frac{K}{2},\frac{J}{2}-\frac{K}{2}-1\right\}-\frac{\check{B}}{2}\,,
    \quad
    \hat{M}_{\dot{k}} = \left\{\frac{\Delta}{2}-\frac{S}{2}-1,-\frac{\Delta}{2}+\frac{S}{2}\right\}+\frac{\check{B}}{2}\,.
\end{align}
From the full set of QQ-relations \eqref{eq:QQrelations} one can derive the following three particularly useful types of identities.

$\bullet$ Using $Q_{a|i}$ one can ``rotate" between $\bP$ and $\bQ$ according to
\beq \label{eq:RaisingLoweringQQ}
Q_{a \mid i}^{ \pm} \mathbf{Q}^i=\mathbf{P}_a\,,\quad
Q_{a \mid i}^{ \pm} \mathbf{P}^a=\mathbf{Q}_i\,,\quad
Q^{a \mid i \pm} \mathbf{Q}_i=\mathbf{P}^a\,,\quad
Q^{a \mid i \pm} \mathbf{P}_a=\mathbf{Q}^i\,.
\eeq

$\bullet$ Using QQ-relations and the fact that $Q_{\emptyset|\emptyset} = Q_{12|12} = 1$ it follows that
\begin{equation}\label{eq:QQPPrelation}
    \bQ_{k}\bQ^{k} = \bP_{a}\bP^{a} = \bQ_{\dot{k}}\bQ^{\dot{k}} = \bP_{\dot{a}}\bP^{\dot{a}} = 0\,.
\end{equation}

$\bullet$ Lastly, the prefactors ${\mathbb A}$ and ${\mathbb B}$ in \eq{Ppow}, as a consequence of the QQ-relations, are constrained to satisfy
\begin{equation}\label{eq:PrefactorsQuantumNumbers}
\begin{split}
{\mathbb A}_1 {\mathbb A}^1 & =-{\mathbb A}_2 {\mathbb A}^2=\frac{i}{4} \frac{(\Delta-J-K+S)(\Delta+J+K+S+2)}{J+K+1}\;,\\
{\mathbb B}_1 {\mathbb B}^1 & =-{\mathbb B}_2 {\mathbb B}^2=\frac{i}{4} \frac{(\Delta-J-K+S)(\Delta+J+K+S+2)}{\Delta+S+1}\;.
\end{split}
\end{equation}

Analogous relations hold for the right Q system. In particular, to find \eqref{eq:PrefactorsQuantumNumbers} for the dotted system, one sends $S\to -S-2$ and $K\to -K-2$, as can be seen from the asymptotics \eqref{Ppow}.

\paragraph{Analyticity and Gluing Conditions.}
In~\cite{Cavaglia:2021eqr,Ekhammar:2021pys} it was postulated that
$\bP_a$ and $\bP^a$ are functions with one short cut $[-2h,2h]$.
As a result, the QQ-relations~\eq{QQsys} imply that $\bQ_k$ and $\bQ^k$ 
have a ladder of cuts in the lower half-plane $[-2h-i n,2h-i n]$ for $n=0,1,2,\dots$.
Since, \textit{a priori} there is no difference between the
upper and lower half-plane in the QSC formalism, 
we can find an alternative set of $\bQ_k^{\uparrow}$, satisfying the same QQ-relations  and the same $u\to +\infty$ asymptotics, which instead have cuts in the upper half-plane. The same is true for the dotted Q-system. 

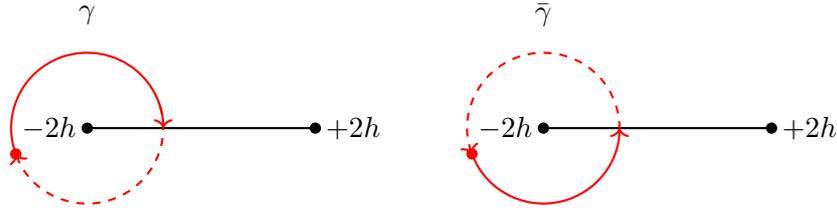
\begin{figure}
    \centering
\begin{center}
\begin{tikzpicture}
    \draw[thick] (0,0) -- (3,0);

    \draw[red,thick,->] ({cos(200)}, {sin(200)}) arc[start angle=200,end angle=0,radius=1];
    \draw[red,thick,<-,dashed] ({cos(200)}, {sin(200)}) arc[start angle=200,end angle=380,radius=1];
    
    \node[fill=red, circle, inner sep=1.5pt] at ({cos(200)}, {sin(200)}) {};
    \node[fill=black, circle, inner sep=1.5pt] at (0, 0) {};
    \node[fill=black, circle, inner sep=1.5pt] at (3, 0) {};

    \node at (3.5,0) {$+2h$};
    \node at (-0.5,0) {$-2h$};
    \node at (0,1.5) {$\gamma$};

    \begin{scope}[xshift=6cm]
        \draw[thick] (0,0) -- (3,0);
    
    \draw[red,thick,->] ({cos(200)}, {sin(200)}) arc[start angle=200,end angle=360,radius=1];
    \draw[red,thick,<-,dashed] ({cos(200)}, {sin(200)}) arc[start angle=200,end angle=0,radius=1];

    \node[fill=red, circle, inner sep=1.5pt] at ({cos(200)}, {sin(200)}) {};
    \node[fill=black, circle, inner sep=1.5pt] at (0, 0) {};
    \node[fill=black, circle, inner sep=1.5pt] at (3, 0) {};

        \node at (3.5,0) {$+2h$};
        \node at (-0.5,0) {$-2h$};
        \node at (0,1.5) {$\bar{\gamma}$};
    \end{scope}
\end{tikzpicture}
\end{center}
    \caption{Two contours for the analytic continuation. $\gamma$ goes around the $-2h$ branch point in the clockwise direction, while $\bar\gamma = \gamma^{-1}$ goes anti-clockwise.}
    \label{fig:g-gbar-contours}
\end{figure}

Gluing conditions establish a connection between the two copies of the Q-systems.
It relates the analytic continuation of $\bQ_k$ to $\bQ^{\uparrow}_{\dot k}$,
and $\bQ_{\dot k}$ to $\bQ^{\uparrow}_{k}$. For
$u \in(-2 h, 2 h)$ the gluing condition reads
\beqa\label{gluingupdown}
\mathbf{Q}_k(u+i 0)=G_k{}^{\dot{n}} 
\mathbf{Q}_{\dot{n}}^{\uparrow}(u-i 0)\,,\qquad
\mathbf{Q}_{\dot{k}}(u+i 0)=G_{\dot{k}}{ }^n 
\mathbf{Q}_n^{\uparrow}(u-i 0)\;,\, 
\eeqa
where $G_k{ }^{\dot{n}}$ and $G_{\dot{k}}{ }^n$ are two independent, analytic matrices. 
One can analytically continue~\eq{gluingupdown} to the vicinity of the cut $[-2h,2h]$ by introducing
a contour $\gamma$, shown in Figure~\ref{fig:g-gbar-contours}, going clockwise around the $u=-2h$ branch point, as
\beqa\label{gluinggamma}
\mathbf{Q}_k^\gamma=G_k{}^{\dot{n}} 
\mathbf{Q}_{\dot{n}}^{\uparrow}\;,\qquad\qquad\qquad
\mathbf{Q}_{\dot{k}}^\gamma=G_{\dot{k}}{ }^n 
\mathbf{Q}_n^{\uparrow}\;. 
\eeqa
When considering only massive excitations,
it was argued that the gluing matrices $G$ are constant and off-diagonal~\cite{Cavaglia:2021eqr,Ekhammar:2021pys}. We will maintain the same assumption in this paper.

\paragraph{$\bQ\omega$-system.} One of the ways to write a closed sub-system of equations from the full Q-system is the $\bQ\omega$-system.
It uses the fact that the lower-half-plane and upper-half-plane Q-functions are related by a periodic function $\Omega_{k}{}^{l}$, namely we have
\begin{equation}
    \bQ^{\uparrow}_k = \Omega_{k}{}^{l}\bQ^{\downarrow}_{l}\,,
    \qquad
    Q^{\uparrow,+}_{a|k} = \Omega_{k}{}^{l}\,Q^{\downarrow,+}_{a|l}\,,
    \qquad
    \left(\Omega_{k}{}^{l}\right)^{-} = Q^{\uparrow}_{a|k}Q^{\downarrow,a|l}\,,
\end{equation}
which in combination with the gluing
condition \eq{gluinggamma} gives rise to the following system of equations
\beq\label{Qomegamain}
(\mathbf{Q}_k)^\gamma=\omega_k{}^{\dot{m}} \mathbf{Q}_{\dot{m}}\;,
\qquad
G_{\dot k}{}^m Q^{\uparrow,+}_{a|m} = \omega_{\dot k}{}^{l}\,Q^{\downarrow,+}_{a|l}\,,
\qquad
(\mathbf{Q}_{\dot k})^\gamma=\omega_{\dot k}{}^{{m}} \mathbf{Q}_{{m}}
\;,
\eeq
where $\omega_{\dot{k}}{}^{l} = G_{\dot{k}}{}^{m}\Omega_{m}{}^{l}$ is an $i$-periodic function of $u$. It has infinitely many cuts at positions $[-2h+i n,2h+i n]$ for $n\in {\mathbb N}$.
The discontinuity of $\omega$ at the cut $[-2h,2h]$  can be expressed in terms of Q-functions as
\beq
(\omega_{\dot{k}}{}^l)^{\bar{\gamma}}-\omega_{\dot{k}}{}^l=\mathbf{Q}_{\dot{k}} (\mathbf{Q}^{l})^{\bar{\gamma}}-
(\mathbf{Q}_{\dot{k}})^\gamma \mathbf{Q}^l\;.
\eeq
\paragraph{$\bP\mu$-system.}
The $\bP\mu$-system is a set of equations similar to the $\bQ\omega$-system, but for $\bP$-functions.
To pass from $\bQ$ to $\bP$ we can use the QQ-relation~\eq{eq:RaisingLoweringQQ}
and define the $\mu$-functions as
\beq\label{muomega}
\mu_a{}^{\dot{b}} \equiv Q_{a \mid c}^{-} \omega^c{ }_{\dot d} Q^{\dot{b} \mid \dot{d},-}\,, \qquad \qquad \qquad
\mu^a{ }_{\dot{b}} \equiv Q^{a \mid c,-} \omega_c{ }^{\dot{d}} Q_{\dot{b} \mid \dot{d}}^{-}\,,
\eeq
where we have raised and lowered indices using the definitions $\omega_{k}{}^{\dot{l}} \omega^{k}{}_{\dot{m}} = \delta^{\dot{l}}_{\dot{m}}$ and $\mu_{a}{}^{\dot{b}} \mu^{a}{}_{\dot{c}} = \delta^{\dot{b}}_{\dot{c}}$ as
\beq\la{mudet}
\det\mu_{a}{}^{\dot b}=\det\mu_{\dot a}{}^{b}=1\;.
\eeq
One can rewrite \eq{muomega} in an alternative way using the lower half-plane Q-functions
\begin{equation}\label{eq:muLowerUpper}
    \mu^{a}{}_{\dot{b}} = Q^{a|k,-}\, G_{k}{}^{\dot{l}}\,Q^{\uparrow,-}_{\dot{b}|\dot{l}}\,.
\end{equation}
The function $\mu$ is ``mirror'' $i$-periodic, meaning that
\beq\label{mugamma}
\mu^{++}=\mu^\gamma\,.
\eeq
The analog of \eq{Qomegamain} reads
\beq\label{Pmumain}
\left(\mathbf{P}_a\right)^{\bar{\gamma}}=
\mathbf{P}_{\dot{b}}\; \mu^{\dot{b}}{ }_a
\,,\qquad
\left(\mathbf{P}_{\dot a}\right)^{\bar{\gamma}}=
\mathbf{P}_{{b}}\; \mu^{{b}}{ }_{\dot a}\;,
\qquad
\left(\bP^{a}\right)^{\bar{\gamma}} = \bP^{\dot{b}}\mu_{\dot{b}}{}^{a}\,,
\qquad
\left(\bP^{\dot{a}}\right)^{\bar{\gamma}} = \bP^{b}\mu_{b}{}^{\dot{a}}\,.
\eeq
and the discontinuity of $\mu$ is given by
\begin{equation}
    \left(\mu_{a}{}^{\dot{b}}\right)^{\gamma} - \mu_{a}{}^{\dot{b}} = \bP_{a}\left(\bP^{\dot{b}}\right)^{\bar{\gamma}}-\left( \bP_{a}\right)^{\gamma} \bP^{\dot{b}}\,.
\end{equation}

\paragraph{Reality.} 
The unitarity of the initial theory and reality of the spectrum manifests itself as a symmetry of QSC under complex conjugation. Complex conjugation maps UHP-analytic Q-functions to LHP-analytic ones. $\bP$ is both UHP-analytic and LHP-analytic and gets mapped to itself up to phases. We fix gauge so that
\begin{equation}\label{eq:PupPconj}
    \overline{\bP}_{a} = (-1)^{a+1}\bP_{a}\,,
    \qquad\qquad
    \overline{\bP}_{\dot{a}} = (-1)^{\dot{a}+1}\bP_{\dot{a}}\,,
\end{equation}
The action on $\bQ$ is slightly more complicated, we fix gauge so that
\beq \label{eq:QupQconj}
\mathbf{Q}^{\uparrow}_k(u) = (-1)^{k+1}\, \overline{\mathbf{Q}}_k(u) \, , \qquad\qquad \mathbf{Q}^{\uparrow}_{\dot{k}}(u) = (-1)^{\dot{k}+1}\, \overline{\mathbf{Q}}_{\dot{k}}(u)\;.
\eeq
As a result, under complex conjugation the $\mu_{ab}$ function behaves as
\beq
\overline{\mu_{12}(u+\tfrac{i}{2})}=\pm \mu_{12}(u+\tfrac{i}{2})\;,
\label{muconj}
\qquad\text{or}\qquad
\bar\mu_{12}(u)=\pm \mu_{12}(u+i)\;.
\eeq
These relations will be useful in what follows.

\paragraph{Gluing with Reality}
We can combine \eqref{eq:QupQconj} and \eqref{gluingupdown} to deduce the following set of gluing equations
\begin{equation}\label{eq:gluingConjugate}
    \bQ_{k}^{\gamma} = N_{k}{}^{\dot{l}}\,\overline{\bQ}_{\dot{l}}\,,
    \qquad\qquad
    \bQ_{\dot{k}}^{\gamma} = N_{\dot{k}}{}^{l}\,\overline{\bQ}_{l}\,,
\end{equation}
where $N_{k}{}^{\dot{l}} = G_{k}{}^{\dot{l}}(-1)^{\dot{l}+1}$. From the consistency of \eqref{eq:gluingConjugate} it follows that $N_{k}{}^{\dot{l}}\bar{N}_{\dot{l}}{}^{m}\bQ_{m} = \bQ_{k}$. In this paper, we will make the assumption that $G$, and hence $N$, are off-diagonal. This assumption is motivated by a similar structure appearing in $\mathcal{N}=4$ SYM and from \cite{Cavaglia:2022xld} which found this to be a consistent choice in the parity symmetric sector. 
One can also see this structure quasi-classically in the properties of the eigenvalues of the classical monodromy matrix and their transformation under the $\mathbb{Z}_4$ symmetry. Ultimately, this assumption needs to be verified by comparing against independent first-principle calculations of the spectrum which are currently lacking. 

Furthermore, we can fix the relation between the two entries of the gluing matrix by demanding that $\det \omega = 1$, which after using $\det \Omega = 1$, implies
\begin{equation}
    G_{k}{}^{\dot{l}} = \begin{pmatrix}
        0 & -\alpha \\
        \frac{1}{\alpha} & 0 
    \end{pmatrix}\,,
    \quad
    G_{\dot{k}}{}^{l} =  \begin{pmatrix}
        0 & -\bar{\alpha} \\
        \frac{1}{\bar{\alpha}} & 0 
    \end{pmatrix}\,,
    \quad
    N_{k}{}^{\dot{l}} = \begin{pmatrix}
        0 & \alpha \\
        \frac{1}{\alpha} & 0
    \end{pmatrix}\,,
    \quad
    N_{\dot{k}}{}^{l} = \begin{pmatrix}
        0 & \bar{\alpha} \\
       \frac{1}{\bar{\alpha}} & 0
    \end{pmatrix}\,.
\end{equation}

\paragraph{Index raising operation.}
Another symmetry of the Q-system is induced by the automorphism of the symmetry factors $\mathfrak{psu}(1,1|2)$.
From $\bP_{a}\bP^{a}=\bQ_{k}\bQ^{k}=0$ it follows that $\bP_{a} = r \epsilon_{ab}\bP^{b}$ and $\bQ_{k}=r'\epsilon_{kl}\bQ^{l}$. Furthermore, consistency with the remaining QQ-relations sets $r'=-\frac{1}{r}$. Q-functions with upper and lower indices are then related as follows 
\beqa\label{Qdot}
\mathbf{Q}^k=+r \epsilon^{k l} \mathbf{Q}_l\,, \quad \mathbf{Q}_k=-\frac{1}{r} \epsilon_{k l} \mathbf{Q}^l\,, \quad \mathbf{P}^a=-\frac{1}{r} \epsilon^{a b} \mathbf{P}_b\,, \quad \mathbf{P}_a=+r \epsilon_{ab} \mathbf{P}^b\;.
\eeqa
An identical statement holds for the right system with $\dot{r}$ instead of $r$. The functions $r,\dot{r}$ have particularly simple analytic properties. Using for example the $\bP\mu$-systems one finds $r^{\bar{\gamma}} = \dot{r}\,,\,\,\dot{r}^{\bar{\gamma}} = r$ implying that $r$ is a rational function of the Zhukovsky variable 
\begin{equation}\label{eq:Zhukovsky}
    x(u)\equiv \frac{u+\sqrt{u-2h}\sqrt{u+2h}}{2h}\;.
\end{equation}
Furthermore, \eqref{Ppow} imposes asymptotics $r \simeq \frac{{\mathbb B}^{1}}{{\mathbb B}_{2}}u^{-\hat{B}}\,,\,\,\dot{r} \simeq \frac{{\mathbb B}^{\dot 1}}{{\mathbb B}_{\dot 2}}u^{-\check{B}}$ fixing
   \beqa
r(u)=\frac{N(x)}{M(x)}\,,
\eeqa
with $N$ and $M$ polynomials constrained at large $u$ by $\frac{N(x)}{M(x)}\simeq \frac{{\mathbb B}^{1}}{{\mathbb B}_{2}}u^{-\hat B}
$ and $\frac{N(1/x)}{M(1/x)}\simeq \frac{{\mathbb B}^{\dot 1}}{{\mathbb B}_{\dot 2}}u^{-\check B}
$.
\paragraph{$\Lambda$-symmetry.}
A symmetry of the QSC, which leaves all the QQ-relations and the gluing condition invariant, is the following
\beq\la{lambdasym}
\bQ_k\mapsto x^{+\Lambda} \,\bQ_k\,,\qquad\qquad\bP_a\mapsto x^{-\Lambda}\,\bP_a\;,
\eeq
while leaving $Q_{a|i}$ invariant. In order to be consistent with the gluing condition \eq{gluinggamma} we have to transform the dotted system in the synchronized way
\beq
\bQ_{\dot k}\mapsto x^{-\Lambda} \,\bQ_{\dot k}\,,\qquad\qquad\bP_{\dot a}\mapsto x^{+\Lambda} \,\bP_{\dot a}\;.
\eeq
Note that this results in shifting $\hat B\mapsto
\hat B+2\,\Lambda$ and $\check B\mapsto
\check B-2\,\Lambda$.

\section{ABA generalities}
\label{sec:aba-gen}
The Asymptotic Bethe Ansatz (ABA) is a powerful method used to determine the spectrum of infinite-volume integrable QFTs. It reduces the complex interactions of particles to a set of algebraic equations, whose solutions, the so-called Bethe roots, parametrise the spectrum of the theory. In the present case there are seven types of Bethe roots~\cite{Borsato:2016xns}:
\beqa
&\text{Auxilary roots: }\qquad\qquad & u_{1,k}\,,\;u_{3,k}\,,\;u_{\dot 1,k}\,,\;u_{\dot 3,k}\,,\\
&\text{Massive roots: }\qquad\qquad & u_{2,k}\,,\;u_{\dot 2,k}\,,\\
&\text{Massless roots: }\qquad\qquad & z_k\,.
\eeqa
We denote the number of auxiliary and massive roots by $K_A$ ($A=1\,,3\,,\dot 1\,,\dot 3$), and the number of massless roots by $K_\circ$. The massive and massless roots  together are also called momentum-carrying, since, as we see below, they appear explicitly in the formula for the energy and momentum of states.

The ABA, like the worldsheet S-matrix, can be written most compactly in terms of Zhukovsky variables $x_{A,k}\equiv x(u_{A,k})$ where $x(u)$ was defined in \eqref{eq:Zhukovsky}. We will use the following notation for shifts of $u$
\beq
x_{A,k}^\pm \equiv x(u_{A,k}\pm\tfrac i2 )\,,
\qquad\qquad
x_{A,k}^{[\pm n]} \equiv x(u_{A,k}\pm\tfrac i2 n)\,, \qquad n\in {\mathbb N}\,.
\eeq
The energy of a state can be found from the momentum carrying Bethe roots as
\beq\label{ABAenergy}
\gamma= 2\ii h\sum_{k=1}^{K_2} \left(\frac{1}{x^{+}_{2,k}} - \frac{1}{x^{-}_{2,k}}\right)
+
2\ii h\sum_{k=1}^{K_{\dot 2}} \left(\frac{1}{x^{+}_{\dot 2,k}} - \frac{1}{x^{-}_{\dot 2,k}}\right)
+
2 \ii h \sum_{k=1}^{K_\circ}\left(\frac{1}{z_k}-z_k\right)\;.
\eeq
In holographic applications, momentum carrying Bethe roots further satisfy the level-matching condition (a.k.a. cyclicity condition)
\beq\label{zeromom}
\prod_{k=1}^{K_\circ}{z_k^2}\prod_{k=1}^{K_2}\frac{x_{2,k}^+}{x_{2,k}^-}
\prod_{k=1}^{K_{\dot 2}}\frac{x_{\dot 2,k}^+}{x_{\dot 2,k}^-}=1\;.
\eeq
Each type of root has a corresponding set of Bethe equations~\cite{Borsato:2016xns} which we summarise below. In addition to various rational terms, momentum-carrying Bethe equations include so-called \textit{dressing factors}. They arise because the symmetry algebra and unitarity fix the S~matrix only up to four scalar functions. These four dressing factors multiply the following S matrices: $\sigma^{\bullet \bullet}$, for two massive excitations of the same chirality; $\widetilde{\sigma}^{\bullet\bullet}$, for two massive excitations of opposite chirality; $\sigma^{\bullet \circ}$ for a massive and massless excitation; and $\sigma^{\circ\circ}
$ for two massless excitations.

To reduce the number of sign-ambiguous square-root factors in the Bethe equations we will modify the equations from \cite{Borsato:2016xns} by normalising the S-matrix slightly differently. This is the same approach taken in \cite{Frolov:2021fmj} and we will follow their conventions, although we will write down the Bethe equations in the spin chain frame as opposed to the string frame, see \cite{Borsato:2016xns} for the relation between the two different frames \footnote{Explicitly we rescale the dressing factor between massive and massless particles according to
\begin{equation*}
\left(\hat{\sigma}_{\text{Here}}^{\bullet \circ}(x,z) \right)^{2} = -\sqrt{\frac{1-\frac{z}{x^-}}{z-\frac{1}{x^+}}}\sqrt{\frac{1-\frac{z}{x^+}}{z-\frac{1}{x^-}}}\left(\sigma_{\text{BOSST}}^{\bullet \circ}(x,z)\right)^{2}
\end{equation*}
where $\sigma_{\text{BOSST}}^{\bullet \circ}(x,y)$ is the dressing phase of
\cite{Borsato:2016xns} and the inverse factor in 
$\hat{\sigma}_{\text{Here}}^{\circ\bullet}$
to preserve the relation
$\left[\sigma^{\bullet \circ}(x,z)\sigma^{\circ \bullet }(z,x)\right]^{2}=1$, following~\cite{Frolov:2021fmj}.
}.
\paragraph{Auxiliary Equations}
\beqa\label{eq:aux-bae}
1 &=& \prod_{j=1}^{K_2} \frac{x_{I,k} - x_{2,j}^-}{x_{I,k} - x_{2,j}^+} \prod_{j=1}^{K_{\dot{2}}} \frac{1-\frac{1}{x_{I,k}x_{\dot 2,j}^+}}{1-\frac{1}{x_{I,k}
x_{\dot 2,j}^-}}
\prod_{j=1}^{K_\circ}\frac{x_{I,k}-1/z_{j}}{
x_{I,k}-z_{j}
}\,,\qquad I=1,3\,,
\eeqa
and similarly for the dotted roots
\beqa\label{eq:aux-bae-dot}
1 &=& \prod_{j=1}^{K_{\dot 2}} \frac{x_{I,k} - x_{\dot 2,j}^-}{x_{I,k} - x_{\dot 2,j}^+} 
\prod_{j=1}^{K_{{2}}} \frac{1-\frac{1}{x_{I,k}x_{2,j}^+}}{1-\frac{1}{x_{I,k}
x_{2,j}^-}}
\prod_{j=1}^{K_\circ}\frac{x_{I,k}-1/z_{j}}{
x_{I,k}-z_{j}
}\,,\qquad I=\dot 1,\dot 3\,.
\eeqa

\paragraph{Left Middle Equation}
\begin{equation}
    \begin{split}\label{eq:BA-2b-left2}
        \left( \frac{x^+_{2,k}}{x^-_{2,k}}\right)^{L} = &
        \prod_{\substack{j=1\\j\neq k}}^{K_2} \frac{x^+_{2,k}-x^-_{2,j}}{x^-_{2,k}-x^+_{2,j}}
        \frac{1-\frac{1}{x^+_{2,k}x^-_{2,j}}}{1-\frac{1}{x^-_{2,k}x^+_{2,j}}}
        \left(\sigma^{\bullet \bullet}\right)^2 (x_{2,k},x_{2,j})
        \prod_{j=1}^{K_{\dot{2}}}\frac{1-\frac{1}{x^+_{2,k} {x}^{+}_{\dot 2,j}}}
        {1-\frac{1}{x^-_{2,k} {x}^{-}_{\dot 2,j}}}
        \frac{1-\frac{1}{x^+_{2,k} {x}^{-}_{\dot 2,j}}}
        {1-\frac{1}{x^-_{2,k} {x}^{+}_{\dot 2,j}}}
        \left(\tilde{\sigma}^{\bullet\bullet}\right)^2(x_{2,k},x_{\dot{2},j})
\\
        &\times \prod_{j=1}^{K_1}\frac{x_{2,k}^- -x_{1,j}}{x^+_{2,k}-x_{1,j}}
        \prod_{j=1}^{K_3}\frac{x_{2,k}^- -x_{3,j}}{x^+_{2,k}-x_{3,j}}
        \prod_{j=1}^{K_{\dot{1}}}\frac{1-\frac{1}{x^-_{2,k} x_{\dot{1},j}}}
        {1-\frac{1}{x^+_{2,k} x_{\dot{1},j}}}
        \prod_{j=1}^{K_{\dot{3}}}\frac{1-\frac{1}{x^-_{2,k} x_{\dot{3},j}}}
        {1-\frac{1}{x^+_{2,k} x_{\dot{3},j}}}        \\ 
        &\times \prod_{j=1}^{K_{\circ}} \frac{1- x^+_{2,k}z_j}{x^-_{2,k}-z_j}
        \left(\sigma^{\bullet\circ}\right)^{2}(x_{2,k},z_{j})\;.
    \end{split}
\end{equation}

\paragraph{Right Middle Equation}
\begin{equation}
    \begin{split}
      \label{eq:BA-2b}
      \left(\frac{x_{\dot 2,k}^+}{x_{\dot 2,k}^-}\right)^L 
      &=
      \prod_{\substack{j=1 \\ j\neq k}}^{K_{\dot{2}}} \frac{x_{\dot 2,k}^- - x_{\dot 2,j}^+}{x_{\dot 2,k}^+ - x_{\dot 2,j}^-} 
      \frac{1- \frac{1}{x_{\dot 2,k}^+ x_{\dot 2,j}^-}}{1- \frac{1}{x_{\dot 2,k}^- x_{\dot 2,j}^+}}  (\sigma^{\bullet\bullet}(x_{\dot{2},k},x_{\dot{2},j}))^2
      \prod_{j=1}^{K_2}  \frac{1- \frac{1}{x_{\dot 2,k}^- x_{2,j}^-}}{1- \frac{1}{x_{\dot 2,k}^+ x_{2,j}^+}} \frac{1- \frac{1}{x_{\dot 2,k}^+ x_{2,j}^-}}{1- \frac{1}{x_{\dot 2,k}^- x_{2,j}^+}}(\widetilde{\sigma}^{\bullet\bullet}(x_{\dot{2},k},x_{2,j}))^2
      \\ & \phantom{= }\times
      \prod_{j=1}^{K_{\dot 1}}\frac{x_{\dot{2},k}^+ -x_{\dot 1,j}}{x^-_{\dot{2},k}-x_{\dot 1,j}}
        \prod_{j=1}^{K_{\dot 3}}\frac{x_{\dot{2},k}^+ -x_{\dot 3,j}}{x^-_{\dot{2},k}-x_{\dot 3,j}}
      \prod_{j=1}^{K_{1}}  \frac{1 - \frac{1}{x_{\dot 2,k}^+ x_{1,j}}}
      {1- \frac{1}{x_{\dot 2,k}^- x_{1,j}}}
      \prod_{j=1}^{K_{3}} \frac{1 - \frac{1}{x_{\dot 2,k}^+ x_{3,j}}}
      {1- \frac{1}{x_{\dot 2,k}^- x_{3,j}}}
      \\ &\phantom{= }\times
      \prod_{j=1}^{K_\circ} \frac{x^+_{\dot{2},k}}{x^-_{\dot{2},k}}\,z^2_j\,\frac{x^-_{\dot{2},k}-z_j}{1-x^+_{\dot{2},k} z_j} 
      (\sigma^{\bullet\circ}(x_{\dot{2},k},z_{j}))^2
      \;
      .
    \end{split}
\end{equation}

\paragraph{Massless Equation}
\begin{align}
    z_k^{2L} = &\prod_{\substack{j=1 \\ j\neq k}}^{K_{\circ}} \left(-\frac{z_k}{z_j} \left(\sigma^{\circ\circ}\right)^2(z_k,z_j)\right)
    \prod_{j=1}^{K_1}\frac{1/z_k -x_{1,j}}{z_k-x_{1,k}}
    \prod_{j=1}^{K_{\dot 1}}\frac{z_k-x_{\dot{1},j}}{1/z_k-x_{\dot{1},j}}
    \prod_{j=1}^{K_3}\frac{1/z_k -x_{3,j}}{z_k-x_{3,j}}
    \prod_{j=1}^{K_{\dot 3}}\frac{z_k-x_{\dot{3},j}}{1/z_k-x_{\dot{3},j}}\nn \\
    &\times \prod_{j=1}^{K_2}
    \frac{z_k-x^-_{2,j}}{z_k x^+_{2,j}-1}
    (\sigma^{\circ\bullet})^2(z_k,x_{2,j})\times \prod_{j=1}^{K_{\dot 2}}
     \frac{x^-_{\dot{2},j}}{z^2_k x^+_{\dot{2},j}}\frac{z_k x^+_{\dot{2},j}-1}{z_k -x^-_{\dot{2},j}}
    (\sigma^{\circ\bullet})^2(z_k, x_{\dot 2,j})\,. \label{masslesABA}
\end{align}

\section{The Large Volume Limit of QSC}\label{sec:MasslessABALimit}
In this section, we analyse the infinite-volume limit of the QSC. In this limit, the QSC is expected to simplify to an algebraic set of equations, similar to those discussed in the previous section. The asymptotic limit of the QSC was first studied in the case of AdS$_5$/CFT$_4$ in \cite{Gromov:2014caa}, and we follow similar steps to those outlined there. However, in AdS$_3$/CFT$_2$ an additional complication arises because the branch cuts are no longer quadratic~\cite{Borsato:2013hoa}, presenting new QSC features that were first addressed in~\cite{Cavaglia:2021eqr,Ekhammar:2021pys}. In general, taking an infinite-volume limit is not entirely straightforward, as this process does not always commute with analytic continuation due to Stokes-like phenomena. These subtleties must be carefully handled to obtain consistent results.

\subsection{Massless excitations in QSC formalism}\la{sec:masslessQSC}
Let us outline the main feature of
the new class of solutions of QSC proposed in this paper, which includes the massless degrees of freedom.  
Massless excitations manifest themselves as zeroes of $\mu_{1}{}^{\dot{2}}$ (and $\mu_{\dot 1}{}^{{2}}$) lying on the cut, which we denote by $z_k$. These massless roots are similar to the one that appears in ${\cal N}=4$ SYM theory with non-integer spin $S$, as was found recently in~\cite{Ekhammar:2024neh}, though massless excitations in $\AdS_3$ backgrounds appear already in the near-BMN perturbative worldsheet theory. To test this proposal, in the next sub-section, we consider the ABA limit of the QSC equations. 

Below, we demonstrate how the full structure of the ABA equations emerges from the QSC formalism, incorporating all the expected features of the massless modes. However, the asymptotic large-volume regime in theories with massless degrees of freedom is subject to additional subtleties. To fully resolve those, a numerical analysis for finite-length operators at finite coupling or analytical treatment in the near-BPS limit is required. We leave such investigations for future work.

\subsection{Derivation of the QSC in Asymptotic Limit}
\la{sec:MasslessABALimitDer}

As was shown in \cite{Ekhammar:2024neh}, the ABA-like regime of the QSC appears when a component of the $\omega$-matrix is large and the magnitude of this component controls the precision of the ABA limit. According to \cite{Ekhammar:2024neh} there could be several ABA regimes, 
depending on which components of $\omega$ are large, such as the  DGLAP or BFKL regimes in ${\cal N}=4$ SYM theory.
In \cite{Ekhammar:2024neh}, it was shown that massless excitations, similar to what we expect in the current case, arise in the Regge (BFKL) regime of ${\cal N}=4$ SYM theory for large quantum numbers states.
We will be guided by this finding in the current case.

To obtain the ABA equations from Section~\ref{sec:aba-gen} from the QSC for ``local'' operators with only massive excitations, we   consider a limit when $\omega^{1}{}_{\dot 2}$ and
$\omega^{\dot 1}{}_{2}$ are large~\cite{Cavaglia:2021eqr,Ekhammar:2021pys}. 
In this paper we will assume that adding massless excitations is not expected to change this property. 
In particular, this implies the following simplification of \eq{muomega}
\begin{equation}\label{ABAomega1}
    \mu_{a}{}^{\dot{b}} \simeq Q^-_{a|1}Q^{\dot{b}|\dot{2}-}\omega^{1}{}_{\dot 2}\;,
    \qquad\qquad
    \mu_{\dot a}{}^{{b}} \simeq Q^-_{\dot a|\dot 1}Q^{{b}|{2}-}\omega^{\dot 1}{}_{2}\;.
\end{equation}
As we know from \cite{Cavaglia:2021eqr,Ekhammar:2021pys}, zeroes of $\mu_{1}{}^{\dot 2}$ and $\mu_{2}{}^{\dot 1}$
play a particularly important role because they become momentum carrying Bethe roots.
For these components \eq{ABAomega1} becomes
\begin{equation}\label{ABAomega}
    \mu_{1}{}^{\dot{2}} \simeq Q^-_{1|1}Q^-_{\dot 1|\dot 1}\omega^{1}{}_{\dot 2}\;,
    \qquad\qquad
    \mu_{\dot 1}{}^{{2}} \simeq Q^-_{1|1}Q^-_{\dot 1|\dot 1}\omega^{\dot 1}{}_{2}\;,
\end{equation}
where we used \eq{Qupdown}. One should treat this relation with care:
let us denote by
$\mu^{\rm ABA}$ the asymptotic limit of $\mu$
taken for some generic point on the main sheet. $\mu^{\rm ABA}$  is an analytic function by itself and 
it approximates $\mu$ well when the distance to the branch cut is $\sim 1$. However, when we approach the branch cut the discrepancy increases and eventually for $u\sim h$ the error may become $\sim 100\%$. That is to say that the ABA limit of $\mu$ does not commute with the analytic continuation to points near the cut.
One of the reasons for that is that usually $\omega^{1}{}_{\dot 2}$ is indeed the dominating component for points $u\sim 1$, but near the cut all components of $\omega$ are of the same order and the approximation \eq{ABAomega} fails.
The same holds true in the much better studied cases of QSCs for ${\cal N}=4$ SYM~\cite{Gromov:2014caa} and ABJM theories~\cite{Bombardelli:2017vhk}.

An important assumption, previously made in \cite{Cavaglia:2021eqr,Ekhammar:2021pys} and adopted here, is that $\mu^{\rm ABA}(u)$, defined as the limit of $\mu$ away from the cut, is an analytic function with \textit{quadratic} branch cuts. It is intriguing to speculate that perhaps one can interpret the breakdown of this analytic property as being due to massless wrapping; we leave such questions for future studies. Additionally, we assume that $\mu^{\rm ABA}$ is mirror $i$-periodic, meaning $\tilde\mu^{\rm ABA}(u) = \mu^{\rm ABA}(u+i)$, where the tilde denotes the analytic continuation, consistent with the properties of the exact $\mu$ \eq{mugamma}. It was observed in \cite{Cavaglia:2021eqr,Ekhammar:2021pys} that these assumptions lead to a consistent ABA limit in the massive sector and successfully reproduce all massive ABA equations, including the crossing equations for the dressing phases. In this paper, we use the same set of assumptions. We currently lack a rigorous proof of these assumptions; future numerical studies are needed to verify their validity. In the following discussion, we omit the ABA superscript for $\mu$ and assume that $\mu$ refers to the ABA limit of $\mu$.

\paragraph{Fixing the ratio of $\mu$'s.}

We now proceed to show that the relation \eq{ABAomega} supplemented with the analyticity conditions discussed above allows us to fix 
$\mu_{1}{}^{\dot{2}}$ and $\mu_{\dot 1}{}^{{2}}$ in terms of its zeroes. 
Below we will focus on $\mu_{1}{}^{\dot{2}}$, with similar arguments applying to $\mu_{\dot 1}{}^{2}$.
$\mu_{1}{}^{\dot{2}}$ can have two different types of zeros: those which are situated on the branch cut $[-2h,2h]$ 
and those which are not.
Let us denote zeroes of $\mu_{1}{}^{\dot{2}}$ which are situated 
on the branch cut $[-2h,2h]$ by $\theta_i$.
As zeroes can be either above or below the cut it 
is better to use the $x$-plane and denote zeroes by 
$z_i$ for $i=1,\dots,N$, so that $\theta_i=h(z_i+1/z_i)$. 
As is shown below, zeroes on other cuts i.e. 
at $[-2h+i n,2h+i n]$ are related to $z_i$.
The remaining zeroes of $\mu_{1}{}^{\dot{2}}(u+i/2)$ will be denoted as $u_{k}$ with $k=1,\dots,M$
\footnote{This also includes possible
 zeros which are accidental on the branch cuts, i.e. it includes all zeros which satisfy
$\mu(u_{k}+i/2+i0)=\mu(u_{k}+i/2-i0)=0$. We will assume $M$ to be a finite number.}.
To keep track of these zeros we introduce the following notation
\begin{align}
    \betheQ = \prod_{k=1}^{M}(u-u_{k})\,,
    \qquad
    \kappa = \prod^{N}_{i=1}(x-z_{i})\,,
    \qquad
    &\bar{\kappa} = \prod^{N}_{i=1}\(x-\frac{1}{z_i}\)\,.
\end{align}
From \eq{muconj} we see that $\betheQ$ is a real polynomial
and $\bar\kappa(x)$ is the complex conjugate of $\kappa(x)$.

Following \cite{Gromov:2014caa,Ekhammar:2024neh}, the key starting point of the ABA derivation is the following combination
\begin{equation}\label{Fdef}
    F^2 \equiv \frac{\mu_{1}{}^{\dot{2}}(u)}{\mu_{1}{}^{\dot{2}}(u+i)} \frac{\betheQ^+}{\betheQ^-} \frac{\bar{\kappa}}{\kappa}\,,\qquad\qquad F(\infty) = \pm 1\;.
\end{equation}
Note that, due to \eq{muconj}
$F^2$ can also be written as
\begin{equation}
        F^2          = \pm \frac{\mu_{1}{}^{\dot{2}}}{\bar\mu_{1}{}^{\dot{2}}} \frac{\betheQ^+}{\betheQ^-} \frac{\bar{\kappa}}{\kappa}\;,
\end{equation}
which in turn implies
\begin{equation}\label{Fcconj}
    \bar F^2 = \pm 1/F^2 \;.
\end{equation}
Due to mirror periodicity of $\mu$, see~\eqref{mugamma}, we immediately find
\begin{equation}
F^2= \frac{\mu_{1}{}^{\dot{2}}}{\tilde\mu_{1}{}^{\dot{2}}} \frac{\betheQ^+}{\betheQ^-} \frac{\bar{\kappa}}{\kappa}\,.
\end{equation}
From this we deduce the following property of $F$ under the analytic continuation:
\begin{align}\label{FFtilda}
    &F \Tilde{F} = \pm \frac{\betheQ^+}{\betheQ^-} \frac{1}{\prod_{i}z_i}\;.
\end{align}
Above, we have used the assumption that $\mu_{1}{}^{\dot{2}}$ have a square-root cut in the ABA-limit.
Furthermore, from \eq{ABAomega} we find
\begin{eqnarray}\label{eq:FFQQ}
F^2=\frac{Q_{1|1}^-Q_{\dot 1|\dot 1}^-}{Q_{1|1}^+Q_{\dot 1|\dot 1}^+} \frac{\betheQ^+}{\betheQ^-} \frac{\bar{\kappa}}{\kappa}\;,  
\end{eqnarray}
where the $\omega^{1}{}_{\dot 2}$ factors cancel out due to i-periodicity. We conclude that $F$ does not have cuts in the 
upper half plane, as that is true for the RHS. 
In addition, \eq{Fcconj} tells us that there could not be cuts in the lower half plane.
From its definition $F$ does not have poles or zeros on the branch cut $[-2h,2h]$. 
Furthermore, due to our definition of ${\mathbb Q}$,  $F$ cannot have any poles or zeros above or below the real axis.
Thus, we conclude that $F$ is an analytic function with no zeroes or poles on the main sheet, and the only singularity is the quadratic branch cut $[-2h,2h]$.
Finally, \eq{FFtilda} defines a scalar Riemann-Hilbert problem for $F$, which, together with the asymptotic condition $F(\infty)=1$, fixes $F$ uniquely.
Furthermore, for~\eq{FFtilda} to have a solution, there is an additional condition on the zeroes of $u_{2,i}$. Indeed, consider
\beq
F_{0}=\frac{1}{\prod_k \sqrt z_k}\frac{B_{(+)}}{B_{(-)}}\;,
\eeq
where we use the standard notation
\begin{equation}\la{Bnonbf}
    B_{(\pm)} = \prod_{k=1}^{M} \sqrt{\frac{h}{x_k^{\mp}}} \(\frac{1}{x}-x^{\mp}_k\)\;,
    \quad
    R_{(\pm)} = \prod_{k=1}^{M} \sqrt{\frac{h}{x_k^{\mp}}} (x-x^{\mp}_k)\;,
    \quad 
    x^{\pm}_k=x(u_{k}\pm \frac{\ii}{2})\,.
\end{equation}
$F_0$ is analytic on the main sheet except for the branch cut $[-2h,2h]$, and satisfies \eq{FFtilda}.
Thus the ratio $G=F/F_0$ should satisfy $G \tilde G=1$ and so is analytic 
on the whole plane of $x$. Since it is constant as $x\to\infty$, it must be that simply $G(x)=\pm 1$ or $F=\pm F_0$.
At the same time, $F(\infty)=1$ implies that there is a non-trivial constraint on $z_k$ and $x_k$
\begin{equation}\label{zeromomentum}
    \prod^{M}_{k=1} \frac{x_k^{+}}{x_k^-}\prod_{k=1}^{N} z_{k} = 1\,,
\end{equation}
which is similar to the level-matching condition~\eqref{zeromom}.
In conclusion, we find that $F = \pm F_0$,
with the sign determined by $F(\infty)=1$. 
To avoid the sign ambiguities we introduce a new notation, which differs by a constant factor from more standard \eq{Bnonbf}
\begin{align}\label{bfBR}
    &\bfB_{(\pm)} = \prod_{k=1}^{M}\left(1-\frac{1}{x\,x^\mp_k}\right)
    \;,\qquad
    \bfR_{(\pm)} = \prod_{i=1}^{M}(x -\,x^{\mp}_k)\;,\qquad\bfB_{(\pm)}\bfR_{(\pm)}=h^{-M}{\mathbb Q}^\pm\;.
\end{align}
Using this notation and~\eqref{zeromomentum} we can also write
\begin{equation}\label{Fsol}
    F = \frac{\mathbf{B}_{(+)}}{\mathbf{B}_{(-)}}\;.
\end{equation}

\paragraph{Finding $\mu$.}
Knowing $F$, we can easily find $\mu_{1}{}^{\dot{2}}$. 
Indeed, from \eq{Fsol} and \eq{Fdef} we get a first-order finite difference equation for $\mu_{1}{}^{\dot{2}}$ 
\begin{equation}\label{eq:mu-over-mu-plus-2}
    \frac{\mu_{1}{}^{\dot{2}}(u)}{\mu_{1}{}^{\dot{2}}(u+i)} = \frac{\betheQ^-}{\betheQ^+} \frac{\kappa}{\bar{\kappa}}\left(\frac{\bfB_{(+)}}{\bfB_{(-)}}\right)^2\;.
\end{equation}
To find a solution of this equation, we introduce
\begin{align}\label{deff}
    &f = \prod_{n=0}^{\infty} \frac{\bfB^{[2n]}_{(+)}}{\bfB^{[2n]}_{(-)}}\;,\qquad
    \bar f = \prod_{n=0}^{\infty} \frac{\bfB^{[-2n]}_{(-)}}{\bfB^{[-2n]}_{(+)}}
    \qquad\text{s.t.}\quad
    \frac{f}{f^{[2]}}= \frac{\bar f^{[-2]}}{\bar f} = \frac{\bfB_{(+)}}{\bfB_{(-)}}    \,.
\end{align}
We find
\begin{equation}\label{musol}
    \mu_{1}{}^{\dot{2}}\propto \mathcal{P}\betheQ^- 
    f \bar{f}^{[-2]}\, \prod_{n=0}^{\infty} {\kappa^{[2n]}} \bar{\kappa}^{[-2n-2]}\,,
\end{equation}
where the infinite product is convergent up to an infinite $u$-independent factor, which is denoted by the $\propto$ symbol. 
Here $\mathcal{P}$ is a periodic function that needs to be determined. Note that complex conjugation \eq{muconj} holds provided $\bar{\cal P}=\pm {\cal P}$. In order to fix the remaining periodic factor ${\cal P}$, we impose the mirror periodicity of $\mu$
e.g.
$\Tilde{\mu}_{1}{}^{\dot{2}}(u) = \mu_{1}{}^{\dot{2}}(u+i)$
which, after a large cancellation of factors, gives
\footnote{The extra $(-1)^N$ factor comes from treating carefully the infinite product by taking a finite but large upper limit $\Lambda$, which creates the 
factor of $\kappa^{[2\Lambda+2]}/\bar\kappa^{[-2\Lambda-2]}$ giving $(-1)^N$ in the limit.}
\begin{equation}\la{PPtilda}
    \frac{\tilde{\mathcal{P}}}{\mathcal{P}} = (-1)^N(-x)^{N} \prod_{i=1}^N \frac{1}{z_i}\prod_k\frac{x_k^-}{x_k^+}=x^{N}\;.
\end{equation}
Since all zeros of $\mu_{1}{}^{\dot{2}}$ are already accounted for in~\eqref{musol}, the function ${\cal P}$, and as a consequence also $\tilde {\cal P}$, are analytic functions without zeroes or poles outside the cuts. Let us show that
this implies that $N$ should be even. Writing  
\begin{equation}
    N=2 K_{\circ}\,,
\end{equation}
\eq{PPtilda} is formally solved by the following infinite product
\begin{equation}\label{Pprod}
    \mathcal{P} \propto p \prod_{n=-\infty}^{\infty} 
    \left(\frac{1}{x^{[2n]}}\right)^{K_\circ}\,,
\end{equation}
where $p$ is a periodic function without cuts.
Due to \eq{PPtilda}, $\tilde p/p = 1$, i.e. it should be a meromorphic function. Furthermore, we see that $K_\circ$ should be an integer; otherwise $\mathcal{P}$ gets a non-trivial monodromy when going around the cut $[-2h,2h]$.

In \eq{Pprod} extra care should be taken in defining the infinite product. To see this, notice that defining it with a symmetric cut-off $n=\pm \Lambda$,
upon shifting $u\to u+i$ the product changes by a factor $\left(\frac{x^{[-2\Lambda]}}{x^{[2\Lambda+2]}}\right)^{K_\circ}$, which in the limit $\Lambda\to\infty$ gives $(-1)^{K_\circ}$. Hence, depending on the parity of $K_\circ$ the extra factor $p$ is either periodic or anti-periodic. To fix $p$, we recall that from our definition of the gluing matrix $G$ it follows that $\mu_{1}{}^{\dot{2}}$ must have power-like asymptotics while the products in \eqref{musol} can exhibit exponential growth, which has thus to be cancelled by $p$. We postpone the task of fixing $p$ until later in this section because it is more convenient to first analyse the other objects in the construction, while keeping the meromorphic $i$-periodic/anti-periodic for even/odd $K_\circ$  factor $p$ unfixed.

\paragraph{Fixing $Q_{1|1}Q_{\dot 1|\dot 1}$.} 
Having fixed $F$ in \eq{Fsol} we can use \eqref{eq:FFQQ} to find $Q_{1|1}Q_{\dot 1|\dot 1}$. We get the following equation
\begin{equation}
    \frac{(Q_{1|1}Q_{\dot{1}|\dot{1}})^-}{(Q_{1|1}Q_{\dot{1}|\dot{1}})^{+}} = \frac{\betheQ^-}{\betheQ^+} \frac{\kappa}{\Bar{\kappa}} \left(\frac{\bfB_{(+)}}{\bfB_{(-)}}\right)^2\;.
\end{equation}
Since
$(Q_{1|1}Q_{\dot{1}|\dot{1}})^-$ is analytic in the UHP and has power-like asymptotics at infinity, the solution is uniquely fixed up to a constant factor as
\begin{equation}\label{Q11Q11sol}
    Q^-_{1|1}\,Q^-_{\dot{1}|\dot{1}} \propto \betheQ^- \,  f^2\,\prod_{n=0}^\infty \frac{\kappa^{[2n]}}{\bar \kappa^{[2n]}}\;.
\end{equation}

\paragraph{Fixing $\omega$.} Knowing $Q_{1|1}Q_{\dot 1|\dot 1}$ in \eq{Q11Q11sol} 
and $\mu_{1}{}^{\dot{2}}$ in \eq{musol}, we  find $\omega^{1}{}_{\dot 2}$  from \eq{ABAomega}:
\begin{equation}\label{eq:OmegaABA}
\omega^{1}{}_{\dot 2} \propto \frac{  \bar{f}^{[-2]}}{f}\, \mathcal{P}\prod_{n=-\infty}^{\infty}\bar{\kappa}^{[2n]}
\end{equation}
where ${\cal P}$ is given in \eq{Pprod}.

\paragraph{Relation between $\mu_{\dot{1}}{}^{2}$ and  $\mu_{{1}}{}^{\dot 2}$.}
All the arguments above can be repeated interchanging dotted and un-dotted indexes. If we look at the equation \eq{Q11Q11sol},
its l.h.s. does not change under this replacement, so neither should the r.h.s. However, the r.h.s. of \eq{Q11Q11sol} is written in terms of the zeroes of $\mu_{\dot{1}}{}^{2}$ meaning that 
the expression for $\mu_{{1}}{}^{\dot 2}$ should coincide with \eq{musol}. Thus essentially these two quantities can only differ by an overall factor, and due to \eq{ABAomega} the same applies to $\omega^{1}{}_{\dot 2}$ and $\omega^{\dot 1}{}_{2}$.
Thus we conclude the following relation, which will be important below
\beq\la{omega12i}
\frac{\mu_1{}^{\dot 2}}{\mu_{\dot 1}{}^{2}}=
\frac{\omega^1{}_{\dot 2}}{\omega^{\dot 1}{}_{2}} = \frac{\omega_{2}{}^{\dot{1}}}{\omega_{\dot{2}}{}^{1}} = \frac{G_{2}{}^{\dot{1}} \Omega_{\dot{1}}{}^{\dot{1}}}{G_{\dot{2}}{}^{1}\Omega_{1}{}^{1}} =  \frac{\bar{\alpha}}{\alpha}\equiv \zeta
\eeq
where we have used that $\frac{\Omega_{\dot{1}}{}^{\dot{1}}}{\Omega_{1}{}^{1}}$ must be a constant from consistency and can therefore be evaluated at an arbitrary point. Since $\Omega_{k}{}^{l} \simeq \delta^l_k$ for large $x$, we conclude that $\Omega_{1}{}^{1}=\Omega_{\dot{1}}{}^{\dot{1}}$ in the ABA approximation. Finally, we have introduced $\zeta$, a unimodular constant factor, to be found later.

\paragraph{Splitting  $Q_{1|1}$ and  $Q_{\dot 1|\dot 1}$.}
In order to split the product 
$Q_{1|1}Q_{\dot 1|\dot 1}$ in \eq{Q11Q11sol} let us go back to the middle equation of \eq{Qomegamain}, which we evaluate for $a=1,\;\dot k=2$ and we take the ABA approximation, which sets the summation index $l=1$\footnote{Recall that in the gauge \eq{eq:QupQconj}, \eq{eq:PupPconj} we have $\bar Q_{1,1}=-Q^{\uparrow}_{1,1}$.}
\beq
-G_{\dot 2}{}^1 \bar Q^{+}_{1|1} = \omega_{\dot 2}{}^{1}\,Q^{+}_{1|1}
\eeq
where in addition we accommodate the off-diagonal form of the gluing matrix and use complex conjugation to get $\bar Q_{1|1} = Q^\uparrow_{1|1}$. Similarly, for the dotted version we have
\beq
-G_{2}{}^{\dot 1} \bar Q^{+}_{\dot 1|\dot 1} = 
\omega_{2}{}^{\dot 1}\,Q^{+}_{\dot 1|\dot 1}\;.
\eeq
By dividing the two equations, we get
\beq
\frac{\bar Q_{1|1}}{\bar Q_{\dot 1|\dot 1}} =\,\frac{Q_{1|1}}{Q_{\dot 1|\dot 1}}\equiv {\cal M}\;.
\eeq
We see that the r.h.s. is meromorphic for all ${\rm Im}\;u>-1/2$,
whereas the l.h.s. is meromorphic for ${\rm Im}\;u<1/2$, meaning that ${\cal M}$ is a meromorphic function (i.e. no cuts) on the whole complex plane with power-like asymptotic.

Next, multiplying and dividing \eq{Q11Q11sol} by ${\cal M}^-=Q_{1|1}^-/Q_{\dot 1|\dot 1}^-$ we find
\begin{equation}\label{Q11solV2}
(Q^-_{1|1})^2 \propto 
{\cal M}^-
\betheQ^- \,  f^2\,\prod_{n=0}^\infty \frac{\kappa^{[2n]}}{\bar \kappa^{[2n]}}
\,,\qquad \qquad
(Q^-_{\dot 1|\dot 1})^2 \propto 
\frac{1}{{\cal M}^-}
\betheQ^- \,  f^2\,\prod_{n=0}^\infty \frac{\kappa^{[2n]}}{\bar \kappa^{[2n]}}
\;.
\end{equation}
We see that both l.h.s. are analytic functions with cuts, meaning that all poles and zeroes of ${\cal M}$ should coincide with subsets of zeros of ${\mathbb Q}$, but also all zeroes of the polynomial ${\cal M}{\mathbb Q}$ should be double degenerate so we denote
\newcommand{\proptoextra}{\mathrel{\stackrel{\propto}{\rule{0.5em}{0.2pt}}}}
\beq
{\cal M}{\mathbb Q} \proptoextra ({\mathbb Q}_2)^2\,,\quad\frac{1}{\cal M}{\mathbb Q} \proptoextra ({\mathbb Q}_{\dot 2})^2\;\;\Rightarrow\;\;{\mathbb Q}={\mathbb Q}_2{\mathbb Q}_{\dot 2}
\eeq
for monic polynomials ${\mathbb Q}_{2}$ and ${\mathbb Q}_{\dot 2}$. Finally, in order to get $Q_{1|1}$ we need to be able to get the square root of the infinite product factor, which is only possible if we assume that the roots of $\kappa$ are twice degenerate, so we define
\beq\la{kappaisvarkappasquare}
\kappa\equiv \varkappa^2\,,\qquad f_\circ \equiv 
\prod_{n=0}^\infty \frac{\varkappa^{[2n]}}{\bar \varkappa^{[2n]}}\,,\qquad\varkappa=\prod_{i=1}^{K_\circ}(x-z_i)\,,\qquad\bar\varkappa=\prod_{i=1}^{K_\circ}(x-1/z_i)
\eeq
so that we get
\begin{equation}\label{Q11sol}
\boxed{
Q^-_{1|1} \propto 
\betheQ_{2}^- \,  f\,f_{\circ}
\,,\quad
Q^-_{\dot 1|\dot 1} \propto 
\betheQ_{\dot 2}^- \,  f\,f_{\circ}}
\;.
\end{equation}
For what follows it is convenient to also split the factors in ${\bf B}$ and ${\bf R}$ accordingly with the splitting of the roots in ${\mathbb Q}$
\begin{equation}\label{Qseparation}
{\bf B}_{(\pm)}={\bf B}_{2,(\pm)}{\bf B}_{\dot 2,(\pm)}\,,\qquad\qquad
{\bf R}_{(\pm)}={\bf R}_{2,(\pm)}{\bf R}_{\dot 2,(\pm)}\,,
\end{equation}
so that $\bfB_{A,(\pm)}\bfR_{A,(\pm)}=h^{-K_A}{\mathbb Q}_A^\pm$ for $A={2,\dot 2}$.

\paragraph{Fixing $p$. }
At this point we return to the question of fixing the periodic, meromorphic function $p(u)$, defined in \eq{Pprod}, by requiring that $\mu$ has polynomial asymptotics as follows from \eqref{eq:muLowerUpper}. In \eq{kappaisvarkappasquare} we understood that all roots of $\kappa$ have a double degeneracy $\kappa = \varkappa^{2}$, using this we can now rewrite \eqref{musol} as
\begin{equation}\label{eq:MuRecalculated}
\begin{split}
    \mu_{1}{}^{\dot{2}} &\propto \mathcal{P}\,\betheQ^{-}f \bar{f}^{[-2]}\prod_{n=0}^{\infty} \(\varkappa^{[2n]} \bar{\varkappa}^{[-2n-2]}\)^2 \\
    &\propto \betheQ^{-}  p f \bar{f}^{[-2]}\prod_{k=1}^{K_{\circ}}\left( \sinh{\pi(u-\theta_k)}\prod_{n=1}^{\infty}\left(\frac{1-\frac{z_k}{x^{[2n-2]}}}{1-\frac{1}{z_k x^{[2n-2]}}}\right) \left(\frac{1-\frac{1}{z_k x^{[-2n]}}}{1-\frac{z_k}{ x^{[-2n]}}}\right)\right)\,,
\end{split}
\end{equation}
where $\theta_k \equiv h(z_k+\frac{1}{z_k})$. In the previously studied case of the BFKL ABBA for AdS$_5$ \cite{Ekhammar:2024neh} the exponential asymptotics from $\sinh$ was expected from analytic continuation in spin \cite{Gromov:2015wca}.
Since the AdS$_3$ QSC presently considered captures local operators, we instead expect polynomial asymptotics. In order to ensure power-like behaviour of $\mu$,
we must have $p\sim e^{-\pi |u| K_\circ}$ at large $u$. Since $p$ is a meromorphic function, it must therefore have $K_\circ$ poles in each periodicity strip, since massless modes appear as zeroes of $\mu_1{}^{\dot{2}}$. However, as we now know, it may actually
have double zeroes at $\theta_k$ due to \eq{eq:MuRecalculated}. With the choice of $p\propto \prod_{k=1}^{K_{\circ}}\frac{1}{\sinh{\pi(u-\theta_k)}}$ we kill both issues (the exponential growth and double zeroes) at the same time! This results in removing the $\sinh$-factor in \eqref{eq:MuRecalculated}\footnote{In this way we also create poles on the lower part of the branch cut, which, however, appear on the next sheet on the natural section of the Riemann surface of $\mu$ where it is $i$-periodic. These poles are thus likely artefacts of the ABA limit and appear due to non-commutativity of the analytic continuation and large volume limit, which is expected. These poles also appear in the AdS$_{5}$ context~\cite{Ekhammar:2024neh}, where the situation is under complete numerical control.}. 
As we will see below, this natural choice of $p$ leads to nice additional simplifications in our expressions.

After fixing $p$ we can now write
\begin{equation}
\boxed{
    \mu_{1}{}^{\dot{2}} \propto \betheQ^- f \bar{f}^{[-2]} f_{\circ}\bar{f}^{[-2]}_{\circ}\;,
    }
\end{equation}
which exhibits a pleasing symmetry between massive and massless modes. It is natural to think about $f_{\circ}$ as the ``massless" limit of $f$ obtained by sending $x_{k}^{\pm} \rightarrow (z_k)^{\pm 1}$. Using this notation we also find for $\omega^1{}_{\dot 2}$
\begin{equation}
\boxed{
    \omega^{1}{}_{\dot{2}} \propto \frac{\bar{f}^{[-2]} \bar{f}^{[-2]}_{\circ}}{f f_{\circ}}\;.
    }
\end{equation}

\paragraph{Energy formula.}
In the QSC formalism, the energy $\gamma$ is encoded in the large $u$ asymptotics, whereas in the ABA it is written in terms of momentum-carrying Bethe roots.
To find the expression for the energy we can use the large $u$ asymptotics of $Q_{1|1}$ and $Q_{\dot 1|\dot 1}$. From~\eq{Q11sol} and \eq{Ppow} we find
\begin{equation}
Q^-_{1|1}Q^-_{\dot 1|\dot 1}\propto    \betheQ^- \,  f^2 f_{\circ}^2 \sim u^{\Delta-J}=u^{K_2+K_{\dot{2}}+\gamma}\,.
\end{equation}
where $\gamma$ is the contribution coming from $f^2f^2_{\circ}$. The large-$u$ asymptotics of $f$ can be found by noting that
\begin{equation}
    \log\frac{f}{f^{[2]}}=\log\frac{\bfB_{(+)}}{\bfB_{(-)}}
    \sim \frac{1}{x}\sum_{k=1}^M\left(\frac{1}{x_k^+}-\frac{1}{x_k^-}\right)\sim \frac{\gamma_\bullet}{2 \ii u}\,,
\qquad
    \gamma_{\bullet} \equiv 2\ii h\sum_{k=1}^{M} \left(\frac{1}{x^{+}_k} - \frac{1}{x^{-}_k}\right)\,,
\end{equation}
from which it follows that
$    f = u^{\frac{\gamma_{\bullet}}{2}}
$.
Analogously, one can show that
\begin{equation}\la{gammacirc1}
    f_{\circ} = \prod_{n=0}^\infty \frac{\varkappa^{[2n]}}{\bar \varkappa^{[2n]}} \sim u^{\frac{\gamma_{\circ}}{2}}\,,\qquad\qquad
    \gamma_{\circ} \equiv 2\ii h \sum_{i=1}^{K_{\circ}}\left(\frac{1}{z_i}-z_i\right)\,.
\end{equation}
As a result, we get the following relation
$\Delta-J =K_2+K_{\dot{2}}+\gamma_{\circ}+\gamma_{\bullet}$. Because this should be valid for all $h$, we can separate the bare and anomalous parts
\begin{align}
    \Delta_{h=0}-J =K_2+K_{\dot{2}}\,,\qquad\qquad \gamma = \gamma_{\circ}+\gamma_{\bullet}\;,
\end{align}
comparing to the ABA expression~\eq{ABAenergy} we find a perfect match for both energy and momentum!

\subsection{Reconstructing $\bP$ and $\bQ$}
\paragraph{Fixing $\bP$.}
To find $\bP$ and $\bQ$, one can use similar  arguments to those used in AdS$_5$~\cite{Gromov:2014caa} and later adopted to AdS$_3$ in \cite{Cavaglia:2022xld}. Like in~\cite{Cavaglia:2021eqr,Ekhammar:2021pys}, without loss of generality, let us define\footnote{Here we use rather inconvenient historic notation for the indices of $R$ and $B$ which contains the auxiliary roots. The operation of dotting amounts thus to the simultaneous introduction/removal of a dot as well as introduction/removal of tilde and the replacement $1\leftrightarrow 3$.}
\begin{align}\label{eq:Q-ide}
\bP_1 &= {\mathbb A}_1\, x^{-L_1/2}R_{\tilde{1} } B_{\tilde{\dot  1}} \,\mcS \,  \,, & 
\bP^2 &= {\mathbb A}^2\, x^{-L_2/2}R_{\tilde 3} B_{\tilde{\dot 3}} \,\mcS'\,  \,,\\
\bP_{\dot{1}} &={\mathbb A}_{\dot 1}\, x^{-L_{\dot{1}}/2}R_{\dot{3} } B_{3} \,\mcSd  \,, 
& 
\bP^{\dot{2}} &= {\mathbb A}^{\dot 2}\, x^{-L_{\dot{2}}/2}R_{\dot{1}} B_{1} \,\mcSd' \,,
\end{align}
with 
\begin{equation}\label{eq:DefRB}
B_\alpha=\prod_{j=1}^{K_\alpha}\left(\frac{1}{x}-y_{\alpha,j}\right)\,,\qquad\qquad
    R_\alpha=\prod_{j=1}^{K_\alpha}(x-y_{\alpha,j})\,,
\end{equation}
and $\alpha={1,3,\dot 1,\dot 3}$, and analogously in the case of dual roots $\tilde\alpha={\tilde 1,\tilde 3,\tilde{\dot 1},\tilde{\dot 3}}$ with all $|y|> 1$.

By definition, the monic $x$-polynomial factors $R$ contain all zeroes of the corresponding $\bP$ on the main sheet, that is outside the cut. The $B$ factors are polynomials in $1/x$, which have roots for $|x|<1$. We define these roots to be related to the roots of ${\bf Q}$ such that $R_{\tilde{\dot 1}}(x)=B_{\tilde{\dot 1}}(1/x)$ contains all roots of ${\bf Q}^{\dot 2}$ on the main sheet, 
$R_{\tilde{\dot 3}}$ contains the roots of ${\bf Q}^{\dot 1}$ and
similarly relate $R_{1}$ to ${\bf Q}_1$ and $R_{3}$ to ${\bf Q}^2$.
The factors ${\cal S}$ are then so far only restricted to be analytic, have no zeroes on the main sheet outside the cut and have unit asymptotics $u\to\infty$. The extra powers of $x$ and the constant ${\mathbb A}$s are there to ensure the correct asymptotics as postulated in \eq{Ppow}. 

Note that, due to \eq{Qdot} ,the ratio of $\bP_1$ and $\bP^2$ is a rational function of $x$
\beq\la{PPr}
\frac{\bP_1}{\bP^2}=\frac{{\mathbb A}_1}{{\mathbb A}^2}x^{-L_1/2+L_2/2}\frac{R_{\tilde 1}B_{\tilde{\dot 1}}}{
R_{\tilde 3}B_{\tilde{\dot 3}}
}\frac{{\cal S}}{{\cal S}'} = r(u)\,,
\eeq
from which we conclude that $\frac{\cal S}{{\cal S}'}$ is also a rational function of $x$.
Note that by definition of ${\cal S}$ and ${\cal S}'$ it has unit asymptotic at large $x$ and has no zeroes or poles for $|x|>1$. Let us now show that it also has no zeroes or poles for $|x|<1$. We use that $r^\gamma=\dot r=-\frac{\bQ^{\dot 2}}{\bQ_{\dot 1}}$, which then gives
\beq
\(\frac{{\cal S'}}{{\cal S}}\)^\gamma=
-\frac{{\mathbb A}_1}{{\mathbb A}^2}x^{+L_1/2-L_2/2}\frac{B_{\tilde 1}R_{\tilde{\dot 1}}}{
B_{\tilde 3}R_{\tilde{\dot 3}}
}\frac{\bQ_{\dot 1}}{\bQ^{\dot 2}}\,.
\eeq
Note that all zeroes and poles on the r.h.s. 
cancel between $R$s and $\bQ$s by definition of $R_{\tilde{\dot 1}}$ and $R_{\tilde {\dot 3}}$ (see the paragraph below \eq{eq:DefRB}) and there are no other singular or vanishing factors on the r.h.s. for $|x|>1$. Hence, the ratio $\frac{{\cal S'}}{{\cal S}}$
is a rational function of $x$, which is regular and has unit asymptotics, which implies that\footnote{We further assume that $r$ has no zeroes or poles at $|x|=1$, which should be the case for generic states, but may require extra care in some fine-tuned cases.}
\beq\la{SSpr}
{\cal S}'={\cal S}\,,\qquad\qquad\dot{\cal S}'=\dot{\cal S}\;,
\eeq
where we applied the same argument to the dotted quantities. With the relation \eq{SSpr},
from \eq{PPr} and its dotted version we find
\beq\la{PPr2}
r(u)=\frac{{\mathbb A}_1}{{\mathbb A}^2}x^{\frac{L_2-L_1}2}\frac{R_{\tilde 1}B_{\tilde{\dot 1}}}{
R_{\tilde 3}B_{\tilde{\dot 3}}
} = \frac{{\mathbb A}_{\dot 1}}{{\mathbb A}^{\dot 2}}x^{\frac{L_{\dot 1}-L_{\dot 2}}2}\frac{R_{{ 3}}B_{\dot 3}}{
R_{{1}}B_{\dot 1}
}\,.
\eeq
Comparing the poles and zeroes for the two sides of the second equality implies
\beq\la{LsandRs}
-L_1+L_2=+L_{\dot 1}-L_{\dot 2}\,,\qquad\qquad
R_{{1}}B_{\dot 1}
{R_{\tilde 1}B_{\tilde{\dot 1}}} =
\frac{{\mathbb A}^2}{{\mathbb A}_1}
\frac{{\mathbb A}_{\dot 1}}{{\mathbb A}^{\dot 2}}
R_{{ 3}}B_{\dot 3}
R_{\tilde 3}B_{\tilde{\dot 3}}
\;.
\eeq
\paragraph{$\Lambda$-gauge fixing.}
To reduce the number of parameters in our solution we can take advantage of the $\Lambda$-symmetry~\eq{lambdasym}. We fix the allowed powers $\Lambda_{Q}$ by requiring consistency of QQ-relations and the $\bP\mu$-system after the transformation, which leads to the following transformation for the $L$s
\begin{align}
    L_{1} \rightarrow L_{1}+\Lambda\,,
    \qquad
    L_{2} \rightarrow L_{2}-\Lambda\,,
    \qquad
    L_{\dot{1}} \rightarrow L_{\dot{1}} - \Lambda\,,
    \qquad
    L_{\dot{2}} \rightarrow L_{\dot{2}}+\Lambda\,.
\end{align}
We can fix the $\Lambda$-gauge $L_2=L_1$, which together with~\eq{LsandRs} implies that 
\beq
L_1=L_2\equiv \dot L\,,\qquad\qquad L_{\dot 1}=L_{\dot 2}\equiv L\;.
\eeq

\paragraph{Fixing $\bQ$.}
Similarly, we start from the following general ansatz, which takes into account zeroes of $\bQ$'s, but otherwise is completely general.
\begin{align}
\bQ_1  &\propto {x^{L/2+K_\circ/2} }\, R_{1} B_{\dot 1}
{\cal T}\frac{{\bf B}_{2,(-)}}{{\bf B}_{\dot 2,(+)}} \frac{f f_{\circ}}{{\cal S}\varkappa}\;, &
\bQ^2 &\propto   {x^{{L}/2+K_\circ/2} } \, R_{3} B_{\dot 3} {\cal T}'\frac{{\bf B}_{2,(-)}}{{\bf B}_{\dot 2,(+)}} \frac{f f_{\circ}}{{\cal S}\varkappa}
\;,\label{eq:Q-ide-last}
\end{align}
and its dotted version
\begin{align}
\bQ_{\dot 1}  &\propto {x^{{\dot L}/2+K_\circ/2} }\, R_{\tilde{\dot 3}} B_{\tilde 3}
\dot{\cal T}\frac{{\bf B}_{\dot 2,(-)}}{{\bf B}_{2,(+)}} \frac{f f_{\circ}}{\dot{\cal S}\varkappa}\;, &
\bQ^{\dot 2} &\propto   {x^{{\dot L}/2+K_\circ/2} } \, R_{\tilde{\dot 1}} B_{\tilde 1} \dot{\cal T}'\frac{{\bf B}_{\dot 2,(-)}}{{\bf B}_{2,(+)}} \frac{f f_{\circ}}{\dot {\cal S}\varkappa}
\;.\label{eq:Q-ide-last-dotted}
\end{align}
Using $\frac{\bQ^2}{\bQ_1}=-r$ and \eq{PPr} we find ${\cal T}={\cal T}'$ and $\dot {\cal T}=\dot {\cal T}'$ for the dotted version.
To further constrain ${\cal T}$ we use the first QQ-relation in \eq{eq:QQtwoeq} i.e. $Q_{1|1}^+-Q_{1|1}^-=\bP_1 \bQ_1$, which gives from \eq{Q11sol}
\beq\la{dualdot1}
{\bf R}_{2,(+)}
{\bf B}_{\dot 2,(-)}
\bar\varkappa-
{\bf R}_{2,(-)}
{\bf B}_{\dot 2,(+)}
\varkappa\,\,
 \propto\,\,
 {x^{\frac{ L-\dot L+K_\circ}{2}} }\, R_{1} B_{\dot 1}
R_{\tilde{1} } B_{\tilde{\dot  1}} \,{\cal T}\;.
\eeq
We see that ${\cal T}$ is a rational function of $x$. The dotted version of this relation is
\beq\la{dualdot}
{\bf R}_{\dot 2,(+)}
{\bf B}_{2,(-)}
\bar\varkappa-
{\bf R}_{\dot 2,(-)}
{\bf B}_{2,(+)}
\varkappa
 \,\,\propto\,\,
 {x^{\frac{\dot L-L+K_\circ}{2}} }\, 
 R_{\dot 3 } B_{3}
 R_{\tilde {\dot 3}} B_{\tilde  3}
 \,\dot{\cal T}\;.
\eeq
Note that the two equations become very similar under $x\to 1/x$. The l.h.s. of
\eq{dualdot} under this transformation becomes
\beq
\frac{{\bf B}_{\dot 2,(+)}
{\bf R}_{2,(-)}
\varkappa\, x^{-K_\circ}}{\prod_{j=1}^{K_{\dot 2}}(-1/x_{\dot 2,j}^-)\prod_{j=1}^{K_{2}}(-x_{ 2,j}^+)\prod_{j=1}^{K_{\circ}}(-z_k)}-
\frac{{\bf B}_{\dot 2,(-)}
{\bf R}_{2,(+)}
\bar\varkappa\;x^{-K_\circ}}{
\prod_{j=1}^{K_{\dot 2}}(-1/x_{\dot 2,j}^+)\prod_{j=1}^{K_{2}}(-x_{ 2,j}^-)\prod_{j=1}^{K_{\circ}}(-1/z_k)
}
\eeq
the factors in the denominators are in fact equal due to the cyclicity condition \eq{zeromom}.
Up to a constant factor and $x^{-K_\circ}$ it coincides with the l.h.s. of \eq{dualdot1} and thus combining the two equations we get
\beq\la{Tcaldot}
\,{\cal T}(x)
\propto
\,\dot{\cal T}(1/x)\;,
\eeq
where we used \eq{LsandRs}. 

Let us summarise what we know about ${\cal T}$ and $\dot{\cal T}$: they are rational functions of $x$, by definition they do not have zeroes or poles for $|x|>1$,
and from \eq{Tcaldot} that is also true for $|x|<1$. From \eq{dualdot1} we see that no poles are possible at $|x|=1$ as well as zeroes could be absorbed into $R$-factors. In this case we conclude that ${\cal T}$ should be a power of $x$ 
\beq
{\cal T}=x^{K_{\cal T}}\,,\qquad\qquad\dot {\cal T}=1/{\cal T}\;.
\eeq
We can choose the above normalization of ${\cal T}$ and $\dot {\cal T}$, since it is defined only up to a constant in~\eqref{eq:Q-ide-last}.

\paragraph{Fixing ${\cal S}$.}
We will now derive discontinuity relations on ${\cal S}$. These relations can be thought of as ``half-crossing," because we will show that applying them twice one can get the usual crossing relation that the massive dressing phases appearing in ABAs must satisfy. Thus, these relations are in principle more constraining.

From the $\bP\mu$-system~\eqref{Pmumain}, using the definition of $\mu$ in terms of $\omega$ in~\eqref{muomega}, as well as the assumption that $\omega^{1}{}_{\dot{2}}$ is the leading component of $\omega$, we find that in the ABA limit
\begin{equation}\label{eq:ACABA0}
    (\bP_{1})^{\gamma} \simeq  Q_{1|1}^+ \omega^{1}{}_{\dot{2}}\bQ^{\dot{2}}
 \,.
\end{equation}
Plugging in the explicit expressions for Q-functions from~\eqref{eq:Q-ide-last}-\eqref{eq:Q-ide-last-dotted},  we get
\beqa
x^{\dot{L}/2}{\mathcal{S}} ^{\gamma}
B_{\tilde{1} } R_{\tilde{\dot  1}}&\propto&
\({\mathbb Q}_2^+ 
f^{++} f_{\circ}^{++}
\)
\(
\frac{\bar f}{
f^{++}
}
\frac{\bar{f}_{\circ}}{f^{++}_{\circ}}
\)
\(
x^{\frac{\dot{L}+K_\circ}{2}-K_{\cal T}}
R_{\tilde {\dot 1}}
B_{\tilde 1}
\frac{{\bf B}_{\dot 2,(-)}}{{\bf B}_{2,(+)}}
\frac{f  f_{\circ}}{{\dot{\cal S}\varkappa}}
\)\;,
\eeqa
which simplifies to
\beq\label{AhalfcrossingV2}
({\mathcal{S}} )^{\gamma} \, \dot{\mathcal{S}}
\propto  \frac{{\bf B}_{\dot 2,(-)}{\bf R}_{2,(+)}}{\varkappa}  \left( 
\;\frac{\bar{f}^{--} f^{++} \bar{f}^{--}_{\circ}f^{++}_{\circ}}{
x^{K_{\cal T}-\frac{K_\circ}{2}}}
\right)\;.
\eeq

To better parameterise $\mathcal{S}$, we introduce $\sigma_{\bullet}$ and $\sigma_{\circ}$, which we take to be real analytic functions with no zeroes on the main sheet, 
satisfying $\sigma_\bullet,\;\sigma_{\circ}\to 1$
when $u\to\infty$ and
\begin{align}\label{sigmahalfcrossing}
&\sigma_{\bullet}^\gamma\sigma_{\bullet} \propto f(u+i)\bar f(u-i)\,,\qquad\qquad
\sigma_{\circ}^\gamma\sigma_{\circ} \propto f_{\circ}(u+i)\bar{f}_{\circ}(u-i)\;.
\end{align}
These objects are natural building blocks of the BES-scattering phase and its ``massless'' limit. Since the left-hand side of \eqref{sigmahalfcrossing} does not have a cut on the real axis, it follows that $\sigma_{\bullet}$ and $\sigma_{\circ}$ are functions with a square-root cut. We present explicit expressions for $\sigma_{\bullet}$ and $\sigma_{\circ}$ in Subsection~\ref{subsec:ExplicitDressing} by solving~\eq{sigmahalfcrossing}. 

Let us now parameterize $\mathcal{S},\dot{\mathcal{S}}$ as
\begin{equation}\label{eq:Srhodef}
    \mathcal{S} = \sqrt{\frac{\bfB_{2,(+)}\bfB_{2,(-)}}{x^{-2K_{\circ}}\varkappa\bar{\varkappa}}} \sigma_{\bullet} \sigma_{\circ} \rho\,,\qquad
    \qquad
    \dot{\mathcal{S}} = \sqrt{\frac{\bfB_{\dot{2},(+)}\bfB_{\dot{2},(-)}}{x^{-2K_{\circ}}\varkappa\bar{\varkappa}}} \sigma_{\bullet} \sigma_{\circ} \dot{\rho}\,,
\end{equation}
where $\rho$ and $\dot\rho$, like ${\cal S}$'s, are real functions, tending to $1$ at infinity and analytic, with no zeros on the main sheet outside the cuts. The factors of $\bfB$, $\varkappa$ and $\bar{\varkappa}$ are included to simplify the comparison with dressing phases from the S-matrix literature and they also approach $1$ at infinity. Furthermore, from \eq{sigmahalfcrossing} we get
\beq\label{eq:lastRH}
( \rho )^{\gamma} \dot\rho \propto \sqrt{\frac{R_{2,(+)}}{R_{2,(-)}}\; \frac{B_{\dot 2,(-)}}{B_{\dot 2, (+)}} }\frac{\bar\varkappa}{x^{\frac{K_{\circ}}2+K_{\cal T}}}\,,\qquad\qquad (\dot\rho )^{\gamma} \,\rho \propto \sqrt{\frac{R_{\dot 2,(+)}}{R_{\dot 2,(-)}}\; \frac{B_{ 2,(-)}}{B_{ 2, (+)}} }\frac{\bar\varkappa}{x^{\frac{K_{\circ}}2-K_{\cal T}}}\;.
\eeq
Note that above we switched back to $R_{2,(\pm)}$ and $B_{2,(\pm)}$ functions \eq{Bnonbf} with the extra constant factors for convenience as they transform into each other in a simple way under the analytic continuation.
Below we will construct the functions, which satisfy the following relations, splitting massive and massless parts in \eq{eq:lastRH}
\begin{align}\label{rhobullet}
    &\left(\rho_{\bullet} \right)^{\gamma}\dot{\rho}_{\bullet} \propto \sqrt{\frac{R_{2,(+)}}{R_{2,(-)}}\frac{B_{\dot{2},(-)}}{B_{\dot{2},(+)}}}\,,
    &
    &\left(\dot{\rho}_{\bullet}\right)^{\gamma} \rho_{\bullet} \propto \sqrt{\frac{R_{\dot{2},(+)}}{R_{\dot{2},(-)}}\frac{B_{2,(-)}}{B_{2,(+)}}}\,,
\end{align}
and
\begin{align}\label{rhocirc2}
    &\rho_{\circ} \left({\rho}_{\circ}\right)^{\gamma}\propto\frac{\bar{\varkappa}}{\prod_{k=1}^{K_{\circ}}\sqrt{ x/ z_k}} \,,
\end{align}
and being analytic, non-zero at $|x|>1$ and approaching $1$ at infinity. In  Appendix~\ref{sec:fixingKT} we show that $K_{\cal T}=0$ and that
\beq\label{rhosplit}
\rho=\rho_\bullet \rho_\circ\;,
\qquad\qquad
\dot{\rho}=\dot{\rho}_\bullet {\rho}_\circ\;.
\eeq
We discuss the solutions to the discontinuity equations \eqref{rhobullet} and \eqref{rhocirc2} in Subsection~\ref{subsec:ExplicitDressing} and give further details in Appendix~\ref{app:solving_crossing}.
    
\paragraph{Summary of $\bP$ and $\bQ$.}
Let us finally summarise the expressions for $\bP$ and $\bQ$ we found in this section:
{\footnotesize
\begin{align}\label{eq:Q-ide2}
\bP_1 &\propto x^{-\dot L/2} \,\sqrt{\frac{\bfB_{2,(+)}\bfB_{2,(-)}}{x^{-2K_{0}}\varkappa\bar{\varkappa}}}\,\sigma_{\bullet}\sigma_{\circ}\rho \, R_{\tilde{1} } B_{\tilde{\dot  1}} \,, & 
\bP^2 &\propto x^{-\dot L/2} \,\sqrt{\frac{\bfB_{2,(+)}\bfB_{2,(-)}}{x^{-2K_{\circ}}\varkappa \bar{\varkappa}}}\,\sigma_{\bullet}\sigma_{\circ}\rho\,  R_{\tilde 3} B_{\tilde{\dot 3}} \,,\\
\bQ_1  &\propto \frac{{x^{L/2-K_{\circ}/2}}}{\sigma_{\bullet} \sigma_{\circ} \rho}   R_{1} B_{\dot 1}
\sqrt{\frac{{\bf B}_{2,(-)}}{{\bf B}_{2,(+)}}} \frac{ff_{\circ}}{{\bf B}_{\dot 2,(+)}} \sqrt{\frac{\bar{\varkappa}}{\varkappa}}
 &
\bQ^2 &\propto  \frac{x^{{L}/2-K_{\circ}/2}}{\sigma_{\bullet}\sigma_\circ\rho}  R_{3} B_{\dot 3} \, 
\sqrt{\frac{{\bf B}_{2,(-)}}{{\bf B}_{2,(+)}}}\frac{f f_{\circ}}{{\bf B}_{\dot 2,(+) } }
\sqrt{\frac{\bar{\varkappa}}{\varkappa}}
\;,\label{eq:Q-ide-last2}
\end{align}}
and for the dotted system
{\footnotesize
\begin{align}\label{eq:Qd-ide3}
\bP_{\dot{1}} &\propto x^{-L/2} \,\sqrt{\frac{\bfB_{\dot{2},(+)}\bfB_{\dot{2},(-)}}{x^{-2K_{\circ}}\varkappa\bar{\varkappa}}}\,\sigma_{\bullet}\sigma_{\circ}\dot{\rho}\, R_{\dot{3}} B_{3} \,,  
&\bP^{\dot{2}} &\propto x^{-L/2} \,\sqrt{\frac{\bfB_{\dot{2},(+)}\bfB_{\dot{2},(-)}}{x^{-2K_{0}}\varkappa\bar{\varkappa}}}\,\sigma_{\bullet}\sigma_{\circ}\dot{\rho}\,  R_{\dot{1}} B_{1} \,,
\\ \bQ_{\dot{1}}  &\propto \frac{{x^{\dot L/2-K_{\circ}/2} }}{\sigma_{\bullet}\sigma_{\circ}\dot{\rho}}   R_{\tilde{\dot{3}}} B_{\tilde 3}
\sqrt{\frac{{\bf B}_{\dot{2},(-)}}{{\bf B}_{\dot{2},(+)}}} \frac{f}{{\bf B}_{2,(+)}} \sqrt{\frac{\bar{\varkappa}}{\varkappa}}
f_{\circ}, &
\bQ^{\dot{2}} &\propto  \frac{x^{\dot L/2-K_{\circ}/2}}{\sigma_{\bullet}\sigma_{\circ}\dot{\rho}}  R_{\tilde{\dot{1}}} B_{\tilde{1}} \, 
\sqrt{\frac{{\bf B}_{\dot 2,(-)}}{{\bf B}_{\dot 2,(+)}}} \frac{f}{{\bf B}_{2,(+)}}
\sqrt{\frac{\bar{\varkappa}}{\varkappa}}
f_{\circ}
\;.\label{eq:Qd-ide-last3}
\end{align}
}

\paragraph{Quantum numbers.}
Finally, let us express the charges in terms of the roots of $L,\gamma$ and root numbers. From \eqref{dualdot1} and \eqref{dualdot} the dual root numbers are constrained as
\beqa
K_{\tilde 1}+K_1&=&K_{\tilde 3}+K_3=K_2-1+\frac{K_{\circ }+\dot L-{L}}{2}\;,\\
K_{\tilde {\dot 1}}+K_{\dot 1}&=&K_{\tilde {\dot 3}}+K_{\dot 3}=K_{\dot 2}-1+\frac{K_{\circ }-\dot L+{L}}{2}\;.
\eeqa
Comparing \eqref{Ppow} with explicit expressions for the Q-functions we deduce
\begin{equation}
\begin{aligned}
&\Delta = \gamma + L + K_{\dot 2} + \frac{1}{2}\left( {K_{1}} + {K_{3}} - {K_{\dot 1}} - {K_{\dot 3}} - {K_\circ} \right)\,,
\\
&J = L - K_{2} + \frac{1}{2}\left( {K_{1}} + {K_{3}} - {K_{\dot 1}} - {K_{\dot 3}} - {K_\circ} \right)\,,
\\
&S = - K_{\dot{2}} + \frac{1}{2}\left( {K_{1}} + {K_{3}} + {K_{\dot 1}} + {K_{\dot 3}} - {K_\circ} \right)\,,
\\
&K = - K_{2} + \frac{1}{2}\left( {K_{1}} + {K_{3}} + {K_{\dot 1}} + {K_{\dot 3}} - {K_\circ} \right)\,,
\\
&\hat{B} = K_{1} - K_{3}\,,
\\
&\check{B} = K_{\dot 1} - K_{\dot 3}\,.
\end{aligned}
\label{asms}
\end{equation}

\subsection{Building Blocks For Dressing Factors}\la{sec:buildingblocks}
In this section, we introduce elementary building blocks for the factors $\sigma$ and $\rho$. We will use these building blocks when discussing solutions of the crossing equations in Subsection~\ref{subsec:ExplicitDressing}.

\paragraph{Introducing $\varrho_\bullet,\;\dot\varrho_{\bullet},\;\sigma_\bullet$.}
We first introduce $\varrho_\bullet(u,v)$
and $\varsigma_\bullet(u,v)$ which we require to be real,  analytic without zeroes on the main sheet, approach $1$ at infinity and finally satisfy
\begin{align}\label{rhobulletblocks}
    &\varrho_{\bullet}(x^\gamma,v)\dot{\varrho}_{\bullet}(x,v) \propto 
    \sqrt{\frac{x-y^-}{x-y^+}}\,,
    &
    &\dot{\varrho}_{\bullet}(x^{\gamma},v)\varrho_{\bullet}(x,v) \propto 
    \sqrt{\frac{1/x-y^+}{1/x-y^-}}\,,
\end{align}
with $y^\pm = x^{\pm}(v)$.
Then $\rho_\bullet$ and $\dot\rho_\bullet$ are built as follows
\beq
\rho_\bullet(x) = \prod_{j=1}^{K_{2}} \varrho_{\bullet}(x,u_{2,j})
\prod_{j=1}^{K_{\dot 2}} \dot\varrho_{\bullet}(x,u_{\dot 2,j})\,,\qquad
\dot\rho_\bullet(x) = \prod_{j=1}^{K_{\dot 2}} \varrho_{\bullet}(x,u_{\dot 2,j})
\prod_{j=1}^{K_{2}} \dot\varrho_{\bullet}(x,u_{2,j})\,,
\eeq
so that \eq{rhobullet} is satisfied as a consequence of \eq{rhobulletblocks}.
Similarly, to parameterise the BES phase we define $\varsigma_\bullet$ 
\beq\la{varsigma-bullet-crossing}
\varsigma_\bullet(x^\gamma,v)
\varsigma_\bullet(x,v) \propto\prod_{n=1}^\infty \frac{1-\frac{1}{x^{[+2n]}y^-}}{1-\frac{1}{x^{[+2n]}y^+}}
\frac{1-\frac{1}{x^{[-2n]}y^+}}{1-\frac{1}{x^{[-2n]}y^-}}\;,
\eeq
writing once again $y^\pm = x^{\pm}(v)$ and
from which $\sigma_\bullet$ is obtained as
\beq\label{eq:IdentificationSigmaUV}
\sigma_\bullet(x) = \prod_{j=1}^{K_{2}} \varsigma_{\bullet}(x,u_{2,j})
\prod_{j=1}^{K_{\dot 2}} \varsigma_{\bullet}(x,u_{\dot 2,j})\;.
\eeq

\paragraph{Introducing $\varrho_\circ,\;\varsigma_\circ$.}
We turn next to the massless objects and introduce two elementary building blocks: $\varsigma_\circ$ and $\varrho_\circ$ which we require to have all good analytic properties on the main sheet. $\varsigma_{\circ}$ is defined to satisfy
\begin{equation}
\begin{split}
\la{varrho-circ-cross}
\varsigma_\circ(x^\gamma,y)
\varsigma_\circ(x,y) &\propto \prod_{n=1}^\infty
\frac{\left(x^{{[+2n]}}-y\right) \left(x^{{[-2n]}} -1/y\right)}{\left(x^{{[-2n]}}-y\right) \left(x^{{[+2n]}}-1/y\right)}\,,
\end{split}
\end{equation}
such that 
\begin{equation}
    \sigma_\circ(x) = \prod_{j=1}^{K_{\circ}} \varsigma_\circ(x,z_j)\,.
\end{equation}
Similarly, we define
\beqa\la{varrho-gamma-circ}
\varrho_\circ(x^\gamma,y)\varrho_\circ(x,y)\propto
\frac{x-1/y}{\sqrt{x/y}}\;,
\eeqa
giving $\rho_\circ$ via 
\begin{equation}
\rho_\circ(x) = \prod_{j=1}^{K_{\circ}} \varrho_{\circ}(x,z_j)\,.
\end{equation}

\subsection{Explicit Expressions for the Dressing Factors}\label{subsec:ExplicitDressing}
In Appendix~\ref{app:solving_crossing}, we derive the solutions of the discontinuity equations \eq{rhobulletblocks}, \eq{varsigma-bullet-crossing}, \eq{varrho-circ-cross} and \eq{varrho-gamma-circ}. Our solutions are in general more constrained compared to those derived from S-matrix considerations, which are known to have multiple solutions and require additional physics input or simplicity arguments in order to pinpoint the correct solution. In the remainder of this subsection we simply present the final expressions. 

Firstly, the expressions for $\varsigma_\bullet$ and $\varsigma_\circ$ are related to the BES dressing phase~\cite{Beisert:2006ez,Dorey:2007xn}. Let us introduce the double integral
\beq
\chi(x,y)=
-\oint \frac{dz}{2\pi \ii} 
    \frac{1}{x-z}
    \oint \frac{dw}{2\pi \ii} \frac{1}{y-w}\log\left( \frac{\Gamma\left(1+\ii u_{z}-\ii u_{w}\right)}{\Gamma\left(1-\ii u_{z}+\ii u_{w}\right)}\right)
\eeq
where we assume the integration contours to be slightly inside the unit circle. In terms of $\chi(x,y)$ we have
\begin{equation}\label{eq:SigmaMassiveMain}
    \log \varsigma_{\bullet}(x,y) = \chi(x,y^+)-\chi(x,y^-)\,,
\end{equation}
and
\begin{equation}\label{eq:SigmaMasslessMain}
    \log \varsigma_{\circ}(x,y) = \chi(x,y)-\chi(x,1/y)\,.
\end{equation}

Similar integral representations for $\varrho_\circ$ and $\varrho_\bullet$ are relegated to Appendix~\ref{app:solving_crossing}. Here we instead present the result in terms of  polylogs, which can be more convenient since they make explicit that each term is analytic outside the unit circle. The polylog expressions are
{\footnotesize
\beqa\la{polylogs1}
\nn{\varrho_\bullet} &\propto&\exp\[
-\frac{i \text{Li}_2\frac{(1-x) \left(y^-+1\right)}{(x+1) \left(y^--1\right)}}{2 \pi }+\frac{i \text{Li}_2\frac{(1-x) \left(y^++1\right)}{(x+1) \left(y^+-1\right)}}{2 \pi }
-\frac{i \log \frac{x+1}{x-\frac{1}{y^-}} \log \frac{y^--1}{y^-+1}}{2 \pi }+\frac{i \log \frac{x+1}{x-\frac{1}{y^+}} \log \frac{y^+-1}{y^++1}}{2 \pi }
\right.\\
&&\left.
+\frac{i \log \frac{x-1}{x+1} \left(\log \left(1-\frac{1}{(y^-)^2}\right)-\log \left(1-\frac{1}{(y^+)^2}\right)
-2 \log \({1-\frac{1}{x y^-}}\)
+2 \log \({1-\frac{1}{x y^+}}\)
\right)}{4 \pi }\]\;,\\
\la{polylogs2}
{\dot\varrho_\bullet} &\propto&\exp\left[
\frac{i \text{Li}_2\left(\frac{2 \left(y^+-x\right)}{(x+1) \left(y^+-1\right)}\right)}{2 \pi }-\frac{i \text{Li}_2\left(\frac{2 \left(y^--x\right)}{(x+1) \left(y^--1\right)}\right)}{2 \pi }
-\frac{i \log \frac{x-1}{x+1} \left(\log 
\frac{y^-+1}{y^--1}-
\log 
\frac{y^++1}{y^+-1}
\right)}{4 \pi }\right]\;.
\eeqa
}
Above, the proportionality coefficient is a real function of $y$, which is uniquely fixed by requiring that $\varrho_\bullet,\;\dot\varrho_\bullet\to 1$ when $x\to\infty$. As this constant will cancel in all the expressions for the full dressing phase which appear in the ABA, we will not write them out explicitly here.
Similarly, for $\varrho_\circ$ we have
\beqa\nn
{\varrho^2_\circ(x,y)}&\propto&
\exp\[
 -\frac{i \text{Li}_2\left(-\frac{(x+1) (y-1)}{(x-1) (y+1)}\right)}{\pi }+\frac{i \text{Li}_2\left(\frac{(x+1) (y-1)}{(x-1) (y+1)}\right)}{\pi }-\frac{i \log \frac{(x-1) (y+1)}{(x+1) (y-1)} \log \frac{x-y}{x y-1}}{\pi }\right.\\
\la{polylog3}& & \qquad \qquad+\left.\log \frac{x -y}{x\sqrt y}\]\;.
\eeqa
The r.h.s. of the above equation can equivalently be written as 
\begin{equation}\la{rhoBog}
    \frac{y-x}{x \sqrt{y}}e^{-\frac{i}{2}\theta_{\mathsf{rel}}(\gamma_1,\gamma_2)-i\frac{3\pi}{4}}\,,
\end{equation}
where $\theta_{\mathsf{rel}}(\gamma_1,\gamma_2)\equiv\theta_{\mathsf{rel}}(\gamma_{12})$, with $\gamma_{12}=\gamma_1-\gamma_2$ defined as an integral in equation (4.7) of~\cite{Bombardelli:2018jkj}, or in terms of dilogs in equation (A.17) of the same paper. 
The massless rapidities $\gamma_i$, used in \eq{rhoBog} are~\cite{Fontanella:2019baq} 
\begin{equation}
    e^{\gamma_i}\equiv \tan\frac{p_i}{4}\,.
\end{equation}
Recall that~\cite{Bombardelli:2018jkj}
\begin{equation}\label{eq:ZZ}
    e^{\frac{i}{2}\theta_{\mathsf{rel}}}
    =S_{ZZ}=\prod_{l=1}^\infty
    \frac{\Gamma^2\!\left(l-\tfrac{\gamma_{12}}{2\pi i}\right)\Gamma\!\left(l+\tfrac{\gamma_{12}}{2\pi i}+\tfrac{1}{2}\right)\Gamma\!\left(l+\tfrac{\gamma_{12}}{2\pi i}-\tfrac{1}{2}\right)}{\Gamma^2\!\left(l+\tfrac{\gamma_{12}}{2\pi i}\right)\Gamma\!\left(l-\tfrac{\gamma_{12}}{2\pi i}+\tfrac{1}{2}\right)\Gamma\!\left(l-\tfrac{\gamma_{12}}{2\pi i}-\tfrac{1}{2}\right)}
\end{equation} 
is the same as the famous Zamolodchikov dressing factor that enters the S matrix for the scattering of a sine-Gordon soliton and anti-soliton~\cite{Zamolodchikov:1978xm}.

\section{Deriving Bethe Equations}\label{sec:DerivingBAE}
Above, we found the key Q-functions in the ABA limit parametrised in terms of a finite number of complex parameters. In this section, we show that these parameters satisfy Bethe equations. The BEs are equations for the zeroes of certain Q-functions and for massive zeros they are consequences of the QQ-relations.
In the case of massless zeros, we will obtain the BEs from the $\bP\mu$-system. The $\bP\mu$-system goes beyond standard QQ-relations, i.e., there is in general no such system for other integrable systems, such as rational spin chains. This shows the particularity of the massless excitations in the integrable holographic setting.

\subsection{Massive Bethe Equations}
When writing BEs in the AdS$_3$ case, it is convenient to explicitly break the symmetry between dotted and undotted Q-systems. The symmetry can be restored by bringing dual roots into play. The standard choice, adopted in \cite{Cavaglia:2021eqr,Ekhammar:2021pys} is to use the zeroes of the following Q-functions as Bethe roots
\beq
\begin{array}{cccc|ccc}
\text{Roots:} & u_{1,k} & u_{2,k} & u_{3,k}& u_{\dot 1,k} & u_{\dot2, k}& u_{\dot 3, k} \\
\text{Q-function:} & \bQ_{1} & Q_{1|1} & \bQ^2 & {\bP}_{ \dot 1} & {Q}_{\dot 1|\dot 1} & \bP^{\dot 2}
\end{array}\;,
\eeq
then the Bethe equations can be arrived at by considering the following QQ-relations:
\begin{multicols}{2}
\noindent
\begin{equation}\label{leftBAE}
\begin{aligned}
&\left. \frac{Q_{1|1}^{+} }{Q_{1|1}^{-} }  \right|_{u \in  \left\{ \text{zeros of } \bQ_1  \right\} } = 1\,, \\
&\left. \frac{Q_{1|1}^{++ } \bQ_1^{-} \bQ^{2\,-} }{Q_{1|1}^{-- } \bQ_1^{+} \bQ^{2\,+}  } \right|_{u \in \left\{\text{zeros of } Q_{1|1} \right\} } = - 1 \,, \\
&\left. \frac{Q_{1|1}^{+}   }{Q_{1|1}^{-}  } \right|_{u \in \left\{\text{zeros of } \bQ^2 \right\} } = 1\,, 
\end{aligned}
\end{equation}
\columnbreak
\noindent
\begin{equation}\label{rightBAE}
\begin{aligned}
&\left. \frac{Q_{\dot 1|\dot 1}^{+}  }{Q_{\dot 1| \dot 1}^{-} }  \right|_{u \in  \left\{ \text{zeros of } \bP_{\dot 1}  \right\} }  = 1 \,, \\
&\left. \frac{Q_{\dot 1|\dot 1}^{++} \bP_{\dot 1}^{-} \bP^{\dot 2 \, -} }{Q_{\dot 1|\dot 1}^{--} \bP_{\dot 1}^{+} \bP^{\dot 2 \, +} }  \right|_{u \in  \left\{ \text{zeros of } Q_{\dot 1|\dot 1}  \right\} }  = - 1 \,,  \\
&\left. \frac{Q_{\dot 1|\dot 1}^{+} }{Q_{\dot 1|\dot 1}^{-}  }  \right|_{u \in  \left\{ \text{zeros of } \bP^{\dot 2}  \right\} }  = 1  \,.
\end{aligned}
\end{equation}

\end{multicols}
\noindent All these equations are simply QQ-relations evaluated at some special points. 
For example, the first one is the first relation 
\eq{eq:QQtwoeq} evaluated at $u_{1,k}$, so that the r.h.s. vanishes, whereas the second one is the second relation in \eq{eq:QQtwoeq} evaluated at $u_{2,k}+i/2$ and divided by itself,
evaluated at $u_{2,k}-i/2$.
Our task is now to plug in the expressions for the Q-functions we derived in the previous section and identify the results with the BAE of Section~\ref{sec:aba-gen}.

\paragraph{Auxiliary equations.}
For convenience, let us reproduce here \eq{Q11sol} from above
\begin{equation}\label{Q11sim}
    Q_{1|1} \propto \betheQ_{2} \, f^+\, f^+_{\circ}
    \,,\qquad\qquad
    Q_{\dot{1}|\dot{1}} \propto \betheQ_{\dot{2}}\,f^+f^+_{\circ}\;.
\end{equation}
Inserting these into the exact BAEs \eq{leftBAE} for the auxiliary roots, we get
\begin{equation}\label{eq:AuxQSC}
1=\frac{Q_{1|1}^{+} }{Q_{1|1}^{-} }=\frac{\betheQ_{2}^+}{\betheQ_{2}^-}\frac{f^{[2]}}{f}
\frac{f_{\circ}^{[+2]}}{f_{\circ}}=
\frac{\betheQ_{2}^+}{\betheQ_{2}^-}
\frac{{\bf B}_{(-)}}{{\bf B}_{(+)}}\frac{{\bar\varkappa}}{\varkappa}=
\frac{{\bf R}_{2,(+)}}{{\bf R}_{2,(-)}}
\frac{{\bf B}_{\dot 2,(-)}}{{\bf B}_{\dot 2,(+)}}\frac{{\bar\varkappa}}{\varkappa}
\,,\qquad u=u_{I,k}\,,\quad I=1,3
\end{equation}
where in the second equality we used \eq{deff} and then \eq{bfBR}.
Recall that $\bB_{(-)}=\bB_{2,(-)}\bB_{\dot 2,(-)}$  is a product over both types of massive momentum-carrying roots \eq{Qseparation}.
Equation \eqref{eq:AuxQSC} is precisely \eq{eq:aux-bae}, written in compact notation.

\paragraph{Left Middle Node Equation.}
For the left massive middle node, we have to evaluate the following combination
\beq
-1=\frac{Q_{1|1}^{++ } \bQ_1^{-} \bQ^{2\,-} }{Q_{1|1}^{-- } \bQ_1^{+} \bQ^{2\,+}  }\;.
\eeq
Using \eqref{eq:Q-ide-last2} we get
\begin{equation}
\begin{split}
-1=
\frac{\betheQ_{2}^{++} f^{[3]} f_{\circ}^{[3]}}{\betheQ_{2}^{--}f^- f_{\circ}^-}
\left(\frac{x^-}{x^+}\right)^{L-K_{\circ}}\left(\frac{\sigma^+_{\bullet}\sigma^+_{\circ}\rho^+}{\sigma^-_{\bullet}\sigma^-_{\circ}\rho^-}\right)^{2} \frac{\bfB^-_{2,(-)}\bfB^+_{2,(+)}}{\bfB^-_{2,(+)}\bfB^+_{2,(-)}}&\left(\frac{\bfB^+_{\dot{2},(+)}}{\bfB^-_{\dot{2},(+)}}\frac{f^-f^-_{\circ}}{f^{+}f^+_{\circ}} \right)^2\frac{\bar{\varkappa}^-}{\varkappa^-}\frac{\varkappa^+}{\bar{\varkappa}^+}\\
&\times\frac{R^-_{1}B^-_{\dot{1}}}{R^+_{1}B^+_{\dot{1}}}
\frac{R^-_{3}B^-_{\dot{3}}}{R^+_{3}B^+_{\dot{3}}}\;.
\end{split}
\end{equation}
After cancelling, simplifying some factors and using  \eq{deff} and \eq{rhosplit}, we find
\beqa\label{QSCmiddleleft}
-1&=&\label{eq:L-mom-carry-BAE-from-QSC}
\left(\frac{x_{2,k}^-}{x_{2,k}^+}\right)^{L-K_\circ}
\frac{\betheQ_{2}^{++} }{\betheQ_{2}^{--}}
\frac{{\bf B}_{\dot 2,(-)}^+{\bf B}_{\dot 2,(+)}^+}{
{\bf B}_{\dot 2,(-)}^-
{\bf B}_{\dot 2,(+)}^-
}
\frac{R^-_{1}B^-_{\dot{1}}}{R^+_{1}B^+_{\dot{1}}}
\frac{R^-_{3}B^-_{\dot{3}}}{R^+_{3}B^+_{\dot{3}}}
\(\frac{\sigma^+_{\bullet}\rho_\bullet^+}{\sigma^-_{\bullet}\rho_\bullet^-}\frac{
{\sigma^+_{\circ}\rho_\circ^+}   
}{
{\sigma^-_{\circ}\rho_\circ^-}   
}\)^2
\eeqa
where we see that in order to match with \eq{eq:BA-2b-left2}, we need to require the following relation:
\beq\label{expectedMM}
\left.
\frac{\sigma^+_{\bullet}\rho_\bullet^+}{\sigma^-_{\bullet}\rho_\bullet^-}\right|_{u=u_{2,k}} =
\prod_{j=1}^{K_{2}}
\sigma^{\bullet\bullet}(u_{2,k},u_{2,j})
\prod_{j=1}^{K_{\dot 2}}
\tilde{\sigma}^{\bullet\bullet}(u_{2,k},u_{\dot 2,j})\;.
\eeq
We will discuss this relation in more detail in the next section, reiterating the discussion in~\cite{Cavaglia:2021eqr,Ekhammar:2021pys}.
In terms of the building blocks from  
Subsection~\ref{sec:buildingblocks},
\eq{expectedMM} reduces to the following elementary relations
\beq\la{eq:sigmaMM}
\boxed{
\sigma^{\bullet\bullet}(x,y)=\frac{\varsigma_{\bullet}(x^+,y)\varrho_\bullet(x^+,y)}{\varsigma_{\bullet}(x^-,y)\varrho_\bullet(x^-,y)}\,,\qquad
\tilde\sigma^{\bullet\bullet}(x,y)=
\frac{\varsigma_{\bullet}(x^+,y)\dot\varrho_\bullet(x^+,y)}{\varsigma_{\bullet}(x^-,y)\dot\varrho_\bullet(x^-,y)}\;.
}
\eeq
Now we match the second line in \eq{QSCmiddleleft} with the last line of \eq{eq:BA-2b-left2}, that is we identify
\begin{equation}\label{eq:massless-phase-identification}
    \prod_{j=1}^{K_{\circ}} \frac{1-x^+ z_j}{x^- -z} (\sigma^{\bullet \circ}(x,z_j))^{2} \stackrel{?}{=} \left( \frac{x^+}{x^-}\right)^{K_{\circ}}\left(\frac{\sigma^+_{\circ}\rho^+_{\circ}}{\sigma^-_{\circ}\rho^-_{\circ}} \right)^{2}
\end{equation}
which can be rewritten, using the building blocks \eq{varrho-circ-cross} and \eq{varrho-gamma-circ}, as
\beq\la{sigmaMm_def}
\boxed{
\left(\sigma^{\bullet\circ}(x,y)\right)^2=
\frac{1-\frac{y}{x^-}}{\frac{1}{x^+}-y}\frac{\varsigma_\circ^2(x^+,y)\varrho_\circ^2(x^+,y)}{\varsigma_\circ^2(x^-,y)\varrho_\circ^2(x^-,y)}}\;.
\eeq
In the next section we will show that the r.h.s. indeed satisfies the same crossing equation as expected for the l.h.s. of \eq{sigmaMm_def}. In this way, we will have arrived at the same crossing equation from a very different perspective compared to the S-matrix bootstrap procedure.

\paragraph{Right Middle node equation.}
For the right middle node (i.e. for the dotted massive roots)
we have to evaluate the following combination
\beq
-1=\frac{Q_{\dot 1|\dot 1}^{--} \bP_{\dot 1}^{+} \bP^{\dot 2 \, +} }{Q_{\dot 1|\dot 1}^{++} \bP_{\dot 1}^{-} \bP^{\dot 2 \, -} }
\;.
\eeq
Using \eq{Q11sim} and \eq{eq:Q-ide2} this can be expressed as
\begin{equation}\label{eq:ABARQSC}
-\left(\frac{x_{\dot 2,k}^+}{x_{\dot 2,k}^-}\right)^{L-2K_{\circ}}=
\frac{{\bf B}_{\dot 2,(+)}^+{\bf R}_{\dot 2,(-)}^-}
{{\bf B}_{\dot 2,(-)}^-{\bf R}_{\dot 2,(+)}^+}
\frac{{\bf B}_{2,(+)}^+{\bf B}_{2,(+)}^-}
{{\bf B}_{2,(-)}^+{\bf B}_{2,(-)}^-}
\frac{R^+_{\dot {3}} B^+_{3}}{R^-_{\dot {3}} B^-_{3}}
\frac{R^+_{\dot {1}} B^+_{1}}{R^-_{\dot {1}} B^-_{1}}
\(\frac{\sigma_\bullet^+\dot\rho_\bullet^+}
{\sigma_\bullet^-\dot\rho_\bullet^-}\)^2
\(\frac{\varkappa^-}{{\bar\varkappa}^+}
\frac{\sigma_\circ^+\dot\rho_\circ^+}
{\sigma_\circ^-\dot\rho_\circ^-}
\)^2 \,.
\end{equation} 
All but the final term on the r.h.s. above match 
the first two lines of \eq{eq:BA-2b}, provided the following holds
\beq
\left.
\frac{\sigma^+_{\bullet}\dot{\rho}_\bullet^+}{\sigma^-_{\bullet}\dot{\rho}_\bullet^-}\right|_{u=u_{\dot 2,k}} \stackrel{?}{=}
\prod_{j=1}^{K_{\dot 2}}
\sigma^{\bullet\bullet}(u_{\dot 2,k},u_{\dot 2,j})
\prod_{j=1}^{K_{ 2}}
\tilde{\sigma}^{\bullet\bullet}(u_{\dot 2,k},u_{2,j})\;,
\eeq
which is compatible with \eq{expectedMM}.

Matching the last term on the rhs of \eqref{eq:BA-2b-left2} with the third line of \eqref{eq:ABARQSC} gives the identification
\begin{equation}\label{eq:IdSigmaBC2}
    \prod_{j=1}^{K_{\circ}}\frac{x^+}{x^-}\,z^2_j\,\frac{x^--z_j}{1-x^+ z_j} 
      (\sigma^{\bullet\circ})^2(x,z_j)  \stackrel{?}{=} \left( \frac{x^+}{x^-}\right)^{2K_{\circ}} \left(\frac{\varkappa^-}{\bar{\varkappa}^+}  \frac{\sigma_{\circ}^+ \rho_{\circ}^+}{\sigma_{\circ}^- \rho_{\circ}^-}\right)^{2}\,.
\end{equation}
which after using the algebraic relation
\begin{equation}
\prod_{j=1}^{K_{\circ}}\frac{x^+}{x^-}\,z^2_j\,\frac{x^--z_j}{1-x^+ z_j}  = \left(\frac{x^+}{x^-}\right)^{K_{\circ}} \left(\frac{\varkappa^-}{\bar{\varkappa}^+} \right)^{2} \prod_{j=1}^{K_{\circ}}\frac{1-x^+ z_j}{x^--z_j}\,,
\end{equation}
 reduces to our previous identification of $\sigma^{\bullet \circ}$ in \eqref{eq:massless-phase-identification}. 

Thus, modulo the two relations \eq{expectedMM} and \eq{eq:massless-phase-identification}, , we have shown that all the massive Bethe equations derived from QSC match those from \cite{Borsato:2016xns}. In the next section, we will show that constructing the S matrix dressing phases using these two relations and their discontinuity equations, leads to the correct crossing equations that follow from the S-matrix bootstrap approach. Before doing that, in the next sub-section we show how the massless Bethe equation arrises from this limit of the QSC.

\subsection{Massless Middle Node Equation}
As we have reviewed above, for massive roots, the Bethe equations follow from standard QQ-relations~\cite{Cavaglia:2021eqr,Ekhammar:2021pys}, updating them with massless contributions. On the other hand, the massless middle node equation is the most interesting, as it goes beyond these standard QQ-relations. To find it, we start from the $\bP\mu$ system \eqref{Pmumain}. Upon inverting the matrix $\mu$, we obtain
\beqa
\bP_1 &=& {\mu_1}^{\dot b}\bP_{\dot b}^{\bar\gamma}
= {\mu_1}^{\dot 1}\bP_{\dot 1}^{\bar\gamma}+{\mu_1}^{\dot 2}\bP_{\dot 2}^{\bar\gamma}\,,\\
\bP_{\dot 1} &=& {\mu_{\dot 1}}^{b}\bP_{b}^{\bar\gamma}
= {\mu_{\dot 1}}^{1}\bP_{1}^{\bar\gamma}+{\mu_{\dot 1}}^{ 2}\bP_{2}^{\bar\gamma}\;.
\eeqa
Since massless modes are defined as zeroes of $\mu_1{}^{\dot 2}$ (and $\mu_{\dot 1}{}^{2}$), evaluating the above equations at $z_k$ gives
\beq\label{P1P1dot}
\bP_1 =  {\mu_1}^{\dot 1}\bP_{\dot 1}^{\bar\gamma}\,,
\qquad
\bP_{\dot 1} = {\mu_{\dot 1}}^{1}\bP_{1}^{\bar\gamma}\,,
\qquad
x=z_k\,.
\eeq
Next we notice the following relations, from \eq{ABAomega1} and \eq{Qupdef}
\beq
\mu_{\dot 2}{}^{2}\simeq
Q_{\dot 2|\dot 1}^- Q_{1|1}^-\omega^{\dot 1}{}_{2}
\,,\qquad\qquad
\mu_{1}{}^{\dot 1}\simeq
-Q_{1|1}^- 
Q_{\dot 2|\dot 1}^-
\omega^1{}_{\dot 2}\,.
\eeq
Using \eq{omega12i}, we obtain
\beq
\frac{\mu_{1}{}^{\dot 1}}{\mu_{\dot 2}{}^{2}}=-\frac{\omega^{1}_{\dot 2}}{\omega^{\dot 1}_2}=- {\zeta}\;,
\eeq
where ${\zeta}$ is a combination of the constants appearing in the gluing matrix~\eqref{omega12i}.
Furthermore, evaluating
 \eq{mudet} at 
$x=z_i$
we get
$1={\rm det}\,\mu_{\dot a}{}^{ b}=
\mu_{\dot 1}{}^{1}
\mu_{\dot 2}{}^{2}
$, and thus we get
\beq
\mu_{1}{}^{\dot 1}\mu_{\dot 1}{}^{1}=- \zeta\;,
\eeq
or using \eq{P1P1dot} we obtain
\beq
- \zeta=
\left.\frac{\bP_1\bP_{\dot 1}}{\bP_{1}^{\bar\gamma}\bP_{\dot 1}^{\bar\gamma}}\right|_{x=z_i}\;.
\eeq
Note that it would be very handy to assume that ${\rm Im}\,z_i>0$ i.e. that the energy of the massless modes is positive, in this case $\bar\gamma$ maps $z_i$ to $1/z_i$, while staying on the main sheet. I.e. in this case we can write
\beqa
- \zeta=\frac{\bP_1(z_k)\bP_{\dot 1}(z_k)}{\bP_1(1/z_k)\bP_{\dot 1}(1/z_k)}\;.
\eeqa
Using the explicit form of $\bP_1$ and $\bP_{\dot 1}$ \eq{eq:Q-ide2} which we repeat here for convenience
\beqa\nn
\bP_1 \propto x^{-\frac{\dot L}{2}}\sqrt{\frac{\bB_{2,(+)}\bB_{2,(-)}}{x^{-2K_{\circ}}\varkappa\bar\varkappa}} \sigma_{\bullet} \sigma_{\circ} \rho R_{\tilde{1}} B_{\tilde{\dot{1}}}\;\;,\;\;
\bP_{\dot 1} \propto x^{-\frac{L}{2}}\sqrt{
\frac{
\bB_{\dot 2,(+)}\bB_{\dot 2,(-)}}{x^{-2K_{\circ}}{\varkappa\bar\varkappa}}} \sigma_{\bullet} \sigma_{\circ} \dot\rho R_{\dot {3}} B_{3}
\eeqa
we get
\beqa\nonumber
-\zeta z_k^{L+\dot L-2K_{\circ}}
&=&
\(\frac{\sigma_{\bullet} \sigma_{\circ}}
{\sigma^{\bar\gamma}_{\bullet} 
\sigma^{\bar\gamma}_{\circ}}\)^2
\frac{
\rho\;\;\dot\rho\;
}
{
\rho^{\bar\gamma}\dot\rho^{\bar\gamma}}
\sqrt{{\prod\limits_{j=1}^{K_2}x_{2,j}^+x_{2,j}^-\prod\limits_{j=1}^{K_{\dot 2}}x_{\dot 2,j}^+
x_{\dot 2,j}^-}\frac{
\bB_{2,(+)}\bB_{2,(-)}\bB_{\dot 2,(+)}\bB_{\dot 2,(-)}
}{
{\bf R}_{2,(+)}{\bf R}_{2,(-)}{\bf R}_{\dot 2,(+)}{\bf R}_{\dot 2,(-)}   
}}\\
\label{masslessmiddle}&\times&\frac{
R_{\tilde{1}} B_{\tilde{\dot{1}}}
R_{\dot{3}} B_{3}}{
B_{\tilde{1}} R_{\tilde{\dot{1}}}
B_{\dot 3} R_{3}}\;\;,\;\;x=z_k\;,
\eeqa
which constitutes the massless middle node equation. However, in order to compare 
it with the literature i.e. \eq{masslesABA} we have to rewrite it in terms of the dual roots $u_{1,j}$
and $u_{\dot 1,k}$ instead of $u_{\tilde 1,k}$ and $u_{\tilde{\dot 1},k}$.
For that we use the QQ-relations \eqref{eq:QQtwoeq} and its analytically continued version. When evaluating at $x=z_k$ we find
\begin{equation}\label{BRBR2}
\begin{split}
{\bf R}_{2,(+)}{{\bf B}_{\dot 2,(-)}}
{\bar\varkappa}&= c
z_k^{\frac{L-\dot L+K_\circ}{2}}
R_{\tilde {1}}B_{\tilde{ \dot 1}}R_{ 1} B_{\dot 1}\;,\\
-
\(\prod_{j=1}^{K_{\dot 2}}\frac{-1}{x_{\dot 2,j}^+}
\prod_{j=1}^{K_{ 2}}(-x_{ 2,j}^-)
\prod_{j=1}^{K_\circ}\frac{1}{-z_k z_j}\)
{\bf B}_{2,(-)}
{{\bf R}_{\dot 2,(+)}}{\bar\varkappa}&= c
z_k^{-\frac{L-\dot L+K_\circ}{2}}
B_{\tilde {1}}R_{\tilde{\dot  1}}B_{ 1} R_{\dot 1}
\;.
\end{split}
\end{equation}
By dividing the two equations in \eq{BRBR2} and using momentum conservation we get
\beqa\label{BRBR2dual}
-
z_k^{\dot L-L}\(\prod_{j=1}^{K_{\dot 2}}(-x_{\dot 2,j}^-)
\prod_{j=1}^{K_{ 2}}(-\frac{1}{x_{2,j}^+})
\prod_{j=1}^{K_\circ}(-\frac{1}{z_j})\)
\frac{\bfR_{ 2,(+)} \bfB_{\dot 2,(-)}}{\bfB_{2,(-)}\bfR_{\dot{2},(+)}}
\frac{B_{ 1} R_{\dot 1}}{R_{ 1} B_{\dot 1}}
&=& 
\frac
{R_{\tilde { 1}}B_{\tilde{\dot 1}}}
{B_{\tilde { 1}}R_{\tilde{\dot  1}}}
\;.
\eeqa
And plugging this into the massless middle node equation \eqref{masslessmiddle} we find
\begin{equation}\label{masslessmiddleV2}
\begin{split}
    \pm \zeta z^{2L}_{k}= &\left(\prod_{j=1}^{K_{\circ}}-\frac{z^2_k}{z_{j}}\right)\left(\frac{\sigma_{\bullet} \sigma_{\circ}}{\sigma^{\bar\gamma}_{\bullet} \sigma^{\bar\gamma}_{\circ}}\right)^2 \frac{\rho \dot{\rho}} {\rho^{\bar{\gamma}}\dot{\rho}^{\bar{\gamma}}} \frac{B_{1}R_{\dot{1}}B_{3}R_{\dot{3}}}{R_{1}B_{\dot{1}}R_{3}B_{\dot{3}}} \\
    &\times\frac{\bfR_{2,(+)}}{\bfB_{2,(-)}} \left(\prod_{j=1}^{K_2} \frac{1}{z_k x^+_{2,j}}\right) \prod_{j=1}^{K_2}\left(-z_k\sqrt{x^{+}_{2,j}x^-_{2,j}}\right)\sqrt{\frac{\bfB_{2,(+)}\bfB_{2,(-)}}{\bfR_{2,(+)}\bfR_{2,(-)}}} \\
    &\times \frac{\bfB_{\dot{2},(-)}}{\bfR_{\dot{2},(+)}} \left(\prod_{j=1}^{K_{\dot{2}}}\frac{x^-_{\dot{2},j}}{z_k} \right)\prod_{j=1}^{K_{\dot{2}}} \left(-z_k \sqrt{x^+_{\dot{2},k}x^-_{\dot{2},k}} \right) \sqrt{\frac{\bfB_{\dot{2},(+)}\bfB_{\dot{2},(-)}}{\bfR_{\dot{2},(+)}\bfR_{\dot{2},(-)}}}\;.
\end{split}
\end{equation}
Note that the constant $\zeta$ can be fixed from the requirement that the product of all massive and massless equations is compatible with the cyclicity condition \eq{zeromomentum}.
In Appendix~\ref{app:FixZeta} we show that $\zeta$ should satisfy
\beq\la{zetaconstr}
\zeta = \pm \left(-\ii\right)^{K_{\circ}} \sqrt[\leftroot{1}\uproot{2}K_{\circ}]{1}
\eeq
for a suitable root of unity. This is a necessary, not sufficient, condition for the selection rule \eq{zeromomentum} of the physical states to be fulfilled, so that should be checked for a particular solution in addition. The presence of a phase $\zeta$ could in principle mean that the counting of states will be different for our Bethe equations compared to the ones in the literature. This point deserves further study, but since there is no available data to compare to, we will not pursue it further in this paper. A similar phase $\zeta$ also makes an appearance in the BFKL regime of $\mathcal{N}=4$ SYM \cite{Ekhammar2025}.

In \eq{masslessmiddleV2} we have admittedly redistributed square roots without care, resulting in an overall uncertain sign, which can be absorbed by redefining the unknown constant $\zeta$. We have furthermore purposefully grouped together terms to reproduce the expressions appearing in the ABA of Section~\ref{sec:aba-gen}, namely recognizing the first terms on the second and third line of \eqref{masslessmiddleV2} as
\begin{align}
    \prod_{j=1}^{K_2} \frac{z-x_{2,j}^{-}}{z x_{2,j}^+-1} = \frac{\bfR_{2,(+)}}{\bfB_{2,(-)}}  \prod_{j=1}^{K_{2}} \frac{1}{z x^{+}_{2,j}}\,,
    \quad
    \prod_{j=1}^{K_{\dot{2}}} \frac{x^-_{\dot{2},j}}{z^2 x^+_{\dot{2},k}}\frac{z x_{\dot{2},j}^+-1}{z-x^{-}_{\dot{2},j}} = \frac{\bfB_{\dot{2},(-)}}{\bfR_{\dot{2},(+)}}\prod_{j=1}^{K_{\dot{2}}} \frac{x^{-}_{\dot{2},j}}{z}\,,
\end{align}
where all $\bfR,\bfB$ have $z$ as an argument. We see that \eqref{masslessmiddleV2} and \eqref{masslesABA} match perfectly up to the constant $-\zeta$, provided the following identification holds for the massless excitations
\beqa
\,z_k^{+K_\circ}\(\frac{\sigma_\circ\rho_\circ}{\sigma_\circ^{\bar\gamma}\rho_\circ^{\bar\gamma}}\)^2=
\prod_{j=1}^{K_{\circ}}(\sigma^{\circ\circ})^2(z_k,z_j)\;,
\label{expectedmlml}
\eeqa
and for the massive
\begin{equation}
\label{expectedmlM}
\begin{split}
\Bigg(\prod_{j=1}^{K_2}\left(-z_k\sqrt{x^{+}_{2,j}x^-_{2,j}}\right)&\sqrt{\frac{\bfB_{2,(+)}\bfB_{2,(-)}}{\bfR_{2,(+)}\bfR_{2,(-)}}} \Bigg)\left(\prod_{j=1}^{K_{\dot{2}}} \left(-z_k \sqrt{x^+_kx^-_k} \right) \sqrt{\frac{\bfB_{\dot{2},(+)}\bfB_{\dot{2},(-)}}{\bfR_{\dot{2},(+)}\bfR_{\dot{2},(-)}}} \frac{(\sigma_\bullet)^2\rho_\bullet\dot\rho_\bullet}{
    (\sigma_\bullet^{\bar\gamma})^2\rho_\bullet^{\bar\gamma}
    \dot\rho_\bullet^{\bar\gamma}}\right) \\
    &=
\prod_{j=1}^{K_{{2}}}(\sigma^{\circ\bullet})^2(z_k,x_{2,j})
\prod_{j=1}^{K_{\dot{2}}}(\sigma^{\circ\bullet})^2(z_k, x_{\dot 2,j})\;.
\end{split}
\end{equation}
In terms of the elementary blocks from Section~\ref{sec:buildingblocks} they become
\beqa\la{def:sigmamm}
\boxed{
(\sigma^{\circ\circ})^2(x,y)= x\frac{\varsigma_\circ^2(x,y)}{
    \varsigma_\circ^2(x^{\bar\gamma},y)
}    \frac{\varrho_\circ^2(x,y)}{\varrho_\circ^2(x^{\bar\gamma},y)}
}
\eeqa
and
\beqa\la{def:sigmamM}
\boxed{
(\sigma^{\circ\bullet})^2(x,y)=
\sqrt{\frac{x y^+ - 1}{x-y^+}}
\sqrt{\frac{x y^- - 1}{x-y^-}}
    \frac{\varsigma_{\bullet}^2(x,y)\;\;\varrho_\bullet(x,y)\;\;\dot\varrho_\bullet(x,y)}
{\varsigma_{\bullet}^2(x^{\bar\gamma},y)
\varrho_\bullet(x^{\bar\gamma},y)
\dot\varrho_\bullet(x^{\bar\gamma},y)
}
}\;.
\eeqa

In the next section we will discuss these relations in more detail. In order to fix the sign we can require that the phase goes to $1$ when the massive argument goes to infinity, as the roots at infinity represent the descendants and should not change the equations for finite roots. 

To conclude, we managed to find Bethe equations up to an identification of the dressing phases as well as an overall phase $\zeta$.
In the next section, we study those relations for the dressing phases, check which crossing equations they satisfy, their unitarity, and relation to the existing expressions in the literature.

\section{Crossing and Dressing Phases}\la{sec:dressingpases}
In this section, we discuss the crossing equations for the dressing phase candidates we found in the previous section. We will show that the phases constructed in Section~\ref{sec:DerivingBAE} satisfy the crossing relations given in Section~\ref{sec:aba-gen} and compare them with expressions in the literature.

\subsection{Crossing Relations}\la{sec:crossing}
We emphasise that in the QSC approach the crossing equations are not a priori given, but are instead \textit{derived} from the more constraining discontinuity relations.

\begin{figure}
    \centering
    \begin{tikzpicture}
        \draw[black,thick] (-2,-2) rectangle (2,2);
        \draw[black,thick] (-1,0.5)--(1,0.5);
        \filldraw (-1,0.5) circle (2pt);
        \filldraw (1,0.5) circle (2pt);
        \draw[black,thick] (-1,-0.5)--(1,-0.5);
        \filldraw (-1,-0.5) circle (2pt);
        \filldraw (1,-0.5) circle (2pt);
        \draw[<-,red] (-0.6,1) .. controls (0,1) and (0,-1) .. (-0.6,-1);
        \draw[red] (-1.5,0) .. controls (-1.2,1) .. (-0.6,1);
        \draw[<-,red] (-0.6,-1) .. controls (-1.2,-1) .. (-1.5,0);
        \filldraw[red] (-1.5,0) circle (2pt);
    \end{tikzpicture}
    \caption{The path $\bar{\gamma}_c$ depicted in the $u$-plane. Note that it first crosses the branch-cut of $x^{+}$ and thereafter the cut of $x^{-}$.}
    \label{fig:enter-label-crossing}
\end{figure}
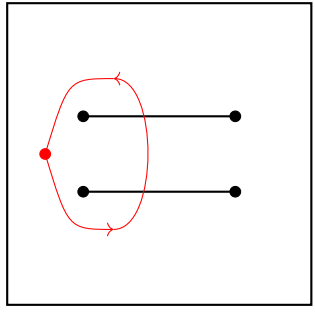

Let us start by formally defining crossing for the ABA phases. 
The definition we adopt here, following \cite{Frolov:2021fmj}, 
differs for the massive and massless case.
For the massive case the dressing phases have two branch cuts on the main sheet and 
crossing involves analytically continuing the dressing phases from below the lower cut then crossing the upper cut and then returning back like in Figure~\ref{fig:enter-label-crossing}.
We denote this path by $\bar{\gamma}_c$. 

For the massless case the phases have only one cut and it was argued 
in \cite{Frolov:2021fmj} that one should analytically continue from above the cut, 
following the path $\gamma$ as defined previously.
Even though that looks at first counter-intuitive, this interpretation appears to be consistent with our QSC-based findings.

\paragraph{Massive-Massive case.}
For $\sigma^{\bullet\bullet}$ and $\tilde\sigma^{\bullet\bullet}$, 
defined in \eq{eq:sigmaMM}, we find the following crossing equations, as derived in Appendix~\ref{app:crossing}
\beqa\la{crossing_mv_mv}
\[\sigma^{\bullet\bullet}(x^{\bar\gamma_c},y)
\tilde\sigma^{\bullet\bullet}(x,y)\]^2&=&
\frac{(y^-)^4 \left(x^--y^+\right) \left(y^+-x^+\right) \left(y^+ x^+-1\right)^2}
{(y^+)^4 \left(x^--y^-\right) \left(y^--x^+\right) \left(y^- x^+-1\right)^2}\,,\\ \la{crossing_mv_mv_tilda}
\[\tilde\sigma^{\bullet\bullet}(x^{\bar\gamma_c},y)
\sigma^{\bullet\bullet}(x,y)\]^2&=&\frac{(y^-)^4 \left(x^--y^+\right)^2 \left(y^+ x^--1\right) \left(y^+ x^+-1\right)}
{(y^+)^4 \left(x^--y^-\right)^2 \left(y^- x^--1\right) \left(y^- x^+-1\right)}\,.
\eeqa
These relations agree with \cite{Borsato:2013hoa,Frolov:2021fmj}. 
As explained in Appendix~\ref{app:crossing} to arrive at 
the above expression we used the definition \eq{eq:sigmaMM}
and the discontinuity equations \eq{varsigma-bullet-crossing} and \eq{rhobulletblocks}.

\paragraph{Massive-Massless case.} 
Similarly for $\sigma^{\bullet\circ}$ from the definition \eq{sigmaMm_def}
and using the discontinuity equations \eq{varsigma-bullet-crossing} and \eq{varrho-gamma-circ} we get
\beqa\label{eq:MmCrossing}
\[\sigma^{\bullet\circ}(x^{\bar\gamma_c},y)
\sigma^{\bullet\circ}(x,y)\]^2&=&
\frac{\left(x^--y\right) \left(y x^+-1\right)}{y^4 \left(y x^--1\right) \left(x^+-y\right)}\;,
\eeqa
in perfect agreement with \cite{Borsato:2016xns,Frolov:2021fmj}.\footnote{As noted in~\cite{Frolov:2021fmj}, there was a typo in~\cite{Borsato:2016xns}: the $y^{-4}$ term is missing.}

\paragraph{Massless-Massive and Massless-Massless cases.}
Using the definition of $\sigma^{\circ\bullet}$ in \eq{def:sigmamM}, $\sigma^{\circ \circ}$ in \eq{def:sigmamm}  
and performing the analytic continuation along $\gamma$ this time we get
\beqa \label{eq:mMCrossing}
\[\sigma^{\circ\bullet}(x^{\gamma},y)
\sigma^{\circ\bullet}(x,y)\]^2&=&\frac{(x-y^-)(1/x-y^+)}{(x-y^+)(1/x-y^-)}\;,\\
\la{sigma_ml_ml_crossing}
\[\sigma^{\circ\circ}(x^{\gamma},y)
\sigma^{\circ\circ}(x,y)\]^2&=&\frac{(xy-1)^2}{(y-x)^2}\;,
\eeqa
which again agrees with \cite{Borsato:2016kbm,Frolov:2021fmj}. Even though the above equations are identical to those in \cite{Frolov:2021fmj},
the solution to crossing equations is not unique. We now turn to an explicit comparison of our dressing phases to those found in~\cite{Frolov:2021fmj}.

\subsection{Explicit Expressions for the Dressing Phases}\label{sec:ExplicitDressing}\la{FSphasesMM}
In this section we  give explicit expressions for our dressing phases and compare them with the literature.
For that one should plug the solutions for $\varrho$'s and $\varsigma$'s into the definitions of the dressing phases e.g. \eq{def:sigmamM} and \eqref{def:sigmamm}.
For simplicity, we  write $\sigma_{\FS}$ for the phases proposed in \cite{Frolov:2021fmj}. The main challenge is thus to rewrite the result in the notations of \cite{Frolov:2021fmj}, allowing for a direct comparison.

A heavily utilised tool in \cite{Frolov:2021fmj} is a rapidity parameterisation inspired by \cite{Fontanella:2019baq}:
\begin{equation*}
\begin{split}
    \gamma^{\pm}(u) &= \log\left( \mp \ii \frac{x^{\pm}-1}{x^{\pm}+1}\right)=\frac{1}{2} \log \frac{u\pm\frac{\ii}{2}-2g}{u\pm\frac{\ii}{2}+2g} \mp \ii \frac{\pi}{2}\,,
    \\
    \gamma^{\circ}(z) &= \log(-\ii \frac{z-1}{z+1})=\frac{1}{2}\log(\frac{z+\frac{1}{z}-2}{z+\frac{1}{z}+2})-\frac{\ii \pi}{2}\,,
\end{split}
\end{equation*}
and we will make use of the following shorthand notation $\gamma_{1}^{a} = \gamma^{a}(u),\gamma_{2}^{a} = \gamma^{a}(v)$ and finally $\gamma_{12}^{ab} = \gamma_{1}^{a}-\gamma_{2}^{b}$ with $a,b=+,-,\circ$.

\paragraph{Massive-Massive.}
The massive-massive phases of \cite{Frolov:2021fmj} are given as 
\begin{equation}\label{eq:FSMM}
    \left(\frac{\sigma^{\bullet \bullet}_{\FS}(u,v)}{\sigma_{\text{BES}}^{\bullet\bullet}(u,v)}\right)^{2} = -\frac{\sinh \frac{\gamma^{+-}_{12}}{2}}{\sinh \frac{\gamma^{-+}_{12}}{2}}\, e^{-\varphi^{\bullet\bullet}(u,v)}\,,\quad
    \left(\frac{\tilde{\sigma}^{\bullet \bullet}_{\FS}(u,v)}{\sigmaBES^{\bullet\bullet}(u,v)}\right)^{2} = \frac{\cosh \frac{\gamma^{-+}_{12}}{2}}{\cosh \frac{\gamma^{+-}_{12}}{2}}\, e^{-\tilde{\varphi}^{\bullet\bullet}(u,v)}\,,
\end{equation}
where the $\varphi$ factors are
\begin{align}\label{eq:varphiFS}
    &\varphi^{\bullet \bullet}(u,v) = \varphi_+^{\bullet\bullet}(\gamma^{--}_{12}) + \varphi^{\bullet \bullet}_{+}(\gamma_{12}^{++}) + \varphi^{\bullet\bullet}_{-}(\gamma_{12}^{-+})+\varphi^{\bullet \bullet}_{-}(\gamma_{12}^{+-})\,, \\\label{eq:varphiTFS}
    &\tilde{\varphi}^{\bullet \bullet}(u,v) = \varphi_-^{\bullet\bullet}(\gamma^{--}_{12}) + \varphi^{\bullet \bullet}_{-}(\gamma_{12}^{++}) + \varphi^{\bullet\bullet}_{+}(\gamma_{12}^{-+})+\varphi^{\bullet \bullet}_{+}(\gamma_{12}^{+-})\,,
\end{align}
and the building blocks are defined as
\begin{align}
    &\varphi_-^{\bullet \bullet}(\gamma) = +\frac{\ii}{\pi} \Li_{2}(+e^{\gamma})- \frac{\ii}{4\pi}\gamma^2+\frac{\ii}{\pi}\gamma \log(1-e^{\gamma})-\frac{\ii \pi}{6}\,, \\
    &\varphi_+^{\bullet\bullet}(\gamma) = -\frac{\ii}{\pi}\Li_{2}(-e^{\gamma})+\frac{\ii}{4\pi} \gamma^2 -\frac{\ii}{\pi}\gamma \log(1+e^{\gamma})-\frac{\ii \pi}{12}\,.
\end{align}
The BES-phase $\sigma^{\bullet\bullet}_{\text{BES}}$ is the standard BES phase and thus in our notation $
\sigma^{\bullet\bullet}_{\rm BES}(x,y)=\frac{\varsigma_\bullet(x^+,y)}{\varsigma_\bullet(x^-,y)}
$, the superscripts $\bullet\bullet$ serve as a reminder that we are considering massive-massive scattering. We find that
\beq
\(\frac{\varrho_\bullet(x^+,y)}
{\varrho_\bullet(x^-,y)}\)^2=-\frac{
    \sinh\tfrac{\gamma^{+-}}{2}
    }{\sinh \tfrac{\gamma^{-+}}{2}}
    e^{-\varphi^{\bullet\bullet}}    
\;\;,\;\;
\(\frac{\dot\varrho_\bullet(x^+,y)}
{\dot\varrho_\bullet(x^-,y)}\)^2=+
\frac{\cosh \tfrac{\gamma^{-+}}{2}}{
\cosh\tfrac{\gamma^{+-}}{2}
}
e^{-\tilde\varphi^{\bullet\bullet}}\;,
\eeq
and thus our results are in perfect agreement with \eqref{eq:FSMM}.

\paragraph{Massless-Massive and Massive-Massless.}
Similarly, we have
\begin{align}
    &\left(\frac{\sigma^{\bullet \circ}_{\FS}}{\sigma^{\bullet \circ}_{\text{BES}}}\right)^{2} = -\ii \frac{\tanh \frac{\gamma^{+\circ}_{12}}{2}}{\tanh \frac{\gamma^{-\circ}_{12}}{2}} \frac{1}{\varPhi(\gamma_{12}^{+\circ}) \varPhi(\gamma_{12}^{-\circ})}\,, \quad
    \left(\frac{\sigma^{\circ \bullet}_{\FS}}{\sigma^{\circ \bullet}_{\text{BES}}} \right)^2 = \ii \frac{\tanh \frac{\gamma^{\circ -}_{12}}{2}}{\tanh \frac{\gamma^{\circ +}_{12}}{2}}\frac{1}{\varPhi(\gamma^{\circ +}_{12})\varPhi(\gamma^{\circ -}_{12})}\,,
    \label{eq:FSmMMm}
\end{align}
with 
\begin{equation}
\begin{split}
    \varphi(\gamma) &= \frac{\ii}{\pi}\Li_2(-e^{-\gamma})-\frac{\ii}{\pi}\Li_{2}(e^{-\gamma})+\frac{\ii \gamma}{\pi}\log(1-e^{-\gamma}) - \frac{\ii \gamma}{\pi}\log(1+e^{-\gamma}) + \frac{\ii \pi}{4}\,,\\
    \varPhi(\gamma) &= e^{\varphi(\gamma)}\;.
\end{split}
\end{equation}
When making the comparison for the massless cases, we have to remember that our expressions are derived assuming the massless argument is above the cut, i.e. $x=e^{ip/2}$ for $p\in (0,2\pi)$.
Assuming this is the case, the BES part matches once again perfectly with our expressions, and so does the remaining part
\begin{align}
    \frac{1-\frac{y}{x^-}}{\frac{1}{x^+}-y}\left(\frac{\varrho_{\circ}(x^+,y)}{\varrho_{\circ}(x^-,y)} \right)^{2} &= -\ii \frac{\tanh \frac{\gamma^{+\circ}_{12}}{2}}{\tanh \frac{\gamma^{-\circ}_{12}}{2}} \frac{1}{\varPhi(\gamma_{12}^{+\circ}) \varPhi(\gamma_{12}^{-\circ})}\,, \\
    \quad
    \sqrt{\frac{x y^+-1}{x-y^+}}
    \sqrt{\frac{x y^- -1}{x-y^-}}\frac{\varrho_{\bullet}(x,y) \dot{\varrho}_{\bullet}(x,y)}{\varrho_{\bullet}(x^{\bar{\gamma}},y)\dot{\varrho}_{\bullet}(x^{\bar{\gamma}},y)} &= \ii \frac{\tanh \frac{\gamma^{\circ -}_{12}}{2}}{\tanh \frac{\gamma^{\circ +}_{12}}{2}}\frac{1}{\varPhi(\gamma^{\circ +}_{12})\varPhi(\gamma^{\circ -}_{12})}\,.
\end{align}

\paragraph{Massless-Massless.}
So far we have perfectly reproduced all phases.
However, in the massless-massless case we find a slight disagreement with~\cite{Frolov:2021fmj}.
For massless-massless scattering we find 
\begin{equation}
    \left(\frac{\sigma^{\circ \circ}(u,v)}{\sigma^{\circ\circ}_{\text{BES}}(u,v)}\right)^{2} = x \frac{\varrho_{\circ}^2(x,y)}{\varrho_{\circ}^{2}(x^{\bar{\gamma}},y)} = -\ii \frac{1}{\varPhi(\gamma^{\circ\circ}_{12})^2}
    =-i e^{-i\theta_{\mathsf{rel}}(\gamma_1,\gamma_2)}\,,
\end{equation}
where the relation to the Zamolodchikovs' dressing factor~\eqref{eq:ZZ} is $e^{\frac{i}{2}\theta_{\mathsf{rel}}}\equiv S_{ZZ}$~\cite{Bombardelli:2018jkj}.
On the other hand,~\cite{Frolov:2021fmj} found that the dressing factor is
\begin{equation}\label{eq:DressingSFMM}
    \left(\frac{\sigma^{\circ \circ}_{\FS}(u,v)}{\sigma^{\circ\circ}_{\text{BES}}(u,v)}\right)^{2} = \frac{1}{a(\gamma^{\circ \circ})(\varPhi(\gamma^{\circ\circ}_{12}))^2}\,, 
\end{equation}
where $a(\gamma)$ is a non-trivial function of $\gamma$ to be discussed below in \eq{agamma}. This is the first and only discrepancy with \cite{Frolov:2021fmj}. In the next sub-section we discuss possible reasons for the discrepancy.

\subsection{Massless-massless phase, its unitarity and crossing}\la{sec:disc}
First, to see the problem clearly, consider the  massless limit i.e. sending $x^+\to x$ and $x^-\to 1/x$.
To define the limit more clearly we replace the shift $x^\pm =x(u\pm i/2)$ by $x(u\pm i\epsilon)$ and send $\epsilon\to 0_+$. We find the phases are related nicely in this limit as follows
\beqa
\[\sigma^{\circ\bullet}(y,x^+,x^-)\]^2
\to
-\[\sigma^{\circ\circ}(y,x)\]^2\;\;,\;\;
\[\sigma^{\bullet\circ}(x^+,x^-,y)\]^2
\to 
+\[\sigma^{\circ\circ}(x,y)\]^2\;.
\eeqa
As $(\sigma^{\circ\bullet})^2$ and $(\sigma^{\bullet\circ})^2$ are related by unitarity with a unit coefficient we can see clearly that for our combination of factors $\sigma^{\circ\circ}$ we should have
\beq\la{uniissue}
[\sigma^{\circ\circ}(x,y)\;\sigma^{\circ\circ}(y,x)]^2
=-1\;.
\eeq
At first one could worry that $\sigma^{\circ\circ}$ as written is not unitary. However, that is not the only difference in the massless BAE equation we encountered. We also have the factor $\zeta$, which can be constrained from the zero momentum condition to be \eq{zetaconstr}. This is similar to what is found in the BFKL regime of ${\cal N}=4$ SYM, where the correct counting of states, according to our preliminary results~\cite{Ekhammar2025}, is reproduced.
Thus, when identifying the dressing phases, we still have the freedom to redefine $\sigma^{\circ\circ}$ by a factor of $\pm\ii$, in order to make it unitary. We see that, such a factor in $(\sigma^{\circ\circ})^2$ changes  the sign in the relation \eq{uniissue} and so restores unitarity. However, the massless crossing relation~\eqref{sigma_ml_ml_crossing} then gets an extra minus sign.

A closely related issue was found by \cite{Frolov:2021fmj}, who instead proposed including a factor of $a(\gamma)$ into the massless dressing phase. This factor was defined to satisfy two properties
\begin{equation}
    a(\gamma)a(-\gamma) = 1\,,
    \qquad\qquad
    a(\gamma)a(\gamma+\ii \pi) = -1\,.
\end{equation}
These properties do not fully fix $a(\gamma)$, since, for example, for any non-trivial solution, the inverse $1/a(\gamma)$ is also a solution, but a potential candidate was proposed in \cite{Frolov:2021fmj}
\begin{equation}\la{agamma}
    a(\gamma) =  -\ii \tanh(\frac{\gamma}{2}-\frac{\ii \pi}{4})\,,
    \qquad\qquad
    a(x,y) = -\frac{{\ii(x-y)+(-1+x y)}}{-\ii(x-y)+(-1+x y)}\,.
\end{equation}
Writing $a$ in terms of $x$ and $y$ suggests that it is unlikely that the QSC could generate such a factor, since it has a rather unnatural singularity structure, having a pole at $x=\frac{1-i y}{y-i}$ (curiously the r.h.s. is mapping the unit circle into a real line). 

In Appendix~\ref{appendix:rhoa}, we investigate the possibility of modifying $\varrho_\circ\to\varrho_\circ\varrho_a$, in order to include the $a(\gamma)$ factor into the massless-massless dressing phase.
The main problem we encounter, in addition to violating some fundamental analyticity assumptions of the QSC ingredients, is that such a modification unavoidably also changes the massive-massless phase by a factor violating its crossing relation, since its expression also contains $\varrho_\circ$. On the other hand, the massless-massive dressing phase does not involve $\varrho_\circ$ and therefore would remain unmodified. Hence, changing $\varrho_\circ$ would also violate unitarity in the mixed-mass sector. We note that this conclusion demonstrates explicitly that the QSC result is more constraining than the S-matrix bootstrap. This is because it imposes \textit{stronger} analyticity constraints on \textit{fewer} building blocks, which are used to construct the S matrix. Let us emphasize again that in the QSC derivation, crossing equations, in particular in the mixed-mass sector, come out as a result of the derivation and provide an independent test of the construction.

Since unitarity is fundamental, our results suggest that a sign could be missing in the massless crossing equation found using the S-matrix bootstrap derivation. Such a possibility could arise, for example, from the action of $\algsu(2)_\circ$ on massless representations which could lead to a modification of the charge conjugation matrix. Similar considerations were analysed in~\cite{Torrielli:2021hnd}, where for certain massless sub-sectors different signs in the crossing equation also appeared. In view of our result, it is worth revisiting this point in the S-matrix derivation. This in turn may lead to a better understanding of the ABA limit of the AdS$_3$ QSC and further numerical and analytical tests in different regimes could provide additional clarity on the issue. We hope to return to this in future work. Finally, as was recently observed~\cite{Copetti:2024dcz}, the crossing relation in fact in some cases can be more subtle than naively expected. In that case, the generalized symmetries were helpful in restoring the correct normalization of the crossing equation. In our case, that could be the QSC!

\section{Conclusions}\la{sec:conclusions}
In this paper, we identified new classes of solutions of the AdS$_3$ QSC~\cite{Cavaglia:2021eqr,Ekhammar:2021pys} which contain all expected types of excitations, including massless ones. We showed how, in the large-volume limit, solutions to the QSC are
parametrized by a finite set of roots that satisfy Bethe equations which are structurally equivalent to those in~\cite{Borsato:2016xns}. Our results considerably extend the original analysis of~\cite{Cavaglia:2021eqr} and~\cite{Ekhammar:2021pys}, 
where only massive excitations were included in the large volume limit.

The most complicated parts in the aforementioned Bethe equations are the dressing phases, which we fixed using QSC analytic constraints. These constraints are generally more restrictive than those that follow from integrable S-matrix bootstrap methods.
They lead to \textit{discontinuity} equations on the building blocks of the dressing phases instead of the more traditional crossing and unitarity relations. We solved these equations and used them to fix all dressing phases, finding perfect agreement with \cite{Frolov:2021fmj} in all cases except the massless-massless phase. 
As such, our results provide further strong evidence for the validity of the AdS$_3$ QSC proposal as well as demystify the role of massless modes in the QSC formalism.

Our findings open up a new avenue for the study of the AdS$_3$/CFT$_2$ correspondence, as we now have a detailed knowledge about all perturbative string states at least in the asymptotic regime. The information of this type is  instrumental for moving to the exact finite volume studies of the spectrum. To this end, the massive sector was already analysed by 
means of the QSC in~\cite{Cavaglia:2022xld}, where the first ever exact result for finite-size operators 
was found in this theory in the small-tension (weak coupling) limit to a high order in perturbation theory,
as well as high precision methods were developed for the numerical studies of the spectrum at finite coupling.
Later on an alternative Thermodynamic Bethe Ansatz was proposed in~\cite{Frolov:2023wji}, which was also studied numerically in~\cite{Brollo:2023rgp}, with a focus on the massless sector.
We hope that with our results one may make a direct comparison between the two approaches.

Massless particles play a novel role in integrable $\AdS_3/\CFT_2$ holography, compared to higher-dimensional models. In particular, it is known that the half-BPS protected spectrum of this AdS$_3$/CFT$_2$ dual pair~\cite{deBoer:1998kjm} is intimately related to zero-momentum massless excitations~\cite{OhlssonSax:2012smh,Baggio:2017kza,Majumder:2021zkr}. Our results show how such states fit into the QSC analysis. It would be interesting to explore this approach more fully, for example by
introducing a deformation of the original theory by a small twist. On general grounds, one may expect significant simplification in the vicinity of the BPS states, allowing for an all-loop analytic treatment. In addition, because of their simplified kinematics~\cite{Fontanella:2019baq}, massless excitations have recently been investigated in the context of form factors and boundary ABAs~\cite{Torrielli:2023uam,Bielli:2024xuv,Bielli:2024bve}. It would be interesting to connect these results with the way massless excitations appear in the QSC analysis presented here. Finally, recent string theory results, such as the AdS$_3 \times $S$^3$ Virasoro-Shapiro amplitude with Ramond-Ramond flux~\cite{Alday:2024rjs,Chester:2024wnb}, offer insights that could complement the QSC analysis and allow for cross-checks.

We further hope that our results give new clues on how to generalize the QSC proposal to mixed-flux AdS$_3$ backgrounds which are known to also be integrable~\cite{Cagnazzo:2012se,Hoare:2013lja,Lloyd:2014bsa}. Amongst these, perhaps the most interesting are the near-horizon limits of NS5-branes and fundamental strings, which at a special point in their moduli space are described by the Maldacena-Ooguri WZW model~\cite{Maldacena:2000hw}. In integrable language~\cite{OhlssonSax:2018hgc}, this point corresponds to zero coupling ($h=0$), with only the $k\in\mathbf{Z}$ WZW-level remaining as a non-trivial parameter of the planar theory. Turning on the axion and other R-R moduli takes one away from the WZW point~\cite{OhlssonSax:2018hgc}. This deformation cannot be easily analyzed using worldsheet CFT technology, due to the well-known issues with R-R vertex operators. Integrable methods do not suffer from these problems and so the mixed-flux QSC should offer a practical and computationally efficient solution to the spectral problem for all values of R-R moduli. While the kinematics of the mixed-flux integrable model is significantly more intricate than that of the R-R theory~\cite{OhlssonSax:2023qrk}, we believe it provides enough  analytic information on the QQ-system to fix the mixed-flux QSC. Solving such a mixed-flux QSC, particularly in the small-$h$ limit would allow for important comparisons to complementary results that have recently appeared in the literature, including the $k=1$ model and its Sym$^N($T$^4)$ dual~\cite{Eberhardt:2018ouy,Gava:2002xb,Gaberdiel:2023lco,Frolov:2023pjw,Aharony:2024fid,Gaberdiel:2024nge,Gaberdiel:2024dfw,McStay:2023thk,McStay:2024dtk,Dei:2023ivl}. It would also be interesting to compare mixed-flux QSC calculations with previous results on the spectrum of R-R deformed WZW models~\cite{Cho:2018nfn}.

There have been interesting recent findings in integrable deformations of $\AdS_3$ backgrounds~\cite{Hoare:2022asa,Hoare:2023zti}. Deformations of QSCs for higher-dimensional duals have been investigated in~\cite{Kazakov:2015efa,Klabbers:2017vtw}, and it would be interesting to see how these constructions can be generalised to those $\AdS_3$ deformations. Further, boundaries and defects have also been explored in the context of integrability, see for example~\cite{Bielli:2024xuv,Bielli:2024bve,Bliard:2024bcz}

Finally, one may hope to generalise the QSC studied here to the $\AdS_3\times\Sphere^3\times \Sphere^3\times \Sphere^1$ backgrounds, which are also known to exhibit integrability~\cite{Babichenko:2009dk,OhlssonSax:2011ms,Borsato:2012ud,Sundin:2012gc,Borsato:2015mma}. For these backgrounds, the QQ-system should be based on the Lie super-algebras $\mathrm{d}(2,1;\alpha)^2$, which depend on a free parameter $\alpha\in [0,1]$. It would be very interesting to understand such novel examples of QSCs explicitly, particularly since the dual $\CFT_2$ has been less well understood~\cite{Gauntlett:1998kc,Boonstra:1998yu,Gukov:2004ym,Tong:2014yna,Eberhardt:2017pty}, with very recent progress in~\cite{Witten:2024yod}.

\section*{Acknowledgements}
We are grateful to A.~Cavagli\`a and D.~Volin for collaboration at 
the initial stage of the project, M.~Preti for work on related topics as well as to K.~Zarembo and  B.~Basso
for fruitful discussions. We would like to thank Olof Ohlsson Sax and Alessandro Torrielli for many discussions and sharing their insights.
N.G. is grateful to LPENS Paris for warm hospitality where a part of this work was done. 
The work of N.G. and S.E. was
supported by the European Research Council (ERC) under the European Union’s Horizon 2020 research and innovation program – 60 – (grant agreement No. 865075) EXACTC. N.G.’s research is supported in part by the Science Technology \& Facilities Council (STFC) under the grants ST/P000258/1 and ST/X000753/1. The work of S.E. was partially conducted with funding awarded
by the Swedish Research Council grant VR 2024-00598. B.S.\@ acknowledges funding support from The Science and Technology Facilities Council through Consolidated Grants ``Theoretical Particle Physics at City, University of London'' ST/X000729/1. 

\newpage

\appendix

\section{Dressing Phases and Crossing}\label{app:crossing}
In this appendix, we expand on some of the computational details summarised in Section~\ref{sec:crossing}, by showing how crossing equations for S-matrix dressing phases follow from the discontinuity equations found in section~\ref{sec:MasslessABALimit}.

\subsection{Massive-Massive Crossing}
For completeness we repeat the calculation of the massive-massive crossing equation of~\cite{Borsato:2013hoa} from the QSC following~\cite{Cavaglia:2021eqr,Ekhammar:2021pys}. The goal of this section is to deduce the crossing relation for the r.h.s. of \eq{eq:sigmaMM} starting from the discontinuity equations found in section~\ref{sec:buildingblocks}.

Crossing involves two analytic continuations. Firstly, we  continue through the cut $(-2h-i/2,2h-i/2)$ from below, i.e. oriented in the same way as $\bar\gamma$. We denote this operation as $\bar\gamma_+$, because it flips $x^+\to 1/x^+$. Secondly, the resulting expression is analytically continued through the cut at $(-2h+i/2,2h+i/2)$, once again approaching it from below, which we denote as $\bar\gamma_-$. 

We start by considering the BES part of the phase. Using~\eqref{varsigma-bullet-crossing}, the crossing equation for the full phase is
\beqa \label{eq:BESCrossing}
\(\(\frac{\varsigma_{\bullet}(x^+,y)}{\varsigma_{\bullet}(x^-,y)}\)^{\bar\gamma_+}\)^{\bar\gamma_-}&=&
\(\frac{1}{\varsigma_{\bullet}(x^-,y)\varsigma_{\bullet}(x^+,y)}
\prod_{n=1}^\infty \frac{1-\frac{1}{x^{[+2n+1]}y^-}}{1-\frac{1}{x^{[+2n+1]}y^+}}
\frac{1-\frac{1}{x^{[-2n+1]}y^+}}{1-\frac{1}{x^{[-2n+1]}y^-}}
\)^{\bar\gamma_-}\\
\nn&=&
\frac{\varsigma_{\bullet}(x^-,y)}{\varsigma_{\bullet}(x^+,y)}
\frac{1-\frac{x^- }{y^+}}{1-\frac{1}{x^+ y^-}}
\frac{1-\frac{1}{x^+ y^+}}{1-\frac{x^-}{ y^-}}\,,
\eeqa
giving precisely Janik's crossing equation \cite{Janik:2006dc}.

Next, we consider the non-quadratic-cut parts of the dressing phases i.e. those containing $\varrho_\bullet$ and $\varrho_{\dot\bullet}$. Using the discontinuity relation~\eqref{rhobulletblocks} for $\varrho_{\bullet}$, we find
\beqa \label{eq:CrossingRho}
\(\(\frac{\varrho_{\bullet}(x^+,y)}{\varrho_{\bullet}(x^-,y)}\)^{\bar\gamma_+}\)^{\bar\gamma_-}&=&
\(\frac{1}{\varrho_{\bullet}(x^-,y)\dot\varrho_{\bullet}(x^+,y)}\sqrt{\frac{x^+-y^+}{x^+-y^-}}\)^{\bar\gamma_-}\\
\nn &=&
\frac{\dot\varrho_{\bullet}(x^-,y)}{\dot\varrho_{\bullet}(x^+,y)}
\sqrt{\frac{x^+-y^+}{x^+-y^-}\frac{x^--y^-}{x^--y^+}}\;.
\eeqa
Combining~\eqref{eq:BESCrossing} and~\eqref{eq:CrossingRho} we recover the massive crossing equation \eq{crossing_mv_mv} for the full dressing phase. Similarly, using~\eqref{rhobulletblocks} for $\dot\varrho_{\bullet}$, we find
\beqa\la{eq:CrossingRho2}
\(\(\frac{\dot\varrho_{\bullet}(x^+,y)}{\dot\varrho_{\bullet}(x^-,y)}\)^{\bar\gamma_+}\)^{\bar\gamma_-}
&=&
\frac{\varrho_{\bullet}(x^-,y)}{\varrho_{\bullet}(x^+,y)}
\sqrt{\frac{1/x^+-y^-}{1/x^+-y^+}\frac{1/x^--y^+}{1/x^--y^-}}\;.
\eeqa
which, together with~\eqref{eq:BESCrossing}, leads to the second massive crossing equation~\eq{crossing_mv_mv_tilda}.

We note that the crossing equations \eqref{crossing_mv_mv} and \eqref{crossing_mv_mv_tilda} might appear slightly different from the ones usually presented in the literature. This is merely a cosmetic difference, using the identity
\beq
\frac{x^- y^- \left(x^+-y^+\right) \left(x^+ y^+-1\right)}{x^+ y^+\left(x^--y^-\right) \left(x^- y^--1\right) }=1\,.
\eeq
they can be brought to, for example, the form presented in \cite{Frolov:2021fmj}

\subsection{Massive-Massless Crossing}
To derive the crossing equations for the massive-massless phases, we follow a similar procedure and will use the same contour for analytic continuation as in the massive case reviewed above. 

The massive-massless dressing factor found in~\eqref{sigmaMm_def} is
\beq\la{expectedMMbits}
\left(\sigma^{\bullet\circ}(x,y)\right)^2=\frac{1-\frac{y}{x^-}}{\frac{1}{x^+}-y}
\frac{\varsigma_\circ^2(x^+,y)\varrho_\circ^2(x^+,y)}{\varsigma_\circ^2(x^-,y)\varrho_\circ^2(x^-,y)}
\;.
\eeq
Analytically continuing as before, the first factor's contribution to the crossing equation is
\beqa\label{eq:Mm-pt1}
\(\(\frac{1-\frac{y}{x^-}}{\frac{1}{x^+}-y}\)^{\bar\gamma_+}\)^{\bar\gamma_-}
\frac{1-\frac{y}{x^-}}{\frac{1}{x^+}-y}
=
\frac{x^+\left(x^--y\right) \left(y x^--1\right) }{x^- \left(x^+-y\right) \left(y x^+-1\right)}\,.
\eeqa
Similarly, the contribution of $\varsigma_\circ$ can be found using the discontinuity relation \eq{varrho-circ-cross}
\beqa\label{eq:Mm-pt2}
\(\(\frac{\varsigma_\circ(x^+,y)}{\varsigma_\circ(x^-,y)}\)^{\bar\gamma_+}\)^{\bar\gamma_-}
&=&
\nn\(\frac{1}{\varsigma_\circ(x^-,y)\varsigma_\circ(x^+,y)}
\prod_{n=1}^\infty
\frac{(x^{[+2n+1]}-y)}{(x^{[-2n+1]}-y)}\frac{(x^{[-2n+1]}-1/y)}{(x^{[+2n+1]}-1/y)}
\)^{\bar\gamma_-}\\
&=&
\frac{\varsigma_\circ(x^-,y)}{\varsigma_\circ(x^+,y)}
\frac{\left(x^--y\right) \left(y x^+-1\right)}{y^2 \left(x^+-y\right)\left(y x^--1\right) }\,,
\eeqa
while for the  $\varrho_\circ$ terms, using~\eqref{varrho-gamma-circ}, we get
\beqa\label{eq:Mm-pt3}
\(\(\frac{\varrho_\circ(x^+,y)}{\varrho_\circ(x^-,y)}\)^{\bar\gamma_+}\)^{\bar\gamma_-}
=
\(
\frac{1}{\varrho_\circ(x^-,y)\varrho_\circ(x^+,y)}
\frac{x^+-y}{\sqrt{x^+ y}}
\)^{\bar\gamma_-}=
\frac{\varrho_\circ(x^-,y)}{\varrho_\circ(x^+,y)}
\sqrt{\frac{x^-}{x^+}}
\frac{x^+-y}{x^--y}\;.
\eeqa
Combining~\eqref{eq:Mm-pt1},~\eqref{eq:Mm-pt2} and~\eqref{eq:Mm-pt3} we find the crossing equation for the full massive-massless dressing factor 
\beq
\[\sigma^{\bullet\circ}(x^{\bar\gamma_c},y)
\sigma^{\bullet\circ}(x,y)\]^2=
\frac{\left(x^--y\right) \left(y x^+-1\right)}{y^4 \left(y x^--1\right) \left(x^+-y\right)}\;.
\eeq
This agrees exactly with \cite{Frolov:2021fmj}; the older paper \cite{Borsato:2016xns} had a typo in the factors of $y$ on the rhs.

\subsection{Massless-Massive Crossing Relation} 
In the QSC approach there is no \textit{a priori} relation 
between massive-massless and massless-massive dressing factors and  
we have to find the crossing relation for the massless-massive 
dressing factor separately. This dressing factor is given in~\eq{def:sigmamM}, which we repeat here
\beqa
(\sigma^{\circ\bullet}(x,y))^2=
\sqrt{\frac{x y^+ - 1}{x-y^+}}
\sqrt{\frac{x y^- - 1}{x-y^-}}
\frac{\varsigma_{\bullet}^2(x,y)\;\;\varrho_\bullet(x,y)\;\;\dot\varrho_\bullet(x,y)}
{\varsigma_{\bullet}^2(x^{\bar\gamma},y)
\varrho_\bullet(x^{\bar\gamma},y)
\dot\varrho_\bullet(x^{\bar\gamma},y)
}\;.
\eeqa
For the massless part, according to \cite{Frolov:2021fmj} the crossing contour is simply $\gamma$ (i.e. going CW around the $-2h$ branch point). We emphasize that in the QSC there are no possible ambiguities in defining a preferred crossing direction, all relations controlling the curve are well-defined discontinuity equations. We need only be careful to use a specific crossing path when comparing with literature.

It is simple to check that the first term above is homogeneous under crossing. Similarly, the BES phase does not have a branch cut on the real axis, see the r.h.s. of \eq{varsigma-bullet-crossing}, and is thus also homogeneous under massless crossing
\beqa
\(\frac{\varsigma_{\bullet}^2(x,y)}
{\varsigma_{\bullet}^2(x^{\bar\gamma},y)}\)^\gamma  
\frac{\varsigma_{\bullet}^2(x,y)}{\varsigma_{\bullet}^2(x^{\bar\gamma},y)}
=1\;.
\eeqa
The only non-trivial contribution to crossing comes from the $\varrho_\bullet$ terms. Using~\eqref{rhobulletblocks} we find
\beqa
\(\frac{\varrho_\bullet(x,y)}{\varrho_\bullet(x^{\bar\gamma},y)}\)^\gamma
\frac{\dot\varrho_\bullet(x,y)}{\dot\varrho_\bullet(x^{\bar\gamma},y)}=
\(\frac{\dot\varrho_\bullet(x,y)}{\dot\varrho_\bullet(x^{\bar\gamma},y)}\)^\gamma
\frac{\varrho_\bullet(x,y)}{\varrho_\bullet(x^{\bar\gamma},y)}
&=&\sqrt{
\frac{(x-y^-)(1/x-y^+)}{(x-y^+)(1/x-y^-)}
}\,.
\eeqa
In summary, the full massless-massive crossing equation is
\beq
\[\sigma^{\circ\bullet}(x^{\gamma},y)\sigma^{\circ\bullet}(x,y)\]^2
=\frac{(x-y^-)(1/x-y^+)}{(x-y^+)(1/x-y^-)}\;,
\eeq
which perfectly agrees with \cite{Borsato:2016xns,Frolov:2021fmj}.

\subsection{Massless-Massless Crossing Relation}
The massless-massless dressing factor is given in~\eq{def:sigmamm}, which we repeat here
\beqa
(\sigma^{\circ\circ})^2(x,y)= x\frac{\varsigma_\circ^2(x,y)}{
    \varsigma_\circ^2(x^{\bar\gamma},y)
}    \frac{\varrho_\circ^2(x,y)}{\varrho_\circ^2(x^{\bar\gamma},y)}
\;.
\eeqa
The BES dressing factor is again homogeneous under massless crossing, for the same reasons as in the previous subsection, as is the factor of $x$.
Crossing for the $\varrho_\circ$  part follows from \eq{varrho-gamma-circ} and gives
\beqa
\(\frac{\varrho_\circ(x,y)}{\varrho_\circ(x^{\bar\gamma},y)}\)^\gamma
\frac{\varrho_\circ(x,y)}{\varrho_\circ(x^{\bar\gamma},y)}=\frac{xy-1}{y-x}
\eeqa
which agrees with~\cite{Borsato:2016kbm,Borsato:2016xns,Frolov:2021fmj}.

\section{Integral Representation and Solution for the Dressing Factors}\la{app:solving_crossing}
Here we discuss in more detail solutions to the discontinuity equations for the scalar factors $\varsigma$ and $\varrho$, their properties, uniquness and explicit presentations.
This Appendix is complementary to the discussion in Section~\ref{subsec:ExplicitDressing}.
\subsection{BES-like Phases}
We begin with $\varsigma_\bullet$ and $\varsigma_\circ$, for which in~\eq{eq:SigmaMassiveMain} and~\eq{eq:SigmaMasslessMain} we gave explicit integral representations by solving the discontinuity equations
\eq{varsigma-bullet-crossing} and
\eq{varrho-circ-cross}.
It is easy to check that \eq{eq:SigmaMassiveMain} solves  equation \eq{varsigma-bullet-crossing}. Let us show that this solution is unique. Indeed, assuming there is another solution we can consider their ratio, denoted $\varsigma_h$, which then has to satisfy the homogeneous equation $\varsigma_h(x^\gamma,y)\varsigma_h(x,y)\propto 1$. Since our solutions have to have neither zeroes nor poles for $|x|>1$ and which tend to $1$ at infinity we conclude that $\varsigma_h(x,y)$ is also analytic and has no zeroes inside the unit circle $\varsigma_h(x^\gamma,y)\propto 1/\varsigma_h(x,y)$. Furthermore, by applying $\gamma$ again and dividing by the initial homogeneous equation we get $\varsigma_h(x^{2\gamma},y)=\varsigma_h(x,y)$ (without a possible proportionality coefficient), implying that this function has a quadratic cut and thus is a rational function in the $x$ variable. Further, since it neither has poles nor zeroes and asymptotes to $1$ at large $x$, we conclude that we must have $\varsigma_h(x,y)=1$, which concludes the uniqueness of the initial solution to the non-homogeneous equation. Note that this means that the discontinuity equation \eq{varsigma-bullet-crossing}, obtained from QSC has a unique solution and thus is more constraining than the usual crossing equation which suffers from the freedom of adding CDD factors~\cite{Castillejo:1955ed}.  A similar argument applies for the uniqueness of $\varsigma_\circ$.

We next turn to fixing the non-square-root pieces of the dressing phases, which are the main novelty of the AdS$_3$ case as compared to AdS$_5$ and AdS$_4$~\cite{Borsato:2013qpa}.

\subsection{Fixing $\rho$ and $\dot\rho$ in terms of the building blocks}\la{sec:fixingKT}
In this part we show that $K_{\cal T}=0$ and that \eq{rhosplit} hold. Denoting $\rho_K = \frac{\rho}{\rho_\bullet \rho_\circ}$
and $\dot\rho_K = \frac{\dot \rho}{\dot\rho_\bullet \rho_\circ}$
from \eq{eq:lastRH} we get
\beq\label{eq:rhoK-rhodotK}
(\rho_K)^\gamma\dot \rho_K\propto x^{-K_{\cal T}}\,,\qquad\qquad
(\dot\rho_K)^\gamma \rho_K\propto x^{+K_{\cal T}}\;.
\eeq
The product of the above two equations gives
\beq\la{Kgamma}
(\dot\rho_K\rho_K)^\gamma\dot \rho_K\rho_K\propto 1\;.
\eeq
Applying $\bar\gamma$ and dividing by the above equation we have
\beq
(\dot\rho_K\rho_K)^\gamma=(\dot\rho_K\rho_K)^{\bar\gamma}\;.
\eeq
i.e. $\dot\rho_K\rho_K$ is a rational function of $x$.
Assuming it has no zeroes or poles for $|x|\geq 1$ from 
\eq{Kgamma} we see there are also no poles or singularities inside the unit circle. Since both $\rho$ and $\dot\rho$ are asymptotically $1$, we must have $\dot\rho_K\rho_K=1$. This, together with ~\eqref{eq:rhoK-rhodotK} then implies that
\beq
\frac{(\rho_K)^\gamma}{\rho_K}\propto x^{-K_{\cal T}}\;.
\eeq
Applying $\bar\gamma$ and computing the product we get
\beq
\frac{(\rho_K)^\gamma}{(\rho_K)^{\bar\gamma}}=c\;,
\eeq
for some constant $c$.
This again implies that $\rho_K$ is a rational function of $x$ up to a new possible factor $(\frac{x-1}{x+1})^{i\frac{\log c}{\pi}}$, which removes the constant $c$ on the r.h.s.
In fact, we will attribute this factor to $\rho_\bullet$ so we can assume, without reducing the generality, that $c=1$. In this case $\rho_K$ is a rational function of $K$ without poles or zeroes with constant asymptotics: in other words it is a constant. Thus we also see that we must have $K_{\cal T}=0$.

\subsection{Fixing $\varrho_{\bullet},\dot{\varrho}_{\bullet}$.}
In Section~\ref{sec:MasslessABALimit} the crossing relations for $\varrho_{\bullet}$ and $\dot{\varrho}_{\bullet}$ were found to satisfy~\eq{rhobulletblocks}. To solve these equations it is useful to consider the product and ratio of $\varrho$ and $\dot{\varrho}$. The product ${\varrho_p}\equiv
{\varrho_{\bullet}}{\dot\varrho_{\bullet}}$ satisfies
\beq
\varrho_p(x^\gamma,y){\varrho_p(x,y)}\propto\sqrt{\frac{x-y^-}{x-y^+}}
\sqrt{\frac{1/x-y^+}{1/x-y^-}}
\eeq
and the solution is given
\beq
\log \frac{\varrho_p}{c_{p}} = 
-\left(\IntUpperCC -\IntLowerCC\right)
\frac{1}{4} \left(\frac{1}{x-z}-\frac{1}{\frac{1}{x}-z}\right) \left(
    \log \left(\frac{z-y^-}{z-y^+}\right)
    -\log \left(\frac{\frac{1}{z}-y^-}{\frac{1}{z}-y^+}\right)
\right)
\frac{dz}{2\pi i}
\eeq
where $c_p(y)$ is again an irrelevant factor, which only depends on $y$. It can be found by requiring unit asymptotics for $x\to\infty$.
Here we assume that $|y|$ is sufficiently large so there is no ambiguity in the branch of $\log\(\frac{z^{\pm 1}-y^-}{z^{\pm 1}-y^+}\)$ by following the branch which tends to zero at large $|y|$. 
Below we give an explicit expression which takes care of the analytic continuation.
Uniqueness of the solution, follows by an almost identical argument to the one for $\sigma_{\bullet}$. For the ratio ${\varrho_r}\equiv
\frac{\varrho_{\bullet}}{\dot\varrho_{\bullet}}
$ we get the following equation\footnote{Writing $\log\varrho_r(x,y)=\chi_r(x,y^+)-\chi_r(x,y^-)$, $\chi_r(x,y)$ is often denoted as $\chi^-$ in the literature~\cite{Borsato:2013qpa}.}
\beq\la{varrhodiscr}
\frac{\varrho_r(x^\gamma,y)}{\varrho_r(x,y)}\propto\sqrt{\frac{x-y^-}{x-y^+}}
\sqrt{\frac{1/x-y^-}{1/x-y^+}}
\eeq
which is solved by
\begin{equation}
\begin{split}\la{rhorsol}
\log \varrho_r &= 
\left(\IntUpperCC -\IntLowerCC\right)
\frac{1}{4} \left(\frac{1}{x-z}+\frac{1}{\frac{1}{x}-z}\right) 
\left(
    \log \left(\frac{z-y^-}{z-y^+}\right)+
    \log \left(\frac{\frac{1}{z}-y^-}{\frac{1}{z}-y^+}\right)
\right)
\frac{dz}{2\pi i}\ \\
&+2\alpha(y) \log(\frac{x-1}{x+1})
\end{split}
\end{equation}
where the last term is added to show that the solution is no longer unique. One can see that the previous uniqueness argument fails because the discontinuity equation is of a ratio-rather than product- form. Indeed, the homogeneous equation has the form $\frac{\varrho_h(x^\gamma,y)}{\varrho_h(x,y)}=f(y)$, where the r.h.s. is an unknown function of $y$. This is because the discontinuity equation \eq{varrhodiscr} is only known up to such an x-independent multiplier, so we no longer have $\varrho_h(x^{2\gamma},y)=\varrho_h(x,y)$, meaning that $\varrho_h$ cannot be rationalized with Zhukovsky variables. At the same time, $\frac{\varrho_h(x^\gamma,y)}{\varrho_h(x,y)}=f(y)$ is solved by $\varrho_h(x^\gamma,y)=\(\frac{x-1}{x+1}\)^{i\log f(y)/\pi}$ and this solution is unique. Thus the last term in \eq{rhorsol} is due to our ignorance of the proportionality coefficient in \eq{varrhodiscr} and has to be fixed from additional requirements, such as unitarity of the dressing phases or cyclicity condition. 

One convenient way to fix the arbitrary function $\alpha(y)$ is to require that $\varrho_r$ has the same power-law divergence  near $x=1$ and $x=-1$. This detrmines $\alpha(v)= -\frac{1}{4\pi \ii}\log(\frac{(y^-)^2 - 1}{(y^+)^2-1})$ uniquely. 
One gets the same value of $\alpha(v)$
by requiring $\varrho_r(x=y^+)=\varrho_r(x=y^-)$, which is equivalent to requiring $\sigma^{\bullet\bullet}(x,x)=1$. In section~\ref{zeromodealpha} we also show that this value of the zero mode is needed to ensure the level matching condition. 

In principle, it should be possible to trace all proportionality coefficients in the discontinuity equations to this particular value of $\alpha$, but it is easier to use either $\sigma^{\bullet\bullet}(x,x)=1$ or the level matching condition to fix this ambiguity.

In what follows, we assume this is the correct value of $\alpha$. Finally, we get the following result (again valid for sufficiently large values of the parameter $|y|$).
Thus
\beqa
\log \frac{\varrho_\bullet}{\sqrt c_{p}} &=& 
+\left(\IntUpperCC -\IntLowerCC\right)
\frac{1}{4} \left(
\frac{1}{x-z}
\log \left(\frac{\frac{1}{z}-y^-}{\frac{1}{z}-y^+}\right)
+
\frac{1}{\frac{1}{x}-z}
\log \left(\frac{z-y^-}{z-y^+}\right)
\right) 
\frac{dz}{2\pi i}\\ \nn
&-&
\frac{1}{4\pi i}\log\frac{(y^-)^2-1}{(y^+)^2-1}
\log\frac{x-1}{x+1}
\;,
\eeqa
and
\beqa
\log \frac{\dot\varrho_\bullet}{\sqrt c_{p}} &=& 
-\left(\IntUpperCC -\IntLowerCC\right)
\frac{1}{4} \left(
\frac{1}{x-z}\log \left(\frac{z-y^-}{z-y^+}\right)+
\frac{1}{\frac{1}{x}-z}\log \left(\frac{\frac{1}{z}-y^-}{\frac{1}{z}-y^+}\right)
\right) 
\frac{dz}{2\pi i}
\\ \nn
&+&
\frac{1}{4\pi i}\log\frac{(y^-)^2-1}{(y^+)^2-1}
\log\frac{x-1}{x+1}\;.
\eeqa
Finally, by evaluating the above integrals and ensuring the correct analytic continuation in $y$ from sufficiently large $|y|$, we get \eq{polylogs1} and \eq{polylogs2}. The expressions
\eq{polylogs1} and \eq{polylogs2} are written in the form which makes it manifest that they have no cuts for $|x|,\;|y^+|,\;|y^-|>1$ and thus provide the correct definition for all relevant values, unlike the integral representation which is only valid for sufficiently large $|y|$.

\subsubsection{$\chi$ decomposition and $c_{r,s}$ expansion}

The discontinuity equations~\eqref{rhobulletblocks} imply that the crossing equations~\eq{eq:CrossingRho} and~\eq{eq:CrossingRho2} are the same as the existing literature (e.g. which is exactly the same as (5.12) in~\cite{OhlssonSax:2023qrk}, with $\propto$ replaced by $=$), 
it would be useful to rewire it using the notations of \cite{OhlssonSax:2023qrk}
using the $\chi$-functions defined in~\cite{OhlssonSax:2023qrk}. We find
\begin{equation}\label{eq:varrho-soln-chis}
    \varrho_\bullet(x,y)=e^{i\,\chi_{\sRL}(x,y^+)-i\,\chi_{\sRL}(x,y^-)}\,,\qquad
    \dot\varrho_\bullet(x,y)=e^{i\,\chi_{\sLL}(x,y^+)-i\,\chi_{\sLL}(x,y^-)}\,,
\end{equation}
where $\chi_{\sLL}$ and $\chi_{\sRL}$ are defined in equation (5.16), or equivalently (5.17) and (5.18) of~\cite{OhlssonSax:2023qrk}, and include the extra ``non-integral" terms, first proposed by Andrea Cavagli\`a and one of us (SE) to remove the unwanted log branch-cuts from~\cite{Borsato:2013hoa}. In order to compare these integral expressions to~\eqref{polylogs1} and~\eqref{polylogs2} we note that for $|x|,\,\,|y|>1$
\begin{align}
    \chi_{\sLL}(x,y)&=-
    \frac{1}{2 \pi} \left[\text{Li}_2\left(\frac{2 (y-x)}{(x+1) (y-1)}\right)-\text{Li}_2\left(\frac{2}{x+1}\right)-\text{Li}_2\left(\frac{2}{1-y}\right)\right.
    \nonumber \\ & \qquad \qquad\qquad\left.
    +\frac{1}{2} \log \left(\frac{x-1}{x+1}\right) \log \left(\frac{y+1}{y-1}\right)\right]
    \\
    \chi_{\sRL}(x,y)&=\frac{1}{2 \pi } \left[-\text{Li}_2\left(\frac{(1-x) (y+1)}{(x+1) (y-1)}\right)+\text{Li}_2\left(\frac{1-x}{1+x}\right)+\text{Li}_2\left(\frac{1+y}{1-y}\right)+\text{Li}_2\left(\frac{1}{y}\right)\right.
    \nonumber \\ & \qquad \qquad
    -\text{Li}_2\left(-\frac{1}{y}\right)-\log \left(\frac{y-1}{y+1}\right) \log \left(\frac{x+1}{x-\frac{1}{y}}\right)
    \nonumber \\ & \qquad \qquad\left.
    +\frac{1}{2} \log \left(\frac{x-1}{x+1}\right) \left(\log \left(1-\frac{1}{y^2}\right)-2 \log \left(1-\frac{1}{xy}\right)\right)\right]+\frac{\pi }{24}\,.
\end{align}
The somewhat baroque combinations of logs and dilogs above are needed to ensure that all cuts are inside the unit $x$ and $y$ circles. Terms that depend only on $x$ cancel out in $\varrho_\bullet$ and $\dot\varrho_\bullet$, while those that depend only on $y$ do not play a role in the continuity equations since the latter  are defined up to $y$-dependent terms. With these points one can verify that~\eqref{eq:varrho-soln-chis} is equivalent to~\eqref{polylogs1} and~\eqref{polylogs2}. The large $x$ and $y$ expansion of these phases, encoded in the so-called $c_{r,s}$ coefficients, is given in (6.9) and (6.10) of~\cite{OhlssonSax:2023qrk}, upon setting $s=1$ there. As we show in Section~\ref{FSphasesMM}, those expressions are also equivalent to the expressions in \cite{Frolov:2021fmj}.

\subsection{Fixing $\varrho_{\circ}$}
Finally, let us solve equation \eq{varrho-gamma-circ}. Using the Sochocki–Plemelj theorem, one picks up a pole when crossing the integration contours, and so it is simple to check the following integral representation is a solution to \eq{varrho-gamma-circ}.
\beq
\log \frac{\varrho^2_\circ(x,y)}{c_\varrho(y)}=\left(\IntUpperCC -\IntLowerCC\right)
\left(\frac{1}{x-z}-\frac{1}{1/x-z}\right) \left(\log \left(1-\frac{y}{z}\right)+\frac{\log z}{2}\right)
\frac{dz}{2\pi i}
+\log \left(\frac{y-\frac{1}{x}}{\sqrt{y}}\right)+\frac{i \pi }{2}
\eeq
with the prescription that $|x|>1$ and $|y|=1$ and the contours are around arches of a circle of the radius slightly  
bigger than $1$,
so that the integration does not encounter any cuts or singularities along the contour of integration.
Here $c_\varrho(y)$ is an (in general irrelevant)  
real constant that can be fixed by the requirement that $\varrho_\circ(x,y)\to 1$ at $x\to\infty$  
\beq  
c_\varrho(y)=\frac{e^{\frac{i \text{Li}_2(-y)}{\pi }-\frac{i \text{Li}_2(y)}{\pi }-\frac{3 i \pi }{4}}}{\sqrt{y}}\;,
\eeq
which is real for $|y|=1$ and ${\rm Im}\;y\geq 0$.
The integrals above can be computed analytically to get \eq{polylog3}.

The uniqueness of this solution is more subtle. We see
from the r.h.s. of equation \eq{varrho-gamma-circ} that we have to allow for singularities at $x=y$ and $x=1/y$. As a result, the homogeneous solution can be a rational function of the form $\varrho_{\circ h}^2(x,y)=(\frac{x-y}{x-1/y})^n$, which satisfies $\varrho_{\circ h}^2(x,y)\varrho_{\circ h}^2(x^\gamma,y)=1$; however, it is easy to see that for $|y|=1$ this function is not real and thus would violate the reality of the initial solution.

\subsection{Fixing Zero-Mode from Bethe Equations}
\la{zeromodealpha}
In this subsection we present an alternative way of fixing $\alpha(v)$ in equation \eqref{rhorsol}. The idea is to take the product of all Bethe equations and use momentum conservation. For simplicity, we  perform this exercise with only $u_{\dot{2},k},u_{2,k}$ present, that is we will assume there to be no massless excitations nor any auxiliary excitations. With this restriction, we have two Bethe equations: \eqref{eq:L-mom-carry-BAE-from-QSC} and \eqref{eq:ABARQSC}. They become
\begin{equation}
    -\left(\frac{x^+_{2,k}}{x^-_{2,k}}\right)^{L} = \frac{\betheQ^{++}_2}{\betheQ^{--}_2}\frac{\bfB^+_{\dot{2},(-)}\bfB^+_{\dot{2},(+)}}{\bfB^{-}_{\dot{2},(-)}\bfB^{-}_{\dot{2},(+)}} \left(\frac{\sigma^+_{\bullet}}{\sigma^-_{\bullet}}\frac{\hat{\rho}^+_{\bullet}}{\hat{\rho}^+_{\bullet}}\right)^2\bigg|_{x=x_{2,k}}
\end{equation}
and 
\begin{equation}
    -\left(\frac{x^+_{\dot{2},k}}{x^-_{\dot{2},k}}\right)^{L} = \frac{\bfB^+_{\dot{2},(+)} \bfR^{-}_{\dot{2},(-)}}{\bfB^-_{\dot{2},(-)}\bfR^+_{\dot{2},(+)}}
    \frac{\bfB^+_{2,(+)} \bfB^{-}_{2,(+)}}{\bfB^+_{2,(-)}\bfB^-_{2,(-)}}
\left(\frac{\sigma^+_{\bullet}}{\sigma^-_{\bullet}}\frac{\hat{\dot{\rho}}^+_{\bullet}}{\hat{\dot{\rho}}^-_{\bullet}}\right)^2\bigg|_{x=x_{\dot{2},k}}
\end{equation}
where we have put additional hats over $\rho$ to emphasize that these are not yet the $\rho$s of the main text. 
Let us take the product over all root $u_{2,k},u_{\dot{2},k}$ and then multiply the two equations above. We use the identities
\begin{equation}
\begin{split}
    &\prod_{j=1}^{K_{2}} \frac{\betheQ^{++}_{2}(u_{2,j})}{\betheQ^{--}_{2}(u_{2,j})} = (-1)^{K_{2}}\,, 
    \qquad
    \prod_{j=1}^{K_{\dot{2}}} \frac{\bfB^+_{\dot{2},(+)}(u_{\dot{2},j})\bfR^-_{\dot{2},(-)}(u_{\dot{2},j})}{\bfB^-_{\dot{2},(-)}(u_{\dot{2},j})\bfR^+_{\dot{2},(+)}(u_{\dot{2},j})} = (-1)^{K_{\dot{2}}}\,,
    \\
    &\prod_{j=1}^{K_2} \bfB^{\pm_1}_{\dot{2},(\pm_2)}(u_{2,j}) = \prod_{j=1}^{K_{\dot{2}}}  \bfB^{\mp_2}_{2,(\mp_1)}(u_{\dot{2},j})\,,
\end{split}
\end{equation}
and momentum conservation, 
\begin{equation}
    \prod_{k=1}^{K_2} \frac{x^+_{2,k}}{x^-_{2,k}}\prod_{k=1}^{K_{\dot{2}}}\frac{x^{+}_{\dot{2},k}}{x^-_{\dot{2},k}}=1\,,
\end{equation}
to obtain
\begin{equation}
    1 = \prod_{j=1}^{K_2}\left( \frac{\sigma^+_{\bullet}}{\sigma^-_{\bullet}}\frac{\hat{\rho}^+_{\bullet}}{\hat{\rho}^-_{\bullet}}\right)^2\bigg|_{x=x_{2,k}} \prod_{j=1}^{K_{\dot{2}}} \left(\frac{\sigma^+_{\bullet}}{\sigma^{-}_{\bullet}}\frac{\hat{\dot{\rho}}^+_{\bullet}}{\hat{\dot{\rho}}^-_{\bullet}} \right)^2\bigg|_{x=x_{\dot{2},k}}\,.
\end{equation}
Let us reintroducing the ambiguity that arises from only keeping proportionalities in the "half"-crossing equations by writing
\begin{equation}
    \hat{\rho}_{\bullet} = \rho_{\bullet} \times \left(\frac{x-1}{x+1}\right)^{\tilde{\alpha}}\,,
    \qquad\qquad
    \hat{\dot{\rho}}_{\bullet} = \dot{\rho}_{\bullet}\times \left(\frac{x+1}{x-1} \right)^{\tilde{\alpha}}\,.
\end{equation}
where $\tilde{\alpha}$ is any function of all Bethe roots. We already established unitarity of $\rho_{\bullet}$ and $\dot{\rho}_{\bullet}$ as well as the BES factor in Section~\ref{sec:ExplicitDressing}. It then follows that  
\begin{equation}
    1 = \left(\left(\prod_{k=1}^{K_{2}} \frac{x^+_{2,k}-1}{x^+_{2,k}+1}\frac{x^-_{2,k}+1}{x^-_{2,k}-1}\right)\left(\prod_{k=1}^{K_{\dot{2}}} \frac{x^+_{\dot{2},k}+1}{x^+_{\dot{2},k}-1}\frac{x^-_{\dot{2},k}-1}{x^-_{\dot{2},k}+1}\right)\right)^{2\tilde{\alpha}}\,.
\end{equation}
For example, at weak coupling the two products are not equal to one for generic states and furthermore they behave as $1+{\cal O}(g)$; thus we must set $\tilde{\alpha}=0$. For suitably fine-tuned states one cannot exclude the possibility that the expression in the large round brackets is $1$; however, one can deform continuously by introducing a twist and then by continuity we must have $\tilde\alpha=0$ for all states. This concludes the exercise. 

\section{Constraining $\zeta$}\label{app:FixZeta}
In this appendix we constrain $\zeta$ introduced in \eqref{omega12i}. Recall that $\zeta$ appears in the massless Bethe equations~\eqref{masslessmiddleV2}. This allows us to take the product of all Bethe equations for all roots and use momentum conservation, 
\begin{equation}
    \prod_{k=1}^{K_{\circ}}z_k^2\prod_{k=1}^{K_{2}}\frac{x^+_{2,k}}{x^-_{2,k}}\prod_{k=1}^{K_{\dot{2}}}\frac{x^+_{\dot{2},k}}{x^-_{\dot{2},k}} = 1
\end{equation}
to constrain $\zeta$. It turns out that the full calculation is equivalent to simply considering the case with massless excitations without any massive or auxiliary excitations. The purely massless Bethe equations are
\begin{equation}
\begin{split}
    \pm\zeta z_k^{2L} = &\prod_{j=1}^{K_{\circ}}\left(-\frac{z_k}{z_j} \left(\sigma^{\circ\circ}\right)^2(z_k,z_j) \right)\,.
\end{split}
\end{equation}
Taking their product over all roots $z_k$ and using $\prod_{k=1}^{K_{\circ}} z_k^{2} = 1$ gives
\begin{equation}
    \left(\pm \zeta\right)^{K_{\circ}}= \prod_{j,k=1}^{K_{\circ}} \left(\sigma^{\circ \circ} \right)^{2}(z_{j},z_k) = (-\ii)^{K_{\circ}}(-1)^{\frac{K_{\circ}(K_{\circ}-1)}{2}}\,.
\end{equation}
From this we get $\zeta = \pm \left(-\ii\right)^{K_{\circ}} e^{2\pi i n/K_\circ}$, for some $n=0,1,\dots,K_\circ-1$.

\section{Exploring the possibility of including an $a(\gamma)$ factor}\la{appendix:rhoa}
In this appendix, we explore the possibility that our $\varrho_\circ$ function has to be modified to accommodate the extra $a(\gamma)$ factor in the massless-massless dressing phase. If such a factor were to be included, the modified $\varrho_\circ$ would in turn satisfy a different discontinuity relation to \eq{varrho-gamma-circ}. Nevertheless, let us consider $\varrho_\circ\to\varrho_\circ\varrho_a$ such that 
\beq\la{rhoacross}
\(\frac{\varrho_a(x,y)}{\varrho_a(x^{\bar\gamma},y)}\)^2=\pm i a(x,y)\;.
\eeq
First, notice that, since $a(1/x,y)=-1/a(x,y)$, we should have
\beq
\(\frac{\varrho_a(x^\gamma,y)}{\varrho_a(x,y)}\)^2=\mp\frac{i}{a(x,y)}\,  
\eeq  
which, in turn, implies that
\beq
\(\frac{\varrho_a(x,y)}{\varrho_a(x^{\bar\gamma},y)}\)^2
\(\frac{\varrho_a(x^\gamma,y)}{\varrho_a(x,y)}\)^2
=1
\eeq
so $\varrho_a(x,y)^2$ has a quadratic cut and thus is a rational function of $x$ and $y$. Since we require $\varrho_a(x,y)$ to have no zeros or poles on the main sheet, we can construct the following solution of \eq{rhoacross}
\beq\la{rhoasol}
\varrho^2_a(x,y)\propto \frac{i (x-y)+x y-1}{\sqrt{x}}
\eeq
which for physical values of $y$ i.e. $|y|=1$ and $0<{\rm arg}\; y< \pi $ has a zero inside the unit circle at $x=\frac{1+i y}{y+i}$. One can easily prove that the solution is unique up to a constant factor -- the homogeneous solution has to be an analytic function of $u$ with no poles or zeros i.e. a constant.

 There are two immediate problems with the $\varrho^2_a(x,y)$ factor \eq{rhoasol} within the ABA limit of the QSC. Firstly, by definition, $\varrho_\circ$ should go to $1$ at $x\to\infty$, which does not hold for \eq{rhoasol}. Secondly, $\varrho_a^\gamma$ has a zero outside the unit circle, which contradicts the definition of the factors of $R$ and $B$ in the ansatz for $\bP_a$ in \eq{eq:Q-ide2} and \eq{eq:Qd-ide3}, which should already contain all zeros.

Furthermore, including such a term in $\varrho_\circ$ would lead to an even more serious discrepancy in the massive-massless dressing phase \eq{sigmaMm_def}, which would become
\beq
\left(\sigma^{\bullet\circ}(x,y)\right)^2\to
\left(\sigma^{\bullet\circ}(x,y)\right)^2\[
\frac{\sqrt{x^-} \left(x^+ y+i x^+-i y-1\right)}{\sqrt{x^+} \left(x^- y+i x^--i y-1\right)}
\]\,.
\eeq
In turn, under crossing, this would produce an extra factor 
$\frac{-2 u_x u_y-i u_y+8}{-2 u_x u_y+i u_y+8}$ on the right-hand side of the crossing equations. Finally, the unitarity relation between
$\sigma^{\bullet\circ}$ with $\sigma^{\circ\bullet}$ would be destroyed by this additional factor, since $\sigma^{\circ\bullet}$ is expressed in terms of $\varrho_\bullet$ in~\eqref{def:sigmamM} and so remains unchanged by this modification.

\newpage

\input{main.bbl}

\end{document}